\documentclass[useAMS,usenatbib,usegraphicx]{mn2e}

\usepackage{amssymb}
\usepackage{graphicx}
\usepackage{wasysym}
\usepackage{pifont}
\newcommand{\sio}{\sigma_{\textrm{\tiny{obs}}}}
\newcommand{\sil}{\sigma_{\textrm{\tiny{lens}}}}
\newcommand{\sih}{\sigma_{\textrm{\tiny{h}}}}
\newcommand{\re}{R_{\textrm{\tiny{eff}}}}
\newcommand{\rl}{R_{\textrm{\tiny{lens}}}}
\newcommand{\rein}{R_{\textrm{\tiny{Ein}}}}
\newcommand{\ml}{M_{\textrm{\tiny{lens}}}}
\newcommand{\mv}{M_{\textrm{\tiny{vir}}}}

\title[A lensing view on the fundamental plane]{A lensing view on the fundamental plane}
\author[Dominik Leier]
{Dominik Leier$^{1}$,$^{2}$\thanks{Electronic address: \tt leier@physik.uzh.ch} \\
$^{1}$Institute for Theoretical Physics, University of Z\"urich, Winterthurerstrasse 190, 8057 Z\"urich, Switzerland \\
$^{2}$Astronomisches Rechen-Institut, Zentrum f\"ur Astronomie der Universit\"at Heidelberg,\\
{} M\"onchhofstrasse 12-14, 69120 Heidelberg, Germany}

\begin{document}
\date{Accepted for publication in MNRAS, 4 August 2009}

\pagerange{\pageref{firstpage}--\pageref{lastpage}} \pubyear{2009}

\label{firstpage}

\maketitle

\begin{abstract}
For lensing galaxies we introduce a formal velocity dispersion $\sil$, based on enclosed mass and the virial theorem. This is calculated from an ensemble of pixelated lens models, and found to be fairly model independent. A sample of 18 well-known early-type lensing galaxies and two clusters is found to be consistent with $\sil=\sio$. Both the early-type lensing galaxies and the clusters can thus be determined as being virialized.  In a second step we calculate the I-band luminosity and the total mass content for the sample of lensing galaxies, which enables us to analyze the mass-to-light relation $L\propto M^\alpha$. We determine $\alpha=0.70\pm 0.08$, excluding constant $M/L$ and consistent with previous studies of the Fundamental Plane. Additionally we verify that this relation does not extrapolate to clusters, which have a much higher $M/L$. The sample used for this analysis comprises 9 lensing galaxies from the Sloan Lens ACS Survey (SLACS) and another 9 from the CfA-Arizona Space Telescope LEns Survey (CASTLES) as well as the lensing clusters ACO 1689 and ACO 2667.
\end{abstract}
%\pacs{98.62.Sb, 98.62.Ve, 95.35.+d, 98.52.Eh}


\section{Introduction}

The Fundamental Plane (FP) for early type galaxies is a well-known scaling relation between the effective radius $\re$, the kinematic velocity dispersion $\sio$ and the surface brightness $I_{<\re}$ \citep{dj87,dr87}. Not so well understood is the mismatch between theoretical predictions for the FP on one hand and observations on the other. Combining the virial theorem 
\begin{equation}
M \propto \re \sigma^2
\label{eq:vt1}
\end{equation}
and the universality of light profiles $L \propto \re^2 I$ while assuming a constant mass-to-light ratio yields
\begin{equation}
\re \propto \sigma^2 I^{-1},
\label{eq:3}
\end{equation}
which we name the Vanilla Plane.
In contrast observations show a relation with slightly different power indices $a$ and $b$, as in
\begin{equation}
\re \propto \sigma^a I^{b},
\label{eq:4}
\end{equation}
with $a\approx 1.2$ and $b\approx -0.8$ \citep[e.g.,][]{jo96}.
The power indices are thus not in agreement with the Vanilla Plane indices $(a,b)=(2,-1)$ for constant $M/L$, which is suggestive of an underlying regularity beyond the above formulas.\\

With gravitational lensing as an independent measure of mass one can suitably analyse the structure of the FP as already done in different approaches. \citet{bo07} linked lensing mass and virial mass, whereas \citet{ru03b}, \citet{tr06}, \citet{ji07} (hereafter [JK07]) and \citet{fe08} analyzed the mass-to-light dependence
\begin{equation}
M^\alpha \propto L,
\label{eq:m2l}
\end{equation}
which is a representation of the FP. 

By repeating the step from Equation (\ref{eq:vt1}) to (\ref{eq:3}) for the more general definition of the mass-to-light relation in Equation (\ref{eq:m2l}), $a$ and $b$ in Equation (\ref{eq:4}) can be expressed in terms of the power index $\alpha$, which is what we need to compare the results with previous FP type studies (listed in Table \ref{tab2}). Equating the now $\alpha$-dependent exponents of $\sigma$ and $I$ yields
\begin{equation}
a(b)=-2(1+2b),
\label{eq:ab}
\end{equation}
which only applies for a not unique mapping from $(a,b)$ to $\alpha$ assuming Equation (\ref{eq:m2l}).\\

In this study we combine both the virial approach and considerations including luminosities by means of lensing masses $\ml$ from 18 early type lensing galaxies and 2 clusters discussed in detail in Section \ref{sec:2}. 

An important role is played by a formal velocity dispersion, which we define as 
\begin{equation}
\sil (R) = \sqrt{\frac{2}{3}\frac{GM(<R)}{\pi R}}.
\label{eq:1}
\end{equation}
For an isothermal sphere this is exactly equal to a line-of-sight velocity dispersion.
A short introduction to the determination of $\sil(R)$ with the mass reconstruction method PixeLens and a detailed description of the lensing sample (see Table \ref{tab1}) will be presented in Section \ref{sec:2}.
We define $\ml$ and $\mv$ by comparison with Equation (\ref{eq:vt1}) as:
\begin{equation}
 \ml=\frac{3\pi}{2G}\re\sil^2
\end{equation}
and $\mv=\frac{3\pi}{2G}\re\sio^2$.\\

This in hand we consider the following questions as a rephrased puzzle of FP:\\
\begin{enumerate}
\renewcommand{\theenumi}{(\arabic{enumi})}
\item Is the lensing inferred velocity dispersion $\sil$ from nonparametric mass reconstruction equal to the kinematic velocity dispersion $\sio$?
\item Is this applicable to cluster scale lensing objects?
\item Are the computed $\ml$ and $\mv$ consistent with the FP?
\item Does the FP relation extend to clusters?
\end{enumerate}

Bearing this in mind we want to give a short overview of previous findings.\\
In the above mentioned approach by \citet{bo07} to the FP problem, they find that $\sil\approx \sio$ without taking advantage of luminosities (see also \citet{bo08}). This result is comparable to the findings presented in Section \ref{sec:svs} of this paper. Furthermore \citet{bo07} take no baryonic information into account, but a different plane is introduced, which emerges from a dimensional change in the FP space from surface brightness $I$ to surface density $\Sigma$, giving
\begin{displaymath}
\re \propto \sigma^{a_m} \Sigma^{b_m}.
\end{displaymath}
But in fact this scaling relation, named a more fundamental or mass plane (MP), can be transformed into the shape $M^{-b_m}/\re^{-(1+2b_m)} \propto \sigma^{a_m}$ which is consistent with our theoretical assumptions of Equation \ref{eq:1} and thus represents the virial theorem. We like to point out that the change to $\Sigma$ introduces basically a redshift dependence, which comes along with a grave selection effect that reduces the significance of the scaling relation. Moreover this relation is compared to the existing FP by introducing a new parameterization of the lensing mass
\begin{displaymath}
M_L = c (G^{-1} \sio^2 (\re/2))^\delta,
\end{displaymath}
where $c$ denotes a structure constant and $\delta$ a newly introduced power index. \citet{bo07} find that by doing so "the tilt relative to the virial relation is essentially eliminated". But basically $c$ and $\delta$ are again consistent with Equation \ref{eq:1}. Upon choosing $\delta=1$, $c$ becomes $\log{3\pi/2}$ and the new parameterization in \citep{bo07} turns into a test of the virial theorem.\\

Thus the decreased scatter for a MP is rather a natural consequence of the added fitting parameter and selection effects than a more fundamental scaling relation. Implications on structure variations are hardly possible. An appropriate treatment on the search for reasons for a tilt in the FP originating in certain structural peculiarities includes more elaborate approaches that allow for a distinction between for instance anisotropy and mass-dynamical structure.\\

\citet{ru03b} introduced a self-similar mass model for early-type galaxies, consisting of two components: a concentrated component, which traces the light distribution and a more extended power-law component, which represents the dark matter. They found a strong $r^{-2}$ dominance and therefore used the velocity dispersion $\sigma_{iso}$ for an isothermal model as a surrogate in the FP yielding a mass-to-light relation of $\ml^{0.88^{+0.10}_{-0.11}}\propto L$, which was the first such result from strong lensing. The error of the slope already excluded a constant $M/L$. While substituting $\sigma_{iso}$ they are effectively assuming the virial theorem.\\

[JK07] constrain the average stellar mass fraction of a halo in favor of adiabatically compressed halo models by taking a sample of early type galaxies which consists partially of lensing galaxies used in this sample. By means of a two-component model stellar and virial mass are fitted separately and an isothermal density profile is assumed. The paper takes advantage of already K-corrected B-band magnitudes and lensing masses and is, because of its common subset of gravitational lenses, directly comparable with our data. Although it is not explicitly calculated in their paper by taking their data we found $\ml^{0.88\pm0.12} \propto L$, which is in perfect agreement with the result from \citet{ru03b}.\\

Switching from the lensing point of view considering $\ml$ to the observational one considering $\mv$  enables one to compare the FP from previous studies which were inferred from lensing with the FP based on stellar dynamics. \citet{tr06} analyze the FP by means of virial mass and find that the velocity dispersions for their SDSS lens sample are well approximated by $\sigma_{iso}$, which holds also for our mixed CASTLES/SLACS sample.\\

There are studies, e.g. from \citet{gr97} and \citet{tb04}, which raised hope that a solution for the FP tilt is at least partially given by broken structural homology leading to strong correlations between S\'ersic index $n$ and photometric-independent galaxy properties. \citet{gr97} fit $R^{1/n}$ profiles and make use of the spatial velocity dispersion at spatial effective radius to show the influence of structural non-homology, whereas \citet{tb04} quantify the contribution to the tilt caused by variations of $n$ for a wide range of B-band selected early-type galaxies. The results always show that taking account of non-homology shifts the FP parameters closer but never fully matches the virial expectations. However by comparing the $M/L-\sigma$ relation of 25 E$/$S0 galaxies from the SAURON sample with predictions and virial estimates \citet{ca06} find that the FP tilt is exclusively due to a real $M/L$-variation, while structural and orbital non-homology has a negligible effect, a result also verified in this study.\\

Furthermore, progress in estimating $M_{tot}/M_{stel}$ was recently made by comparing stellar population models with the nonparametric mass profiles also used in this paper, which allow for scanning the dark matter distribution within a galaxy \citep{fe05,fe08}. They found that low-mass galaxies have only little dark matter content at all observed radii. On the contrary high-mass galaxies have little or no dark matter inside the effective radius but at large radii they are clearly dark matter dominated. No kinematic and virial assumptions were required.\\

In this paper we are using combination of kinematic, photometric and lensing inferred data to answer the aforementioned puzzle. The comparison of $\sio$ and $\sil$, which is proportional to a comparison between virial mass and lens mass, is adequate for answering the above questions 1 and 2 as will be shown in Section \ref{sec:svs}. In Section \ref{sec:m2l} we compute the luminosities of the lenses, $\ml$ and $\mv$ needed for $L \propto M^\alpha$ and check the consistency with other FP studies in particular with data from [JK07], who use a common subset of lens systems. Subsequently the a-b-parameter plane is generated including a wide range of recent FP studies. In Section \ref{sec:5} the conclusions are presented.

\section{Lenses and Lens Models}
\label{sec:2}

In this section we introduce the lensing sample and subsequently discuss the lens modelling.\\
Our sample consists of 9 lensing galaxies from SLACS\footnote{www.slacs.org - The full set includes about 70 lenses, but image data was only made available for a small subset.} data, 9 from CASTLES\footnote{cfa-www.harvard.edu$/$glensdata$/$} and 2 lensing clusters. We select these galaxy lenses using two criteria:

\begin{enumerate}
\renewcommand{\theenumi}{(\arabic{enumi})}
\item the lensed images were either point-like sources or contain nearly point-like features, and
\item the availability of $\sio$ data.
\end{enumerate}

Two cluster lenses with such properties are also included for comparison and contrast, since previous to this paper FP studies were carried out for small scale and large scale objects combined \citep[e.g.,][]{sch93}. As an additional motivation it is worth mentioning that \citet{zw37} originally introduced gravitational lensing as a method to estimate masses of galaxy clusters.\\

The 9 CASTLES lenses turn out to be a relatively inhomogeneous sample, a consequence of the fact that they spread over a large range in redshift and effective radii as well as lens radii.\\

The doubly imaged systems among the CASTLES lenses are \textbf{CFRS03.1077}, \textbf{HST15433} and \textbf{MG2016+112}. The effective radius of CFRS03.1077 is not known; hence it is used for the analysis in Section \ref{sec:svs} but not Section \ref{sec:m2l}. In HST15433 there is a neighboring galaxy, but this is thought to only modestly perturb the estimated mass, according to [JK07].\\

\begin{figure}
\begin{center}
\begin{minipage}[l]{0.206\textwidth}
\includegraphics[width=95pt, bb=10 -76 320 320]{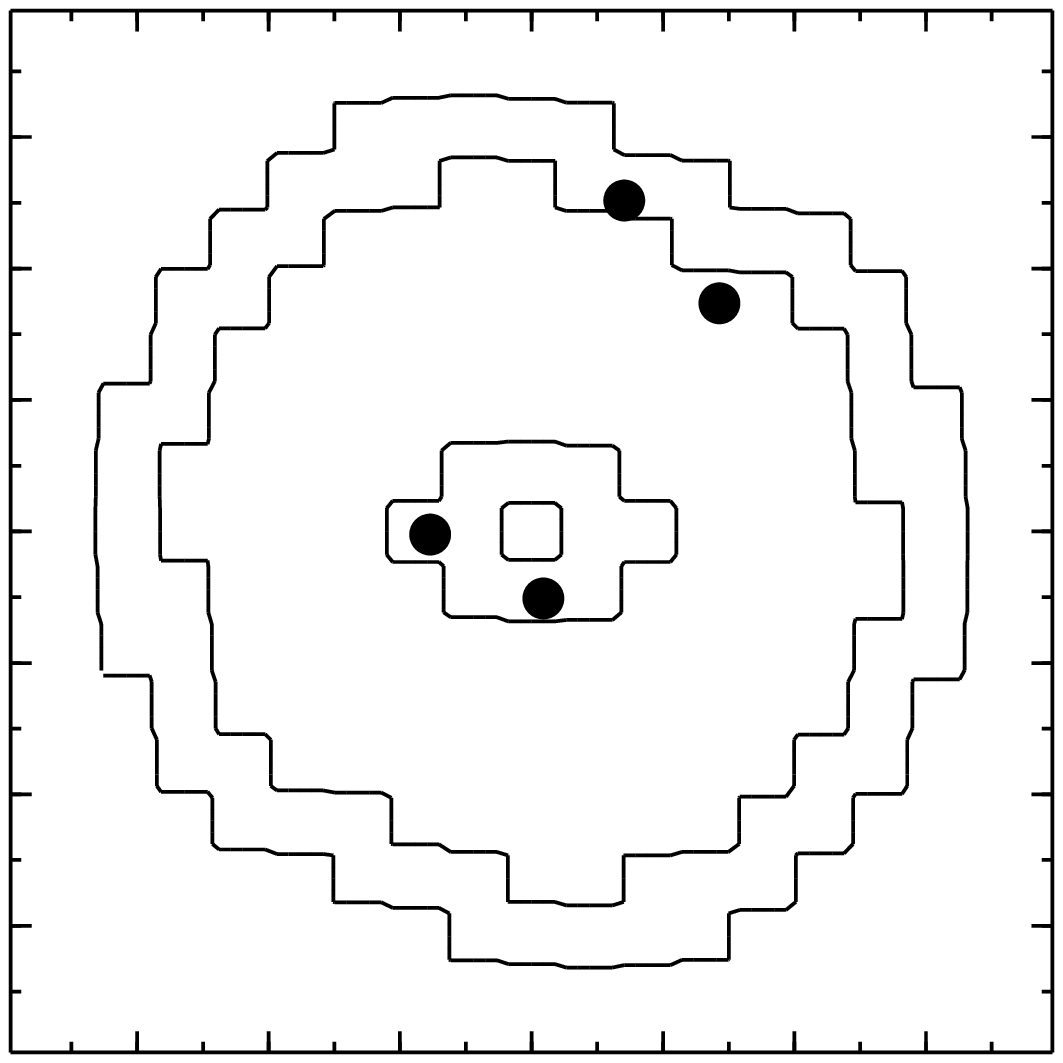} 
\end{minipage}\begin{minipage}[l]{0.9\textwidth}
\includegraphics[width=140pt]{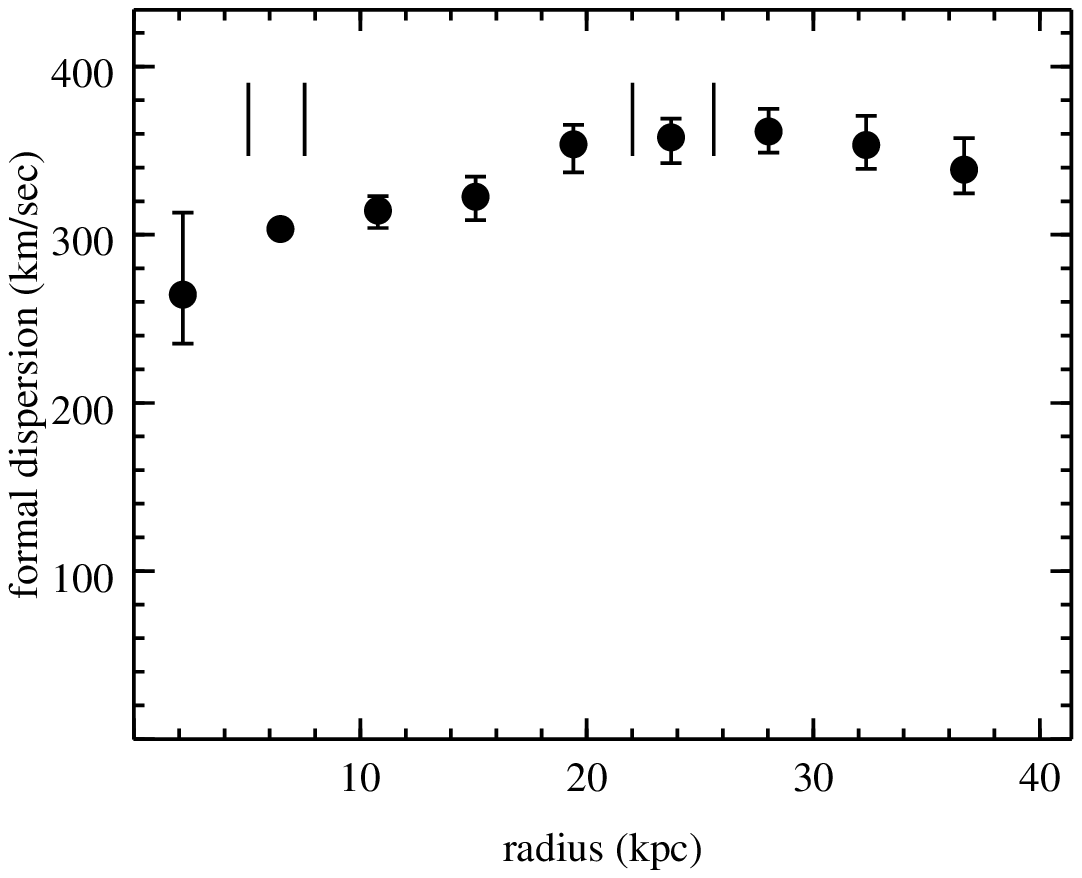}\\ 
\end{minipage}
\end{center}
\caption{Left panel: projected mass distribution of the CASTLES lens Q0957+561. The box size is $4 \times 4$ arcsec$^2$. The dots mark two doubly imaged systems. Right panel: formal velocity dispersion $\sil$. The vertical dashes mark the radial position of the lensed images. The same curve can be seen as second from top in Fig. \ref{vd}.}
\label{fig1}
\end{figure}

\textbf{Q0957+561} is a special case, as there is a doubly imaged galaxy component in addition to the famous double quasar. The lensing galaxy is part of a cluster that contributes significantly to the large image separation \citep{ga92}. Consequently a position below the general trend in a mass-to-light analysis is expected. This lens is an excellent example for the consequences of possibly yet unknown image systems. The considerations following in Section \ref{sec:svs} for $\sil$-$\sio$ are carried out for two different image configurations: on one hand the 2 double image system shown in Figure \ref{fig1} and on the other hand a single double system.\\

The quads from the CASTLES catalogue are \textbf{B0047-2808}, \textbf{PG1115+080}, \textbf{HST14176}, \textbf{B1608+656} and \textbf{Q2237+030}.
B0047-2808 appears to have a double component source but this is probably not important for macro models. PG1115+080 has measured time delays and it is also part of a group, which contributes with a significant external shear. Neither of these were used for the models of this paper; if these are included the lens models tend to become rounder, but $\sil$ changes only by $5\%$ to $10\%$, which is insignificant for the present study. For the complex lens B1608+656 the measured time delays are used and make a more significant difference to the lens models. The lens Q2237+030 is actually the bulge of a barred galaxy. In the present study the bulge is treated as an early type galaxy. HST14176 is a part of a cluster, which is not included in the models. Another possible problem is a large uncertainty in the effective radius.\\

Several of these lenses have been studied individually in great detail. Different papers sometimes disagree on the slope of the profile \citep[e.g.,][for PG1115+080]{tk02b,re07}, but agree on the enclosed mass. Hence the effect on $\sil$ would be small.\\

SLACS lenses populate a redshift range from $0.05$ to $0.5$. Due to smaller mean effective radii and lens redshifts, as a consequence of a limited aperture ($3''$) radius of SDSS fibers, the sample of 9 SLACS lenses appears to be more clustered in mass-to-light plots than the CASTLES sample and has therefore a smaller RMS. The SLACS lenses we use are a subset of the full SLACS sample for which point-like features are identified \citep{fe08}. \\ 

The doubly imaged systems among the SLACS lenses are \textbf{J0037-094}, \textbf{J0912+002}, \textbf{J1330-014} and \textbf{J2303+142}. 
All of them are quite typical with biases according to their observation method. J1330-014 is the nearest lens with $z=0.08$ and shows the smallest $\sio$.
Among these lenses J0912+002 takes a special position. It consists of two long arcs which are represented in this work as four doubles. Moreover, this lensing galaxy has the highest $\sio$ among all lenses of our sample.\\

Quadruply imaged systems are found for the SLACS lenses \textbf{J1205+491}, \textbf{J1636+470} and \textbf{J2300+002}.
Apart from J1636+470 all SLACS lenses have a larger kinematic velocity dispersion than lens velocity dispersion. The mean kinematic velocity dispersion $\langle \sio \rangle$ of our SLACS sample is $10 \%$ higher than $\langle \sio \rangle$ of CASTLES lenses.\\

Also \textbf{J0737+321} and \textbf{J0956+510} are thought to be quads, but in each of these only 3 images are used, as the astrometry of the faintest image was too uncertain. J0737+321 is with $z=0.32$ the most distant in our SLACS subsample and belongs to higher $z$ lenses in the whole catalogue.\\ 

Finally we consider the two lensing clusters \textbf{ACO 1689} and \textbf{ACO 2667}.\\

ACO 1689 has a very large number of multiply imaged systems found by \citet{br05b}. In the present work this cluster is modeled by a set of two 5-image systems, six 3-image systems and one double. The additional systems are known to affect only details \citep{sa07}. Note that there are many more imaged sources, but adding those to the model does not change $\ml$, i.e. the mass model is tightly constrained by this set of image systems. The kinematic line-of-sight velocity dispersion $\sio = 1400$ km s$^{-1}$ of galaxies within the cluster was taken from \citet{lo08} for a subset of 130 galaxies in the inner region of the cluster with velocities $|v| < 3000$ km s$^{-1}$, which contains most likely the biggest mass fraction responsible for the lensed images. This average value applies for a radius of around 400 kpc, a region where the formal velocity dispersion seems to be sufficiently flat and in which roughly half of the projected radii of the 130 galaxies considered in \citet{lo08} are to be found. Furthermore the value is not too far away from the Einstein radius or outermost image position of around 240 kpc. In order to estimate the I-band magnitude of the cluster, the 130 brightest out of 840 galaxies are taken from a cluster survey of \citet{mo96} for which the Gunn g,r and i magnitudes were provided. Together with K-correction, evolution correction and galactic extinction we obtain $L_I = 2.82\times 10^{12} L_{\astrosun}$.\\

ACO 2667: For this cluster three 3-image systems and one double were known to derive the formal mass-only related velocity dispersion curve. The kinematic velocity dispersion $\sio=960^{+190}_{-120}$ km s$^{-1}$ of this lens was determined by \citet{co06} from a sample of 21 galaxies in the inner region of the lensing cluster with a radius of $110 h^{-1}_{70}$ kpc, which is in the same order of magnitude as $\rl = 98 h^{-1}_{72}$ kpc. However, since photometric data for estimating the total flux of galaxies within the cluster was not available, ACO 2667 is not included in the mass-to-light plots of Section \ref{sec:m2l} and consequently there was no need to determine $\re$ for the mass estimate.\\

The above lensing data are modeled using the PixeLens program\footnote{Available from www.qgd.uzh.ch/projects/pixelens/} \citep{sa04,co08}. PixeLens reconstructs the projected mass in a pixelated manner by solving a set of linear constraint lensing equations on the mass distribution by means of the given image positions, the redshifts of lens and source, the Hubble time (herein $h=0.72$ is always assumed) and optionally the time delays between the lensed images. There are certain requirements for the mass distribution to be fulfilled. It has to be non-negative, centrally concentrated, with a local density gradient pointing less than $45^\circ$ away from the center, inversion symmetric (optional), it must not have a pixel which is twice the sum of its neighbors except possibly the central pixel and the circularly averaged mass profile needs to be steeper than $R^{-0.5}$, where $R$ is the projected radius.
Hence an underdetermined set of equalities and inequalities is obtained. Subsequently a Monte-Carlo approach is used to sample over the mass map in order to determine an ensemble of lens models, for which the ensemble-average appears to be the best single model representation. Of course all the uncertainties on any parameter can be derived from this model ensemble. \\

Since PixeLens has been extensively tested in other papers, we won't get into details, but two points are worth noting:
\begin{enumerate}
\renewcommand{\theenumi}{(\arabic{enumi})}
\item For tests of the recovery of simulated galaxy lenses, see \citet{re07}.
\item For the recovery of gross features of even extended lens structures from the information encoded in the image positions of lensed objects, see \citet{sa01,fe08}.
\end{enumerate}
 
For each lens an ensemble of 100 mass maps with $21 \times 21$ pixels each has been computed, from which the mass-profile and therewith the formal velocity dispersion $\sil$ is derived with an $90 \%$ uncertainty, as one can see e.g. for the lens Q0957+561 in Figure \ref{fig1}. In Appendix \ref{sec:B} we compare the average of an ensemble containing 100 models with larger ensembles containing up to 10000 models to find that already small ensembles are sufficient to determine a fairly exact velocity dispersion. Since $\sil(R)$ is not sensitive to ensemble enlargement the number of models is fixed to 100 throughout this analysis.\\

Two points shall be emphasized here. Firstly, the error bars in the right panel of Figure \ref{fig1} represent the model dependence for an ensemble of 100 models and, as one can see, it is not large. Secondly, it is sometimes stated that the enclosed mass $M(<R)$ is known for $R = \rein$, the Einstein radius and unknown for any other $R$, but this is oversimplified. In fact $M(<R)$ has some model dependence at all $R$, but is minimal at $\rein$. $\sil(R\neq\rein)$ has a larger uncertainty than $\sil(\rein)$, but is still fairly well constrained, as one can see in Figure \ref{fig1}. The velocity dispersion at the radial position of the outermost image $\sil(\rl)$ as a quantity, which is as well constrained as $\rein$, is basic to this paper.\\

The PixeLens input files, mass maps as well as the formal velocity dispersion curves can be found in Appendix \ref{sec:B} of the online version of this paper. Note that for lensing clusters the velocity dispersion of the galaxies on their orbit around the center of the cluster is considered instead of the stellar velocity dispersion as in the case of lensing galaxies. The references for all the lenses and further details can be seen in Table \ref{tab1}.\\

\section{$\mathbf{\sil}$ versus $\mathbf{\sio}$}
\label{sec:svs}

\begin{figure}
\begin{center}\includegraphics[width=282pt, bb=28 0 669 499]{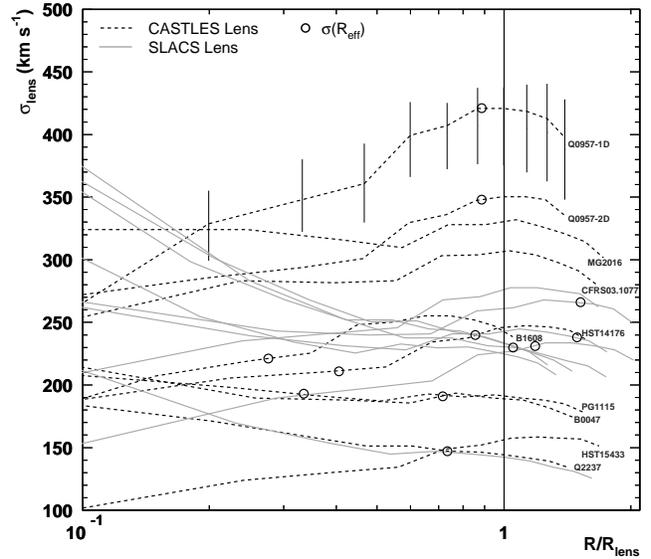}\end{center}
\caption{Formal velocity dispersion curves for all the galaxies. The grey solid lines denote SLACS lenses, the black dashed lines CASTLES lenses. For the sake of readability the error bars, indicating the range of ensemble models, are shown only for one velocity dispersion curve. The radial scale is normalized to the radius of the respective outermost image $\rl$ which is indicated by the horizontal solid line. Open circles denote the effective radius provided it is located in a fairly flat region of the velocity dispersion curve. Q0957 is shown twice in this plot: Q0957-1D models one doubly imaged source while Q0957-2D models two doubly imaged sources.}
\label{vd}
\end{figure}

The formal velocity dispersion curves $\sil(R)$ illustrated in Figure \ref{fig1} for the lens Q0957+561 are now computed for all the lenses. Figure \ref{vd} shows all these curves except for the cluster, which are excluded for the sake of readability and their comparatively high $\sil$ values.\\

In the following we concentrate on the formal velocity dispersion at a radius of the outermost image position $\sil(\rl)$ and at the effective radius $\sil(\re)$. Concerning the latter we cannot take for granted that the velocity dispersion curve at effective radii is still sufficiently flat. Because of this, when considering $\sil(\re)$ we exclude the lenses Q2237, HST15433, J0737, J0912, CFRS03 and J0956, for which this condition is \emph{not} fulfilled.\\

In terms of absolute values the curves for CASTLES lenses extend in average to larger radii whereas the curves of SLACS lenses are smaller due to a limited aperture of the SDSS fibers. Note that 7 SLACS lenses and 3 CASTLES lenses show a clear cuspy shape of the formal velocity dispersion curve towards inner radii as it is the case for the majority of early-type galaxies also in other velocity dispersion field studies \citep[e.g.,][]{co09}. However, in some cases anomalous galaxies exhibit a rising velocity dispersion profile, which might be related to the presence of a disk according to \citet{co09}. Additionally, the pixelated approach causes a variety of differently shaped velocity dispersion profiles differing especially in central regions. This leads consequently to large error bars and a decreased sensitivity in the center, rendering an interpretation of the profiles at smaller radii rather difficult.\\

It should be emphasized that the comparison of either $\sil(\rl)$ or $\sil(\re)$ with $\sio$ measured within an aperture, is a proper procedure, since the $\rl$ is in average less than a factor of 2 different from the aperture radius and for most lenses $\sil$ remains unchanged. For J0737, J1205, J1330 and J2300 the formal velocity dispersion curve ends before reaching the radius of 3 arcsec, that is, the mass does not contribute to the lensing effect, but nevertheless $\sio$ can be taken as an indicator for the real velocity dispersion. With other words the velocity dispersion measurements at aperture radius are probably not representetive since the main mass of the lens is smaller.\\

Comparing the curves labeled Q0957-1D and Q0957-2D shows the probable effect of adding formerly undiscovered image systems. As for Q0957 $\sil (R)$ varies considerably when a formerly unseen doubly imaged system is added. This also affects the relation between $\sil (\rl)$ or $\sil (\re)$ and the kinematic velocity dispersion $\sio$.\\

\begin{figure}
\begin{center}\includegraphics[width=282pt, bb=28 0 669 499]{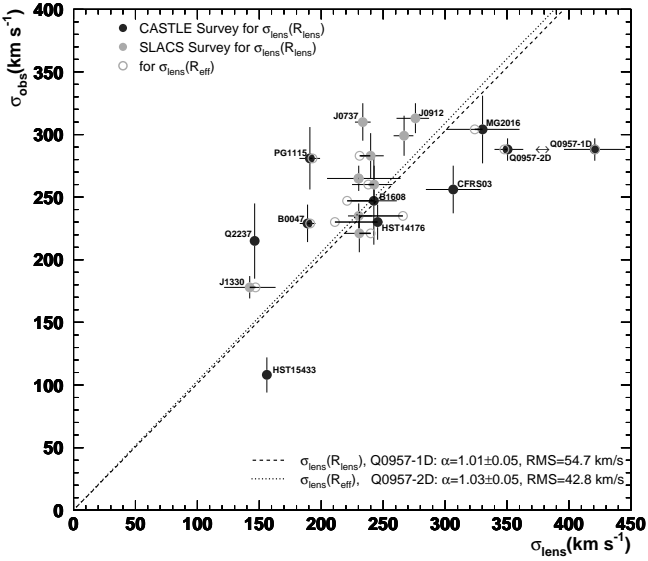}\end{center} 
\caption{$\sil$-$\sio$-plot for all the galaxy lenses. The filled circles refer to formel velocity dispersions $\sil$ measured at outermost image $\rl$. The open circles show $\sil(\re)$. Note that as in Figure \ref{vd} Q0957 is shown twice for different image systems. The dashed (dotted) line represents the fit for the solid (open) circles including Q0957-1D (Q0957-2D).}
\label{s2}
\end{figure}

Both $\sil(\rl)$ and $\sil(\re)$ plotted against $\sio$ can be seen in Figure \ref{s2}.
The comparison between the observed kinematic velocity dispersions and the mass-only related velocity dispersions reveals how virialized the lenses are, because $\sil\approx \sio$ is another representation of the virial theorem in Equation \ref{eq:1}. We constrain the fit by fixing it to the $(0,0)$-point, because a bias would have no physical relevance.  For $\sil$-values at effective radius instead of the radius of the outermost image position $\rl$ the scatter around the best fit decreases considerably. Although all $\sil(\re)$ are within the error bars of $\sil(\rl)$, changing the radii for the determination of the $\sil$-$\sio$-relation might consequently be the right thing to do, since therewith the relation is build on a common basis.\\

Furthermore we included the two-double (2D) and the one-double (1D) system of lens Q0957 in Figure \ref{s2} to demonstrate the grave difference in $\sil$ of the former outlier, reducing the RMS in the $\sil$-$\sio$-plot from 55 km s$^{-1}$ for the $\sil(\rl)|_{1D}$-fit to 43 km s$^{-1}$ for the $\sil(\re)|_{2D}$-fit. We conclude that generally a more complete lens system is to be favored and henceforth we only consider Q0957-2D. The linear best fits fixed to the origin for $\sil(\re)$ and $\sil(\rl)$ yield
\begin{eqnarray}
\sio &=& (1.03 \pm 0.05) \times \sil(\re), \label{fit1}\\
\sio &=& (1.04 \pm 0.04) \times \sil(\rl). \label{fit2}
\end{eqnarray}
As an aside the $y$ error bars plotted in Figure \ref{s2} are the observational errors taken from \citet{ko03,tr04,to99,to98,oh02,kt03,kt02} and \citet{fo92} for the CASTLES lenses and \citet{bo06} for SLACS lenses. The $x$ error bars represent the statistical errors of the formal velocity dispersion for an ensemble of 100 models of possible mass distributions. Thus the rather small error bars can be understood as a relatively model independent lensing mass and formal velocity dispersion. The errors are taken from a radius closest to $\rl$ since the pixelated approach only allows for discrete steps in radius. One could argue about the significance of these errors, because changes in the image positions or lost information like additional image systems or mass contamination of the light path can lead to fairly different results.\\

However, the fits for $\sio(\sil)$ (Equations (\ref{fit1},\ref{fit2})) make clear that a one-to-one correlation between $\ml$ and $\mv$ of the lensing galaxy is probable. It is important to know whether our sample is dominated by a certain kind of model far from $\rho(r) \sim r^{-2}$ corresponding to a constant $\sil$. For that we can study the correlation between the ratios $\sil/\sio$ and $\rl/\re$. In consideration of the virial theorem one can state:\\

\noindent If there is an (anti-)correlation between $\sil/\sio$ and $\rl/\re$ the density profile $\rho(r)$ of the lens should be (flatter) steeper than $r^{-2}$.\\

\begin{figure}
\begin{center}\includegraphics[width=282pt, bb=28 0 669 499]{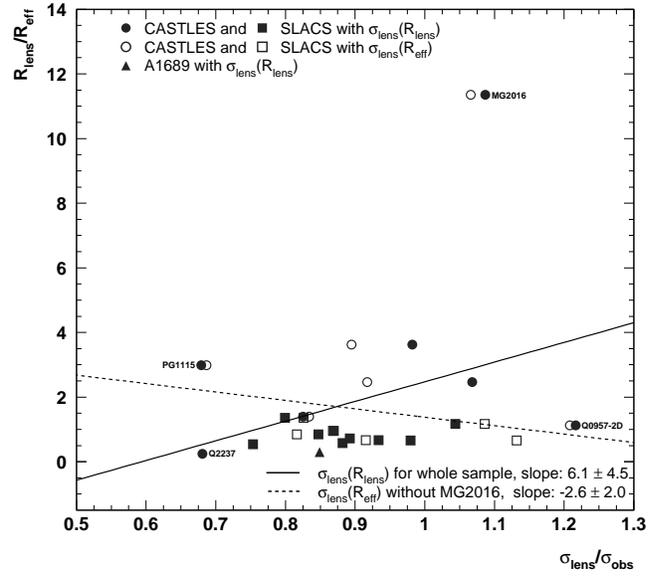}\end{center} 
\caption{Plot of $\rl/\re$ against $\sil/\sio$. The lines indicate extreme scenarios of formal fits for our sample of early-type galaxies which show large errors. For the $\sil(\re)$-fit of the whole sample the positive trend is insignificant. The trend inverts when excluding MG2016. In other words there is neither correlation nor anti-correlation, meaning that in average the density profile for all lenses is consistent with an isothermal ellipsoid.}
\label{ssrr}
\end{figure}

Figure \ref{ssrr} shows this relation for both $\sil(\rl)$ and $\sil(\re)$.
As for the first, the best fit shows a positive trend with large error bars.
For $\sil(\re)$ the positive trend is insignificant and the opposite result is \emph{not} excluded by the error bars. By neglecting the outlier MG2016 with a possibly underestimated $\re$, as will be discussed in Section \ref{sec:m2l}, one finds the inverse trend to be likewise significant. Nevertheless it needs to be emphasized that by excluding only one of the labeled outliers in Figure \ref{ssrr} the slope is strongly affected and can change its algebraic sign. Thus we cannot retrieve a strongly significant statement. In such exclusion scenarios we obtain slopes consistent with constant $\sigma$-ratio. Our sample of early type lensing galaxies for $\sil(\rl)$ ($\sil(\re)$) is clustering around a mean of $1.3 \pm 0.3$ ($1.6 \pm 0.3$) in $\rl/\re$ and around $0.91 \pm 0.04$ ($0.96 \pm 0.06$) in $\sil/\sio$ excluding MG2016 because of its extraordinarily high $\rl/\re$-ratio.
Since we cannot find any type of correlation throughout our sample we can summarize that $\sil$ is model independent.\\

\begin{figure}
\begin{center}\includegraphics[width=282pt, bb=28 0 669 499]{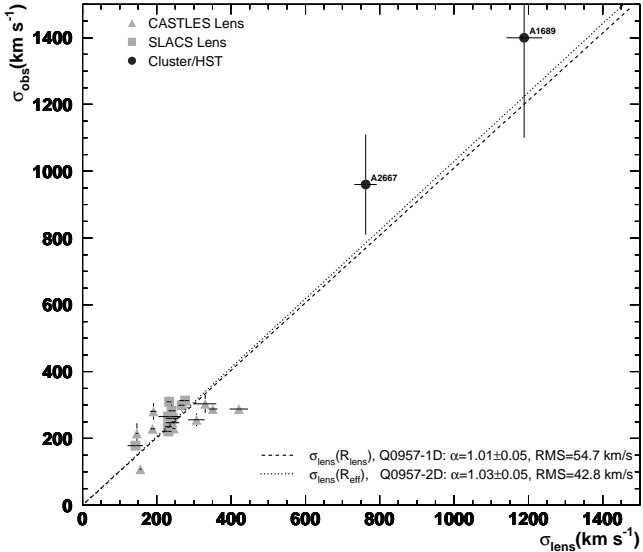}\end{center}
\caption{Like Figure \ref{s2} but with the two clusters ACO 1689 and ACO 2667 included. The straight line fits do not include the clusters. The relation between $\sio$ and $\sil$ extends to cluster scales.}
\label{sscl}
\end{figure}

Extending the $\sio$-$\sil$-plot in Figure \ref{s2} to 100 kpc scale, as to be seen in Figure \ref{sscl}, we can find that lensing clusters fit quite well to the previously found correlations for $\sio(\sil)$ (Equations (\ref{fit1},\ref{fit2})).\\

Going from an assumed isothermal $r^{-2}$ profile to a Hernquist profile makes the velocity dispersion of Equation (\ref{eq:1}) change to $\sih$ shown in Equation (\ref{eq:A1}) of Appendix \ref{sec:A} along the lines of \citet{he90}. This step yields a change of less than $19 \%$ of $\sil$ for most lenses, apart from few exceptions like P1115, which turned out to be an outlier already in Figure \ref{ssrr}.
Furthermore cluster ACO 2667 shows a velocity dispersion increased by $33 \%$. In general central regions of galaxy clusters are best fitted by a Hernquist profile \citep{he90} for the stellar component of the inner cD galaxy and a NFW model \citep{nfw96} for the dark matter component, shown by e.g. \citet{pa04}. That is why we can expect significant changes going from $\sil$ to $\sih$ on larger scales. However, fitting the $\sil$-$\sio$ relation for a Hernquist profile as done before with an isothermal model for $\sil$ reveals a slightly steepened slope compared to Equation (\ref{fit2}) of $(1.13 \pm 0.04)$. The clusters still agree to this relation within the error bars.\\

The dynamical state of galaxy clusters is hard to determine. There are many contradictory investigations on this topic. Optical and X-ray data on one hand indicate ongoing formation processes on substructure level \citep[e.g.,][]{st97,so99}, which should be considered in estimates of $\mv$. On the other hand statistical comparisons of different mass estimates from optical and X-ray observations and weak lensing show perfect agreement on scales much greater than the core radius $R_{\textrm{\tiny{core}}}$ \cite[e.g.,][]{wu97}. Still, on scales of core radii there are discrepancies between X-ray and mass measurements by means of weak lensing. \citet{al98} suggests to consider substructure and line-of-sight alignments of material towards the cluster cores since they will increase the lensing masses without affecting X-ray data and to take account of the dynamical activity which might cause the X-ray analyses to overestimate $R_{\textrm{\tiny{core}}}$. \citet{xu00} take this apparent dichotomy as an indicator of the transition from previrialization to virialization. In this paper however we can probe the virialization state for the two clusters at $\rl$, which is in both cases not far away from $R_{\textrm{\tiny{core}}}$. The core radii of the X-ray selected ACO 2667 and ACO 1689 are about ($76 \pm 8$) kpc \citep{co06} and ($80 \pm 15$) kpc \cite{al98} respectively. Thus with $\sil$ at $\re = 98$ kpc for ACO 2667 we already probe the core region. For ACO 1689 $\re$ is roughly $238$ kpc, which is 3 times the given core radius. By adjusting to smaller scales $\sil(R_{\textrm{\tiny{core}}})$ becomes $\sim 1000$ km s$^{-1}$ and marginally fails the relations (\ref{fit1}) and (\ref{fit2}). It should be emphasized that unlike the sample of lensing galaxies ACO 1689 $R_{\textrm{\tiny{core}}}$ is \emph{not} in a sufficiently flat region of $\sil$ and thus not comparable with the relations for which this was a requirement. Since strong lensing unveils mass regardless of underlying dynamics one can summarize that also in view of findings from previous studies clusters in a wide range of radii can be regarded as virialized.

Nevertheless, the correlation between the kinematic velocity dispersion $\sio$ and $\sil$ is hard to decipher. First the scatter around a best fit that is smaller (larger) than the scatter around the FP in the ($\re$,$\sigma$,$I$) parameter space can be understood as a hint on a basically mass dependent (stellar dynamics dependent) $\sio$. Of course it can also be seen as a merely statistical scatter that is influenced by a possibly biased lens sample.
This allows for drawing the following conclusions:

\begin{enumerate}
\renewcommand{\theenumi}{(\arabic{enumi})}
\item The small scatter and the slope of the best fit of $\sim 1$ makes $\sil$ a good surrogate for $\sio$, which is independent of a particular density profile model.
\item The included elliptical galaxies are thus virialized and
\item the relation can be extended to larger scale objects like clusters, as we can see in Figure \ref{sscl}.
\end{enumerate}

With this in hand we now want to analyze the mass-to-light relationship for the given sample and compare it to the governing FP of early type galaxies.

\section{Mass-To-Light Ratio and the Fundamental Plane} 
\label{sec:m2l}
As a first step to a mass-to-light relation for this sample of early type lensing galaxies we K-correct given I-band magnitudes (centered on 814 nm) and SDSS-i-band magnitudes (centered on 753 nm) in AB units to rest frame I-band since they provide the most complete set of magnitudes for our sample. These are taken from the CASTLE Survey homepage\footnote{cfa-www.harvard.edu$/$glensdata$/$} and \citet{bo06}. In the case of the galaxy cluster ACO 1689 we obtain the overall magnitude by summing over the fluxes of the galaxy content using the catalogue of \citet{mo96}. Hence the K-correction is based on SDSS, HST and ESO spectral templates\footnote{The spectral templates for WFPC2, SDSS ACS and ESO telescopes are taken from the STSCI homepage www-int.stsci.edu$/$instruments$/$wfpc2$/$Wfpc2\_thru$/$, stsci.edu$/$hst$/$acs$/$analysis$/$reference\_files$/$synphot\_tables.html and filters.ls.eso.org$/$efs$/$efs\_fi.htm}.\\

\begin{figure}
\begin{center}
\includegraphics[width=282pt, bb=28 0 669 499]{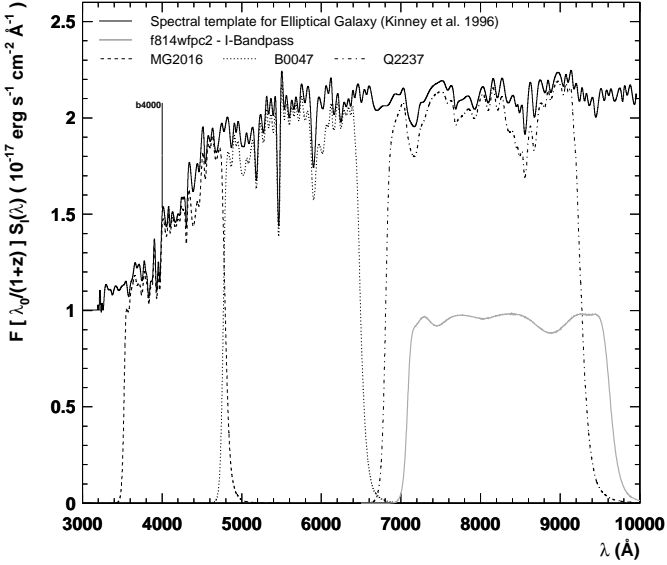}
\end{center}
\caption{Visualization of K Correction: The black solid curve shows the flux template of an elliptical galaxy. The grey solid line represents the  HST WFPC2 I-bandpass taken from www-int.stsci.edu/instruments/wfpc2/Wfpc2\textunderscore{}thru/. The dashed curves are showing the denominator of the integrand in Equation (\ref{eq:kcorr}) for 3 lenses: MG2016 ($z=1.01$, dashed line), B0047 ($z=0.485$, dotted line) and Q2237 ($z=0.04$, dash-dotted line).}
\label{bpxflux}
\end{figure}

We carry out the K-correction from first-principles in preference to a black box program. Following \citet{ok68} we compute the K-corrected flux according to
\begin{eqnarray}
K_x& = & 2.5 \log{(1+z)} \nonumber \\
& &+ 2.5 \log{\left\{ \frac{\int_0^\infty F(\lambda_0) S_x(\lambda) ~d\lambda}{\int_0^\infty F(\lambda_0/(1+z)) S_x(\lambda) ~d\lambda} \right\}},
\label{eq:kcorr}
\end{eqnarray}
where $K_x$ denotes the K-correction for the x-Band expressed in magnitudes. The band width is smaller in the redshifted galaxy, which leads to the first term in (\ref{eq:kcorr}). A source spectrum $F(\lambda)$ is redshifted through fixed spectral-response bands $S_x$ or bandpasses respectively of the detector. The flux at an effective wavelength in the rest frame of a galaxy of redshift $z$, transformed from the effective wavelength $\lambda_0$ of the detector by $\lambda_0 / (1+z)$, will differ from the flux of a galaxy at rest. This leads to the second term in \ref{eq:kcorr}.
Figure \ref{bpxflux} visualizes the denominator of the integrand in Equation (\ref{eq:kcorr}), where the I-bandpass is multiplied by the redshifted flux template of an elliptical galaxy taken from \citet{ki96}. As an aside the apparent SDSS magnitudes are on an AB basis within $3 \%$, which only leads to minor corrections and is therefore neglected in the following analysis. Note that the K-correction is realized with the exact template for $\lambda < 570$ nm. For higher wavelengths we assumed a constant flux for the sake of simplicity. The deviations resulting from this approximation are even in the worst case of a hardly redshifted galaxy in the upper $\lambda$-range like Q2237 of only $0.3 \%$ for $L_I$. This leads to negligible corrections for all following quantities. Furthermore galactic extinction corrections according to \citet{sc98} are applied to the fluxes. The luminosities are calculated in units of solar luminosities according to an AB magnitude\footnote{Listed on www.ucolick.org/$\sim$cnaw/sun.html} $I_{\astrosun} = 4.57$ for WFPC2 data and $i_{\astrosun}=4.48$ for SDSS data calculated along the lines of \citet{fu95}. Subsequently we correct for passive $M/L$-evolution with a slope of 
\begin{displaymath}
\frac{d\log{M/L_I}}{dz}=-0.397
\end{displaymath}
inferred by stellar population synthesis models taken from \citet{br03}.\\

\begin{figure}
\begin{center}\includegraphics[width=282pt, bb=28 0 669 499]{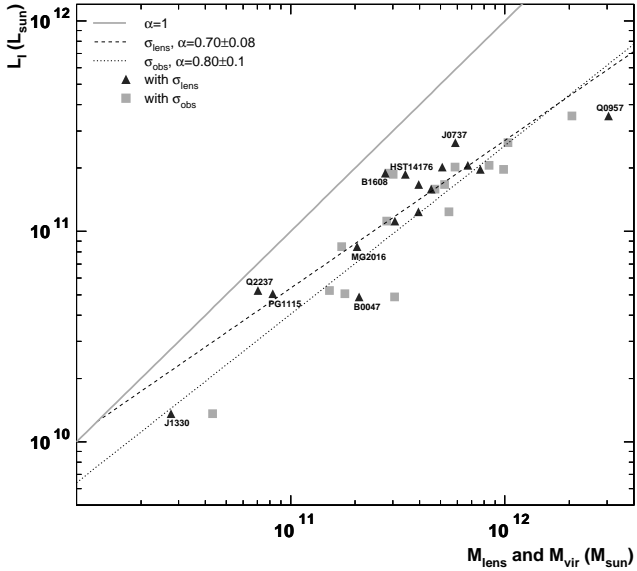}\end{center} 
\caption{Lensing mass and virial mass against I-band luminosity for all galaxies. The triangles denote masses calculated using $\sil$. The squares refer to masses calculated with $\sio$. The best fits for $M^\alpha \propto L$ are plotted for  $\sil(\rl)$ (dashed line) and for $\sio$ (dotted line). The solid line refers to a constant $M/L$-ratio.}
\label{l2m}
\end{figure}

Having the I-band luminosities $L_I$ of all lenses in units of solar luminosities $L_{\astrosun}$ and the velocity dispersions from Section \ref{sec:svs} we can analyze the underlying mass-to-light relation. Figure \ref{l2m} shows the lensing mass $\ml = \re \sil^2(\rl)$ and the virial mass $\mv = \re \sio^2$ plotted against I-band luminosity. The plot also provides a curve representing a constant $M/L$ or according to Equation (\ref{eq:m2l}) a $(\alpha=1)$-line respectively.\\

A closer look at the V-band luminosities for selected galaxies reveals that HST14176, B1608 and MG2016 emerge as outliers with mass-to-light ratios $\lesssim 1$. This can be explained by nearby groups and clusters (e.g. in the case of HST14176) or mass-contamination influencing the path of light. Another reason can be uncertainties in the effective radii, as already mentioned in Section \ref{sec:svs}. If we take for HST14176 (MG2016) $\re=1.06$ ($0.31$) arcsec \citep{tr04} instead of the used $0.71$ (0.22) arcsec \citep{ru03} then $\ml=\re\sil^2$ would increase by a factor of $\sim 1.5$ ($1.4$), since no grave changes in $\sigma$ for a flat formal velocity dispersion curve are expected. This leads for HST14176 to a lensing mass of $5.14 \times 10^{11}$ $M_{\astrosun}$ instead of the former $3.43 \times 10^{11}$ $M_{\astrosun}$ which is then also in the V-band clearly below the ($\alpha=1$)-line. If such uncertainties are the true cause for comparatively high luminosities, then we also need to adjust $\ml$ and $\mv$ in Figure \ref{l2m}. However, changing $\re$ or excluding the problematic lenses from the fit has a negligible impact on the slope $\alpha$ using $\ml$ and only small impact using $\mv$, changing $\alpha$ from ($0.80 \pm 0.10$) to ($0.84 \pm 0.10$). It should be emphasized that we hold on to the dataset of \citet{ru03}, because it provides the effective radii computed on a common basis for the whole CASTLES subset of our lensing objects.

Both sets of data points for $\sio$ and for $\sil (\rl)$ are fitted for the whole sample and reveal the slopes:
\begin{eqnarray}
 \alpha &=&(0.70\pm 0.08) ~\textrm{~for $\ml$,} \nonumber \\
 \alpha &=&(0.80\pm 0.14) ~\textrm{~for $\mv$.} \nonumber 
\end{eqnarray}

It shows that $\alpha=1$ is in any case clearly excluded. Figure \ref{l2m} shows the best fit for both $\ml$ and $\mv$. Note that the fits in the plot cannot be extrapolated to lower masses, which would mean that judging by the intersection with the $(\alpha=1)$-line the luminous mass would overtake the total mass content. The plot and therewith also the FP of nearby lenses show that more massive galaxies have a larger dark matter fraction.\\

\begin{figure}
\begin{center}\includegraphics[width=282pt,bb=28 0 669 499]{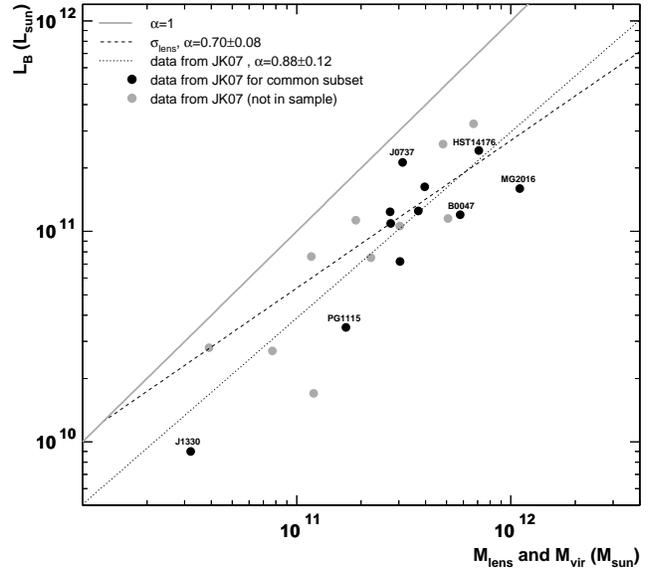}\end{center} 
\caption{Mass-versus-light plot for data from [JK07]. The black circles denote a subset of lenses included in the lens sample of this paper. The grey circles are residual lenses. The dotted line represents the best fit for the whole dataset taken from [JK07]. The dashed line refers to the $\sil$-fit as seen in Figure \ref{l2m}.}
\label{l2mj}
\end{figure}

In Figure \ref{l2mj} the lens sample from [JK07] together with a best fit is shown. Their data from stellar-dynamical measurements on 22 early-type models contains a common subset with the present study. Note that the data in [JK07] was given in B-band luminosities, which explains the shift of the data points towards lower luminosities in most cases. The fit for the whole [JK07] sample yields a slope of $\alpha=0.88\pm 0.12$. As in Figure \ref{l2m} a curve indicating a constant mass-to-light relation is included.\\

From these plots we can summarize, that

\begin{enumerate}
\renewcommand{\theenumi}{(\arabic{enumi})}
\item the slope of the best fit for $\sio$ is consistent with the one for the lensing sample of [JK07] within error bars,
\item the slope of the fit for $\sil$ is not consistent with the fit for data from [JK07], although the error bars do overlap,
\item only the fit for the \citet{ji07} sample is consistent with $\alpha=1$,
\item the slopes of the $\sil$- and $\sio$-fits (for the whole dataset and for a reduced or, due to uncertainties in $\re$, changed dataset) are clearly excluding $\alpha=1$ within their error bars and thus do not agree with a constant $M/L$ ratio.
\end{enumerate}

\begin{figure}
\begin{center}\includegraphics[width=282pt, bb=28 0 669 499]{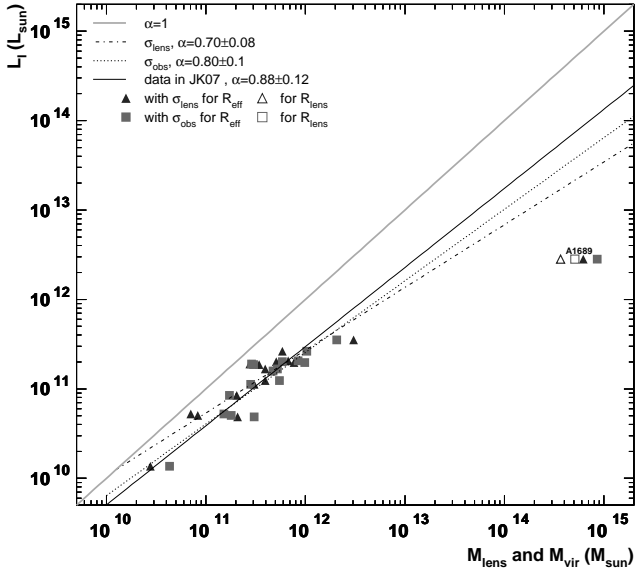}\end{center}
\caption{Like Figure \ref{l2m} but extended by cluster ACO 1689. As expected clusters are not on the FP, because they have a higher mass-to-light ratio. Solid symbols denote masses calculated with $\sigma(\re)$, open symbols denote masses calculated with $\sigma(\rl)$.}
\label{l2mc}
\end{figure}

In Figure \ref{l2mc} we extend determined mass-to-light relations to larger scales. For cluster size objects like ACO 1689, $\re$ is of course not defined. 
Nevertheless, one can still use the mass quantity $R\sigma^2$ to compare the mass-to-light behavior of early type galaxies and clusters. The kinematic line-of-sight velocity dispersion $\sio = 1400$ km s$^{-1}$ of galaxies within the cluster was taken from \citet{lo08} for a subset of 130 galaxies in the inner region of the cluster with velocities $|v| < 3000$ km s$^{-1}$, which contains most likely the biggest mass fraction responsible for the lensed images. This average value applies for a radius of around 400 kpc, a region where the formal velocity dispersion seems to be sufficiently flat and in which roughly half of the projected radii of the 130 galaxies considered in \citet{lo08} are to be found. Furthermore the value is not too far away from the outermost image position of around 240 kpc. Therewith $\re\sio^2(\re)$ and $\re\sio^2(\rl)$ are determined.\\

The data points for the cluster deduced from the formal velocity dispersion $\sil$ at $\rl$ and $\re$ are included in Figure \ref{l2mc}. As expected neither the relation $M^\alpha\propto L$ with $\alpha\approx 0.70$ nor with any other slope presented above does extend to clusters. As shown by \citet{sch93} galaxy clusters follow indeed a different FP relation. We can make up a region in the mass-to-light plot for cluster sized objects, which lies far below all previous lines and matches the findings of \citet{sch93} for a FP consisting of 16 clusters. One should keep in mind that for early-type galaxies $M/L$ can be a suitable dark matter versus baryon estimator because $L$ tracks pretty much all baryons. But this is not a good approximation for clusters, whose total baryonic mass is generally believed to be made of $80 \%$ hot diffuse gas and only $20 \%$ galaxies \cite{fu98}. In order to correct for this discrepancy one might add the missing $80 \%$ expressed in terms of luminosity. Hence the luminosity of ACO 1689 is shifted to $1.4\times 10^{13} L_{\astrosun}$. Despite of this correction we obtain a value significantly below the given fits. Thus clusters can nonetheless be regarded as highly dark matter dominated.\\

\begin{figure*}
\begin{minipage}[l]{0.51\textwidth}
\includegraphics[width=290pt, bb=46 -22 669 499]{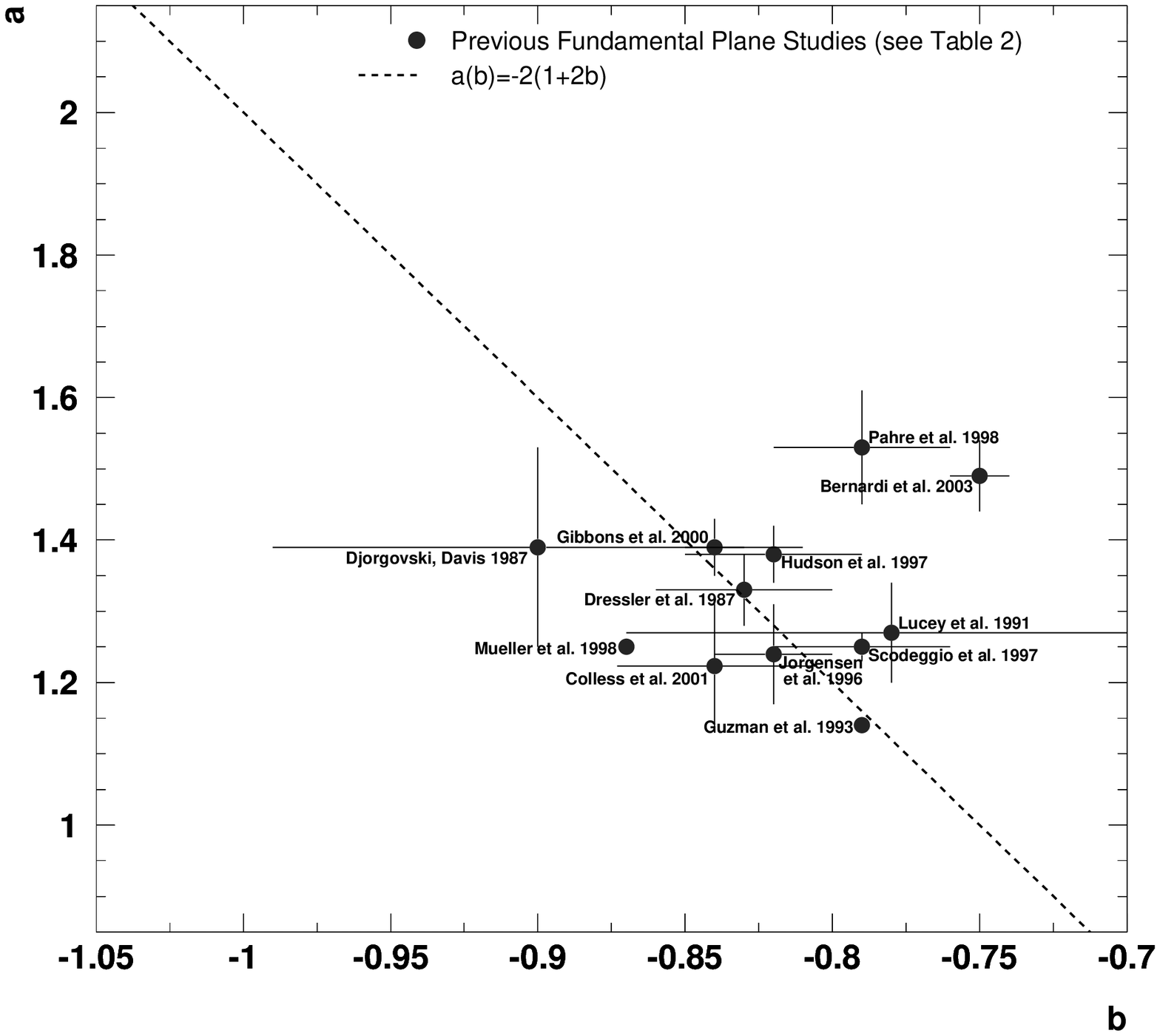}
\end{minipage}\begin{minipage}[l]{0.49\textwidth}
\includegraphics[width=290pt, bb=46 0 669 499]{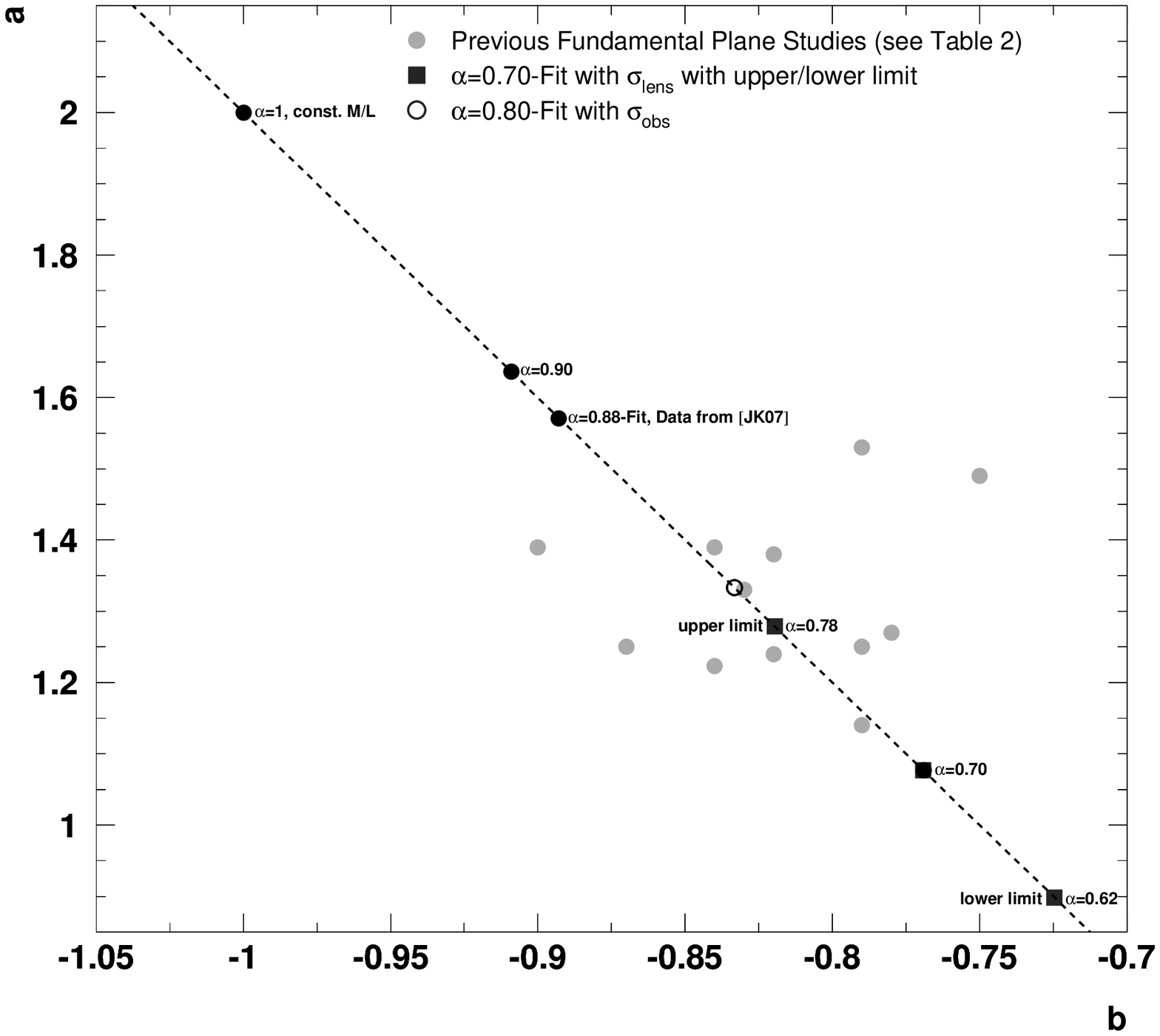}\\ 
\end{minipage}
\caption{Left panel: the a-b-parameter-space according to Equation (\ref{eq:4}). The dashed line represents the mass-to-light power index $\alpha$ for related $a$ and $b$ according to Equation (\ref{eq:ab}).
The plot shows FP parameter from previous studies referenced in Table \ref{tab2}. The error bars are included as far as provided in the references.
Right panel: like left plot but with results from this paper. The squares mark upper, lower and mean value of the fit using $\sil$ for the whole sample of early-type galaxies. The open circle denotes the fit using $\sio$. The black filled circles denote other $\alpha$-values, like e.g. the Vanilla Plane or the fit for data from [JK07]. For the sake of readability and comparison the grey filled circles corresponding to the data shown in the left panel are included.}
\label{figab}
\end{figure*}

We can summarize that our results are in good agreement with most of the recent FP-type studies, as one can see in Figure \ref{figab}. In the two plots the FP parameter study results of the references listed in Table \ref{tab2} are presented (left panel) together with the results of this paper (right panel). Recovering the FP of early-type galaxies by means of the photometric-independent $\sil$ shows that non-homologies like structural and orbital anisotropies, which might change the photometrically determined central velocity dispersion, have small to negligible impact on the FP tilt, as also shown by \citet{ca06}.\\

The FP parameters of our analysis are determined in consideration of the relations $a=2\alpha(2-\alpha)^{-1}$ and $b=-(2-\alpha)^{-1}$:
\begin{eqnarray}
a=1.08,& ~b=-0.77 &~\textrm{for} ~\sil,\nonumber\\
a=1.33,& ~b=-0.83 &~\textrm{for} ~\sio,\nonumber
\end{eqnarray}
corresponding to $\alpha=0.70 \pm 0.08$ and $\alpha=0.80\pm 0.10$ respectively.
Upper and lower limit of the $\sil$-fit are also drawn into the plot and exclude plainly the $M \propto L$ case of the Vanilla Plane.\\

Moreover, the FP parameters found in this study are conspicuously surrounded by the ones found in other studies (see Table \ref{tab2}).
For example in recent SDSS results for nearly 9000 early-type galaxies in a redshift range of $0.01 < z < 0.3$ the parameters are determined to $a=1.49 \pm 0.05$ and $b=-0.75 \pm 0.01$ \citep{be03}, and as an aside have no common $\alpha$, since Equation (\ref{eq:ab}) does not hold. On the other hand \citet{dr87} in one of the first FP parameter studies present parameters, which are almost perfectly in agreement with the fixed $a$ to $b$ relation and a common $\alpha$ of $\sim 0.80$, although measured separately. This value is verified in this paper by the mass-to-light relation found for $\mv$. It can be seen that $(a,b)$ for the slope of the $\sil$-fit is close to the results of \citet{gu93,co01,jo96,sc97,lu91} and \citet{dr87} in ascending order of distance in $(a,b)$-space. Except for \citet{hu97,pa98,gi00} and \citet{be03} the errors of previous $(a,b)$-studies, as far as they were given, overlap with the error bars in this study. In particular the results of \citet{jo96} and \citet{co01} agree with the upper limit of $\alpha$-values from the $\sil$-fit. However, the $\alpha$ estimate from the dataset of [JK07], which matches the result from \citet{ru03} can be excluded. Since for all previous FP type studies kinematic velocity dispersion measurements are used, our findings suggest that the real underlying $(a,b)$ values are even closer to the lower right corner of Figure \ref{figab}.

\section{Conclusion}
\label{sec:5}
We can summarize the findings of this paper as follows:
\begin{enumerate}
\renewcommand{\theenumi}{(\arabic{enumi})}
\item Independent of the details of lens models the lensing masses and virial masses basically agree, since $\sil \approx \sio$, as demonstrated in Section \ref{sec:svs}. This verifies the virial theorem.
\item The relation between the lensing inferred velocity dispersion $\sil$ and the observed kinematic velocity dispersion $\sio$ extends to cluster sized lensing objects within rather large uncertainties originating from a poorly defined scale radius $\re$ as shown up for the two galaxy clusters ACO 1689 and ACO 2667.
\item Using the results for $\sil$ ($\sio$) in Section \ref{sec:m2l} the lensing mass (virial mass) is calculated according to $M \approx \re \sigma^2$. We find the mass-to-light relation $\ml^{0.70\pm0.08} \propto L$ for the whole sample and $\mv^{0.80\pm0.10} \propto L$ to be consistent with most other FP type studies. We point out that the FP defined by using $\sil(\rl)$ is based on lensing velocity dispersions within $\rl$, which is not correlated to the effective radius. In order to render the used quantities unequivocal we analyze the change in $\sil$-$\sio$ switching from $\rl$ to $\re$ and find only a marginally different slope, though a reduced scatter in the $\sil$-$\sio$-plot can be seen. A few lenses are problematic outliers due to observational uncertainties but excluding these does not effectively change the result. With $R\propto \sil^{1.08} I^{-0.77}$ the FP of early type galaxies is recovered, excluding clearly the Vanilla Plane. Thus also non-homology as a reason for the FP tilt can be excluded.
\item As shown for ACO 1689, clusters are far from the FP since they have a much higher dark matter fraction than early type galaxies.
\end{enumerate}

The FP tilt discovered by \citet{dr87} and recovered in this study using $\sil$ as a surrogate is an often discussed matter (see Table \ref{tab2}) in astrophysics. The reasons for the deviation from the Vanilla Plane are hard to resolve, because neither the mass-structure, the mass-to-light ratio nor the dark matter fraction are directly and independently observable. Until a consensus on the explanation for the FP is found it is necessary to focus on quantities which are unequivocally related to a certain physical entity. For this purpose $\sil$ is proposed in this paper, since it fulfils the necessary condition of preserving the viriality for both elliptical galaxies and clusters.

\section*{Acknowledgments}
I would like to thank Prasenjit Saha for patiently answering my many lensing questions and giving useful suggestions on making the paper clearer. Many thanks to Ignacio Ferreras and Andrea Macci\'o for all helpful suggestions, to Jonathan Coles for giving further insights in PixeLens and Justin Read for useful discussions. Furthermore I thank the Swiss National Science Foundation for financial support.

\bibliographystyle{mn2e}
\bibliography{lib}

\appendix
\section{}
\label{sec:A}
For a Hernquist profile the velocity dispersion $\sih$ of a projected distribution is
\begin{eqnarray}
\sih^2(\rl) &\sim& \frac{G\mv^2}{a \ml} \frac{(1-s^2)^2}{[(2+s^2)X(s)-3]}  \Big\{ \frac{1}{2} \frac{1}{(1-s^2)^3} \nonumber \\
& & \times \left[ -3s^2X(s) (8s^6-28s^4+35s^2-20) \right.  \nonumber \\
& & \left.  - 24s^6+68s^4+65s^2+6\right] -6\pi s \Big\}
\label{eq:A1}
\end{eqnarray}
with $s=R/a$, whereas $a$ denotes a scale length and
\begin{displaymath}
X(s) = \left\{
\begin{array}{lcl}
 \frac{\mathrm{sech}^{-1}{s}}{\sqrt{s^2-1}} & \textrm{for} & 0 \leq s \leq 1 \\
 \frac{\cos^{-1}{s^{-1}}}{\sqrt{s^2-1}} & \textrm{for} & 1 \leq s \leq \infty
\end{array} \right.
\end{displaymath}
according to \citet{he90}.

\section{}
\label{sec:B}

\begin{figure}
\begin{center}
\includegraphics[width=282pt, bb=28 0 669 499]{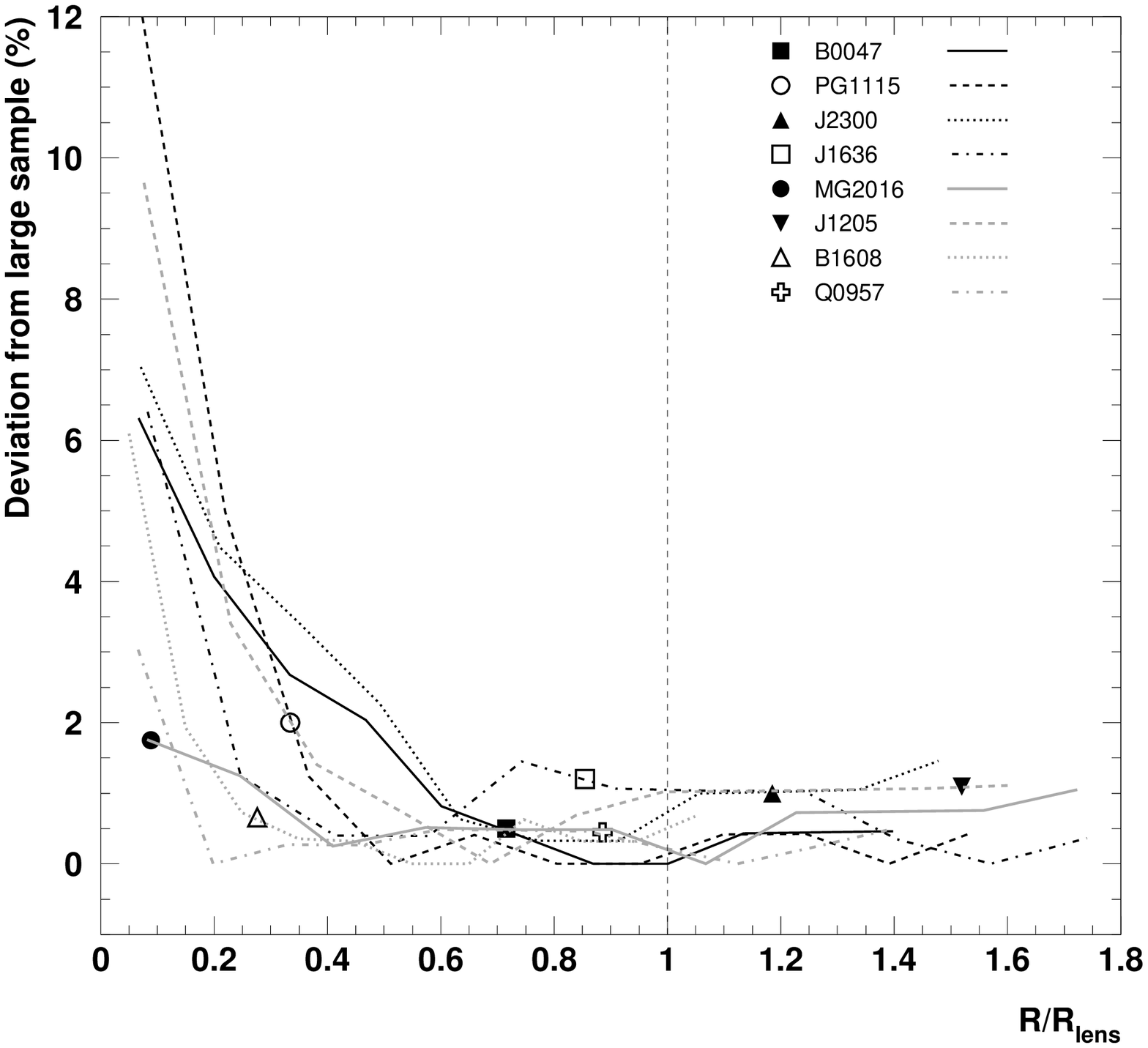}
\end{center}
\caption{Absolute deviation of formal velocity dispersion curve of a small ensemble from the one of a large ensemble in terms of percent plotted against the radius in terms of $\rl$. The dashed vertical line denotes $\rl$. The listed markers indicate the deviation of $\sil$ at the effective radius.}
\label{enstest}
\end{figure}

In Figure B1 the absolute deviation between the average formal velocity dispersion of an ensemble of 100 models and one with about 10000 models, which we take as a close to exact representation of the lens model, is shown in terms of percentage. The analysis is done for a subsample of doubles, quads and multiple object systems of SLACS and CASTLES lenses. We finde that $\sil(\rl)$ ($\sil(\re)$) of the smaller ensemble deviates less than $\sim 1 \%$ ($\sim 2 \%$) from the corresponding velocity dispersion for a larger ensemble.

\begin{table*}
\begin{scriptsize}
\hspace{-1.4cm}\begin{tabular}{@{} lcccccccccl @{}}
\hline
\textbf{Lens} & $\mathbf{z_L}$  &$\mathbf{\sio}$& $\mathbf{\sil}$ & $\mathbf{\re}$ & $\mathbf{\rl}$ & $\mathbf{\ml}$ & $\mathbf{\mv}$ & $\mathbf{L_I}$  & $\mathbf{\ml/L_I}$ & \textbf{Ref.}\\
& &$\mathbf{[\textbf{km~s}^{-1}]}$& $\mathbf{[\textbf{km~s}^{-1}}]$ & $\mathbf{[\textbf{kpc}]}$ & $\mathbf{[\textbf{kpc}]}$ & $\mathbf{[10^{11} \times M_{\astrosun}]}$ & $\mathbf{[10^{11} \times M_{\astrosun}]}$ & $\mathbf{[10^{11}\times L_{\astrosun}]}$ & $\mathbf{[M_{\astrosun}/L_{\astrosun}]}$ & \\
\hline \hline
B0047-2808\vspace{0.5mm}& 0.4850 & $229\pm15$ & $189.0\pm6.4\phantom{0}$ & 5.324 & 7.440 & $2.084\pm0.141$ & $3.059\pm0.401$ & 0.760 & 2.74 & \ding{172}\ding{173}\ding{172}\\
CFRS03.1077\vspace{0.5mm} & 0.9380 & $256\pm19$ & $306.6\pm22.0$ & -     & 15.905& - & -& -     & -    & \ding{172}\ding{174}\ding{172}\\
Q0957 (2D)\vspace{0.5mm}  & 0.3600 & $288\pm9\phantom{0}$& $351.6^{+12.8}_{-10.4}$ & 22.64 & 25.529& $30.458^{+2.477}_{-1.583}$ & $20.58\pm1.29\phantom{0}$ & 4.922 & 6.19 & \ding{172}\ding{175}\ding{181}\\
PG1115+080\vspace{0.5mm}  & 0.3100 & $281\pm25$ & $190.8\pm8.4\phantom{0}$ & 2.072 & 6.188 & $0.826\pm0.073$ & $1.792\pm0.319$ & 0.671 & 1.23 & \ding{172}\ding{176}\ding{172}\\
HST14176\vspace{0.5mm}		& 0.8100 & $230\pm14$ & $245.6\pm4.0\phantom{0}$ & 5.190 & 12.780& $3.430\pm0.112$ & $3.008\pm0.366$ & 3.910 & 0.88 & \ding{172}\ding{177}\ding{172}\\
HST15433\vspace{0.5mm}		& 0.4970 & $108\pm14$ & $156.2\pm2.8\phantom{0}$ & -     & 4.601 & -     & -     & -     & -    & \ding{172}\ding{174}\ding{172}\\
B1608+656\vspace{0.5mm}		& 0.6300 & $247\pm35$ & $242.6\pm18.4$ & 4.291 & 15.542& $2.766\pm0.420$ & $2.868\pm0.813$ & 3.354 & 0.82 & \ding{172}\ding{178}\ding{172}\\
MG2016+112\vspace{0.5mm}  & 1.0100 & $304\pm27$ & $330.4\pm29.6$ & 1.707 & 19.388& $2.041\pm0.366$ & $1.729\pm0.307$ & 2.128 & 0.96 & \ding{172}\ding{179}\ding{172}\\
Q2237+030\vspace{0.5mm}		& 0.0400 & $215\pm30$ & $146.4\pm2.8\phantom{0}$ & 2.993 & 0.743 & $0.703\pm0.027$ & $1.516\pm0.423$ & 0.542 & 1.30 & \ding{172}\ding{180}\ding{172}\\
\hline
J0037-094\vspace{0.5mm}		& 0.1954 & $265\pm10$ & $230.4^{+33.6}_{-36.0}$ & 6.804 & 6.470 & $3.956^{+1.239}_{-1.139}$ & $5.234\pm0.395$ & 1.990 & 1.99 & \ding{182}\\
J0737+321\vspace{0.5mm}		& 0.3223 & $310\pm15$ & $233.6\pm 6.4\phantom{0}$ & 9.823 & 5.333 & $5.874\pm0.322$ & $10.34\pm1.00\phantom{0}$ & 3.544 & 1.66 & \ding{182}\\
J0912+002\vspace{0.5mm} 	& 0.1642 & $313\pm12$ & $276.0^{+11.2}_{-15.2}$  & 9.203 & 5.271 & $7.679^{+0.637}_{-0.821}$ & $9.877\pm0.757$ & 2.289 & 3.35 & \ding{182}\\
J0956+510\vspace{0.5mm}		& 0.2405 & $299\pm16$ & $266.8^{+7.6\phantom{0}}_{-8.4\phantom{0}}$ & 8.607 & 6.201 & $6.710^{+0.390}_{-0.414}$ & $8.430\pm0.902$ & 2.572 & 2.61 & \ding{182}\\
J1205+491\vspace{0.5mm}		& 0.2150 & $235\pm10$ & $230.2^{+10.6}_{-8.6}$ & 7.805 & 5.138 & $4.531^{+0.427}_{-0.332}$ & $4.722\pm0.402$ & 1.936 & 2.34 & \ding{182}\\
J1330-014\vspace{0.5mm}		& 0.0808 & $178\pm9\phantom{0}$ & $142.3^{+20.9}_{-20.7}$ & 1.244 & 1.696 & $0.276^{+0.087}_{-0.074}$ & $0.432\pm0.044$ & 0.147 & 1.88 & \ding{182}\\
J1636+470\vspace{0.5mm}		& 0.2282 & $221\pm15$ & $230.7^{+9.3}_{-12.3}$ & 5.256 & 6.150 & $3.065^{+0.252}_{-0.318}$ & $2.812\pm0.382$ & 1.376 & 2.23 & \ding{182}\\
J2300+002\vspace{0.5mm}		& 0.2285 & $283\pm18$ & $239.8^{+10.6}_{-7.8}$ & 6.256 & 5.278 & $3.942^{+0.355}_{-0.253}$ & $5.489\pm0.698$ & 1.522 & 2.59 & \ding{182}\\
J2303+142\vspace{0.5mm}		& 0.1553 & $260\pm15$ & $242.7^{+14.9}_{-17.9}$ & 7.901 & 5.303 & $5.098^{+0.646}_{-0.724}$ & $5.851\pm0.675$ & 2.333 & 2.19 & \ding{182}\\
\hline
ACO 1689\vspace{0.5mm}	      & 0.1830 & $1400\pm 300$ & $1188.7^{+40.0}_{-56.0}$ & 400.0 & 237.6 & $6192^{+424}_{-570}$  & $8589\pm2187$ & 33.48 & 185 & \ding{183} \\  
ACO 2667\vspace{0.5mm}	      & 0.2330 & $960^{+190}_{-120}$ &  $762.0^{+7.2}_{-8.0}$ & -  & 98.01 & -     & -    & -      & -     & \ding{184}\\
\hline\hline
\end{tabular}
\end{scriptsize}
\caption{Full set of gravitational lenses used for this analysis. The first 9 lenses are from CASTLES, the following 9 from SLACS and the last two are clusters. The image positions and flux data have been taken from HST data (www.cfa.harvard.edu/glensdata/), \citet{bo06} and \citet{co06}. The symbols mark the references for the data in the columns $z_L$, $\sio$ and $\re$ and refer to the following publications: \ding{172} \citep{ru03}, \ding{173} \citep{ko03}, \ding{174} \citep{tr04}, \ding{175} \citep{to99}, \ding{176} \citep{to98}, \ding{177} \citep{oh02}, \ding{178} \citep{kt03}, \ding{179} \citep{kt02}, \ding{180} \citep{fo92}, \ding{181} \citep{ke98}, \ding{182} \citep{bo06}, \ding{183} \citep{lo08} and \ding{184} \citep{co06}. Note that $\sio$ is the kinematic central velocity dispersion, which is in the case of SLACS lenses the line of sight stellar velocity dispersion measured by the $3"$ diameter SDSS spectroscopic fiber. The $\sil$ values are determined for the projected distance $\rl$ from the outermost lensing image to the central lensing mass. The effective radii given in arcseconds in \citet{ru03} and \citet{bo06} have been transformed into kpc. All quantities in the table assume $H_0 = 72$ km s$^{-1}$ Mpc$^{-1}$, $\Omega_{m}=0.3$ and $\Omega_{\Lambda}=0.7$.}
\label{tab1}
\end{table*}

\begin{center}
\begin{table*}
\begin{scriptsize}
\begin{center}
\begin{tabular}{@{} lll @{}}
\hline
Reference & $a$ & $b$ \\
\hline\hline
\citep{dr87} & $1.33\pm 0.05$ & $-0.83 \pm 0.03$ \\
\citep{dj87} & $1.39\pm 0.14$ & $-0.90 \pm 0.09$ \\
\citep{lu91} & $1.27\pm 0.07$ & $-0.78 \pm 0.09$ \\
\citep{gu93} & $1.14$         & $-0.79$ \\
\citep{jo96} & $1.24\pm 0.07$ & $-0.82 \pm 0.02$ \\
\citep{hu97} & $1.38\pm 0.04$ & $-0.82 \pm 0.03$ \\
\citep{sc97} & $1.25\pm 0.02$ & $-0.79 \pm 0.03$ \\
\citep{pa98} & $1.53\pm 0.08$ & $-0.79 \pm 0.03$ \\
\citep{mu98} & $1.25$         & $-0.87$ \\
\citep{gi00} & $1.39\pm 0.04$ & $-0.84 \pm 0.01$ \\
\citep{co01} & $1.22\pm 0.09$ & $-0.84 \pm 0.03$ \\
\citep{be03} & $1.49\pm 0.05$ & $-0.75 \pm 0.01$ \\
\hline
this paper: & &\\
for $\sil$  & $1.08$ & $-0.77$ \\
~(upper limit)  & $1.28$ & $-0.82$ \\
~(lower limit)  & $0.90$ & $-0.72$ \\
for $\sio$  & $1.33$ & $-0.83$ \\
~(upper limit)  & $1.64$ & $-0.91$ \\
~(lower limit)  & $1.08$ & $-0.77$ \\
\hline\hline
\end{tabular}
\end{center}
\end{scriptsize}
\caption{List of previously found FP parameters and the results of this paper.}
\label{tab2}
\end{table*}
\end{center}
\label{lastpage}

\clearpage

\begin{figure*}
\begin{flushleft}
\textbf{The following Figures \ref{B2}, \ref{B3} and \ref{B4} are for online publication only.}
\end{flushleft}
\caption{CASTLES lenses used in this analysis. First column includes the Pixelens input files, the second column shows the formal velocity dispersion curves and in the third column the projected mass distributions (red dots mark the image positions, blue dots mark the source position) are presented. IMPORTANT NOTE: The y-axes of the velocity dispersion plots need to be multiplied by $\sqrt{2/\pi}\approx 0.8$ to yield the true $\sil$ values.}
\label{B2}
\vspace{0.5cm}
\begin{minipage}[l]{0.18\textwidth}
\small
\begin{tiny}
\begin{verbatim}
object B0047-2808
symm pixrad 10
redshifts 0.485 3.6
quad
 1.270  0.105
-0.630 -0.995  0
 0.520 -1.045  0
-0.730  0.705  0
g 13.7
\end{verbatim}
\end{tiny}\end{minipage}
\begin{minipage}[l]{0.15\textwidth}
\includegraphics[width=75pt]{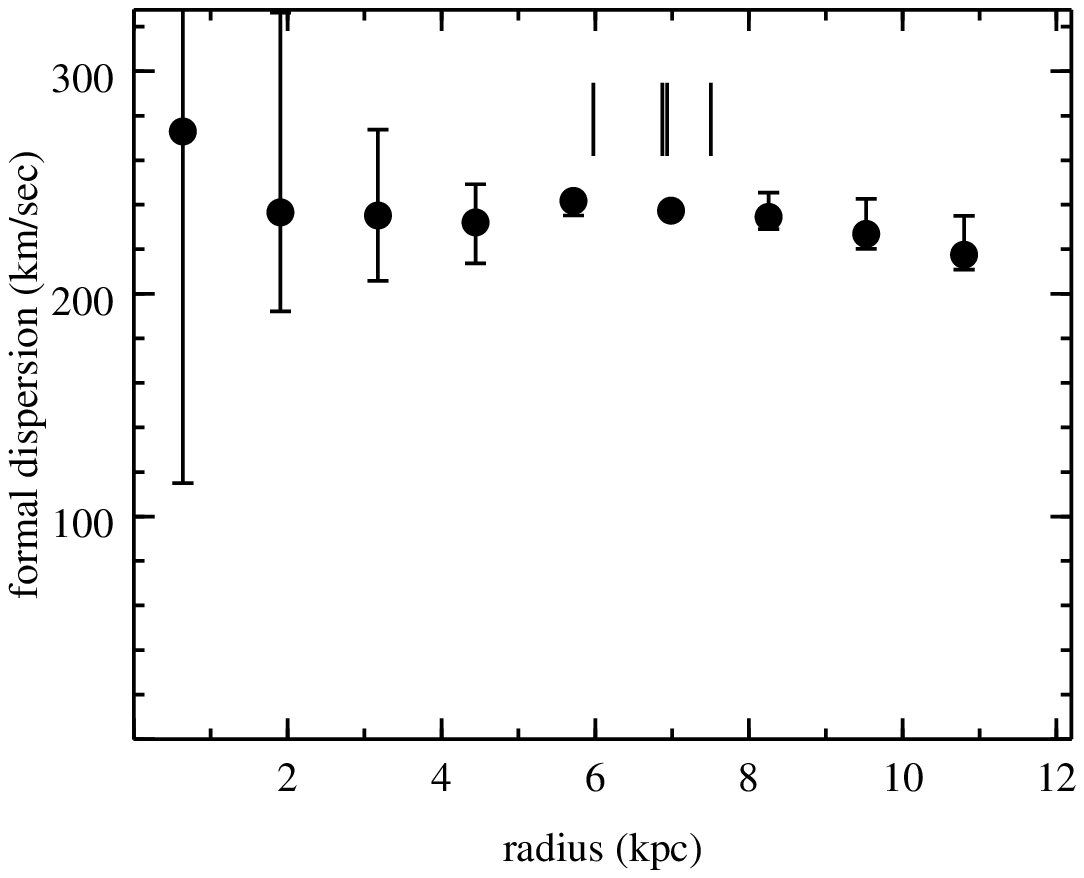}
\end{minipage}\begin{minipage}[r]{0.15\textwidth}
\includegraphics[width=52pt, bb = 0 15 320 320]{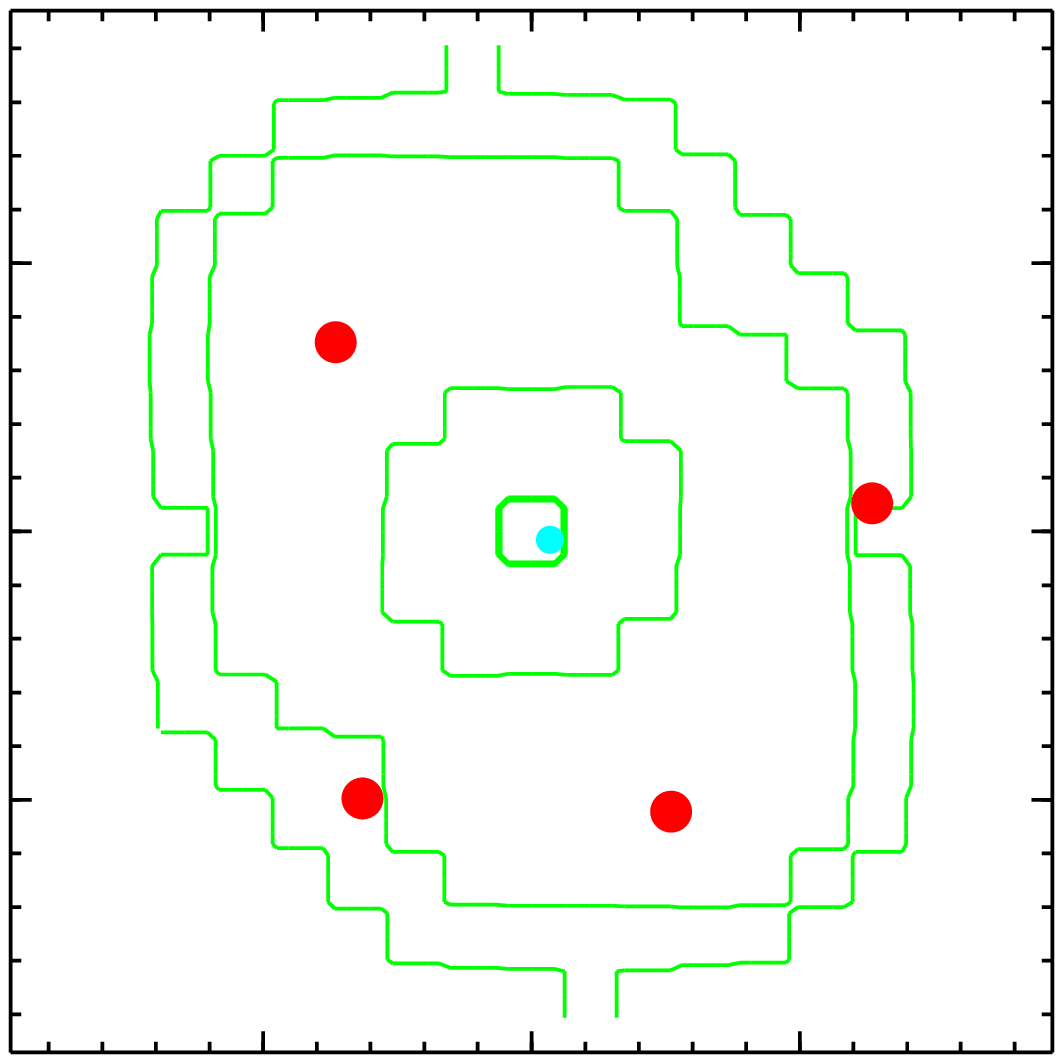}\\
\end{minipage}\begin{minipage}[l]{0.18\textwidth}
\small
\begin{tiny}
\begin{verbatim}
object CFRS03.1077
pixrad 10 
redshifts 0.938 2.941
quad
-2.085 0.383
-1.979 -0.521 0
-1.415 -1.213 0
0.713 0.340 0
g 13.7
\end{verbatim}
\end{tiny}\end{minipage}
\begin{minipage}[l]{0.15\textwidth}
\includegraphics[width=75pt]{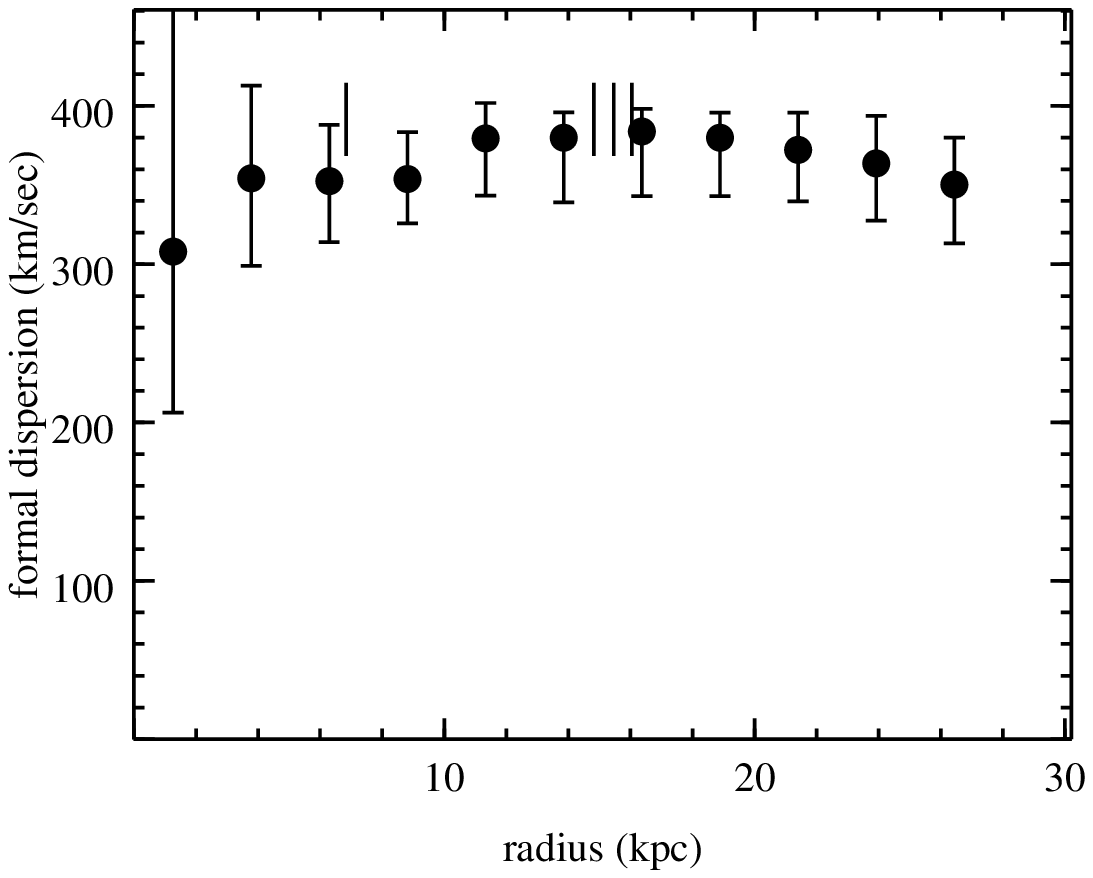}
\end{minipage}\begin{minipage}[r]{0.15\textwidth}
\includegraphics[width=52pt, bb = 0 15 320 320]{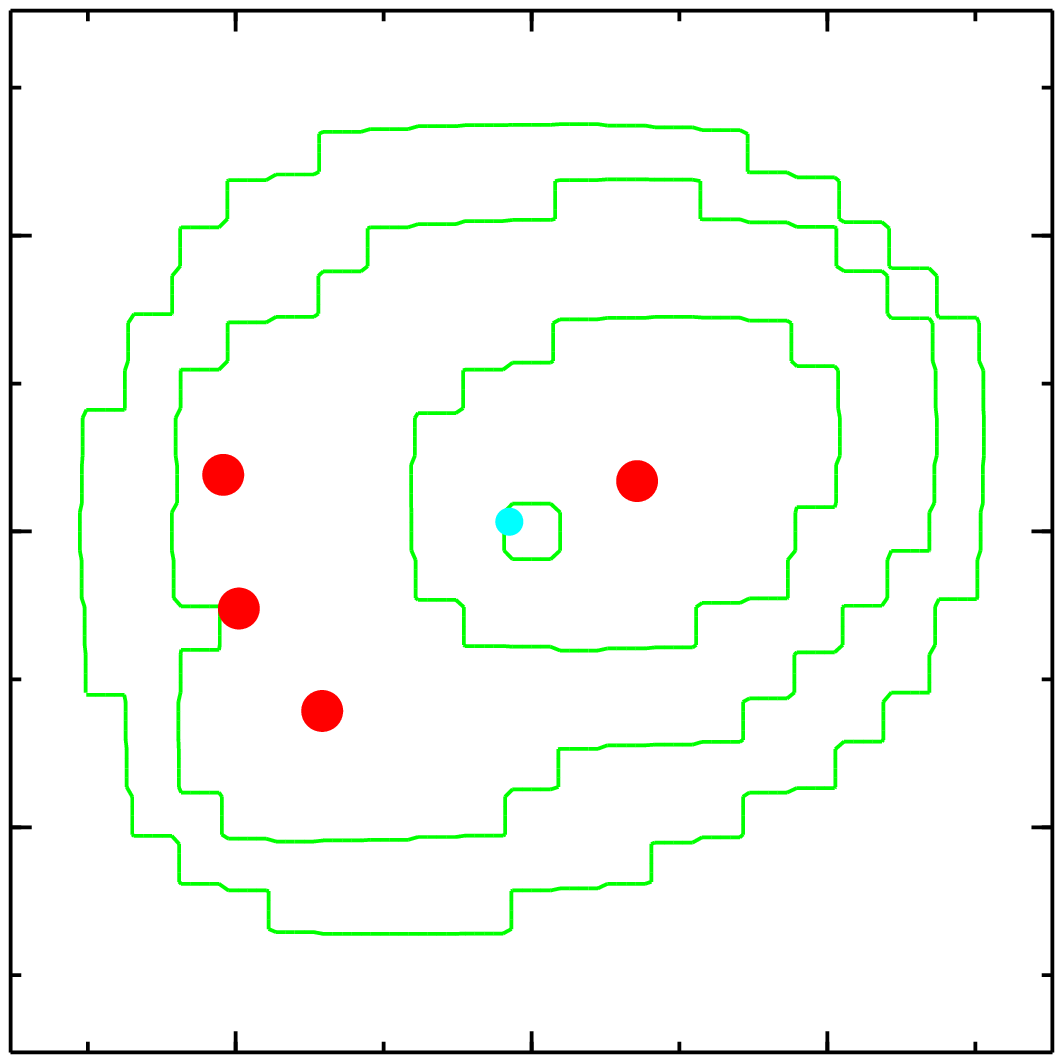}\\
\end{minipage}
\end{figure*}

\begin{figure*}
\begin{minipage}[l]{0.18\textwidth}
\small
\begin{tiny}
\begin{verbatim}
object Q0957+561 (2D)
symm pixrad 10
redshifts 0.356  1.41
shear -30
double
1.408  5.034
0.182 -1.018  423
double
2.860  3.470
-1.540 -0.050 0
g 13.7
\end{verbatim}
\end{tiny}\end{minipage}
\begin{minipage}[l]{0.15\textwidth}
\includegraphics[width=75pt]{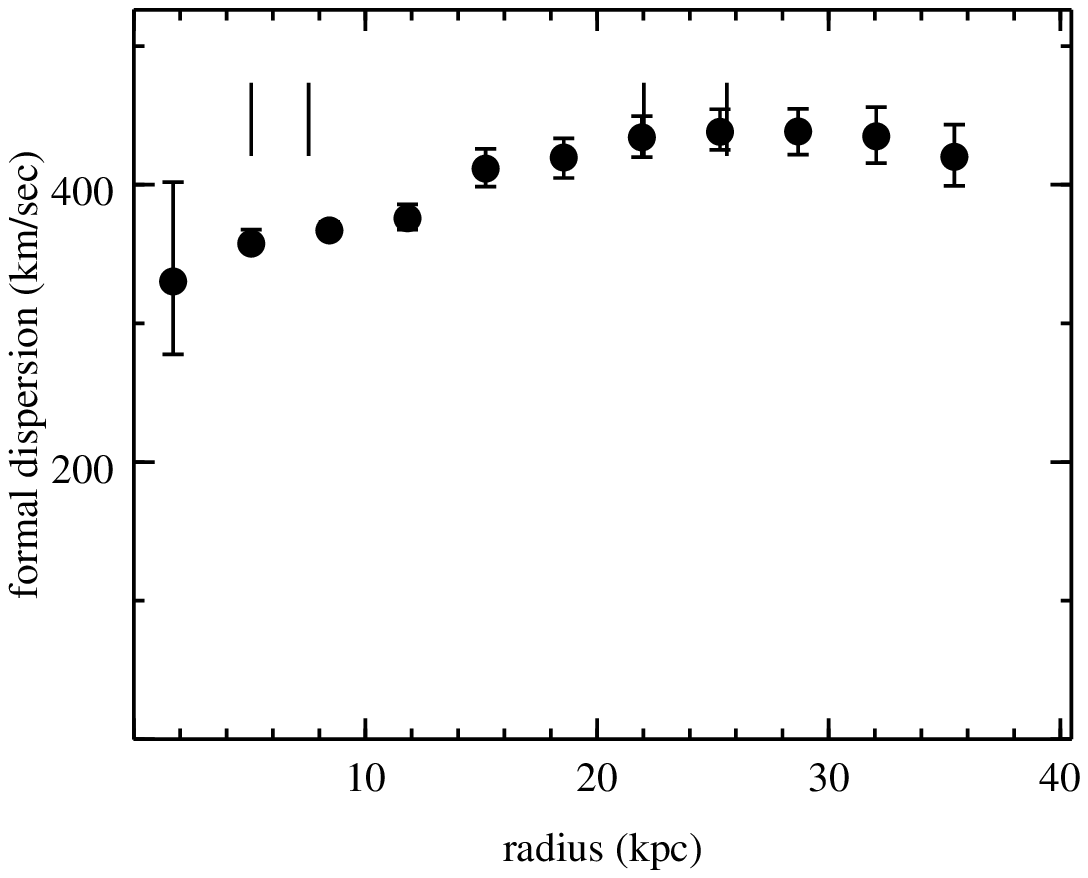}
\end{minipage}\begin{minipage}[r]{0.15\textwidth}
\includegraphics[width=52pt, bb = 0 15 320 320]{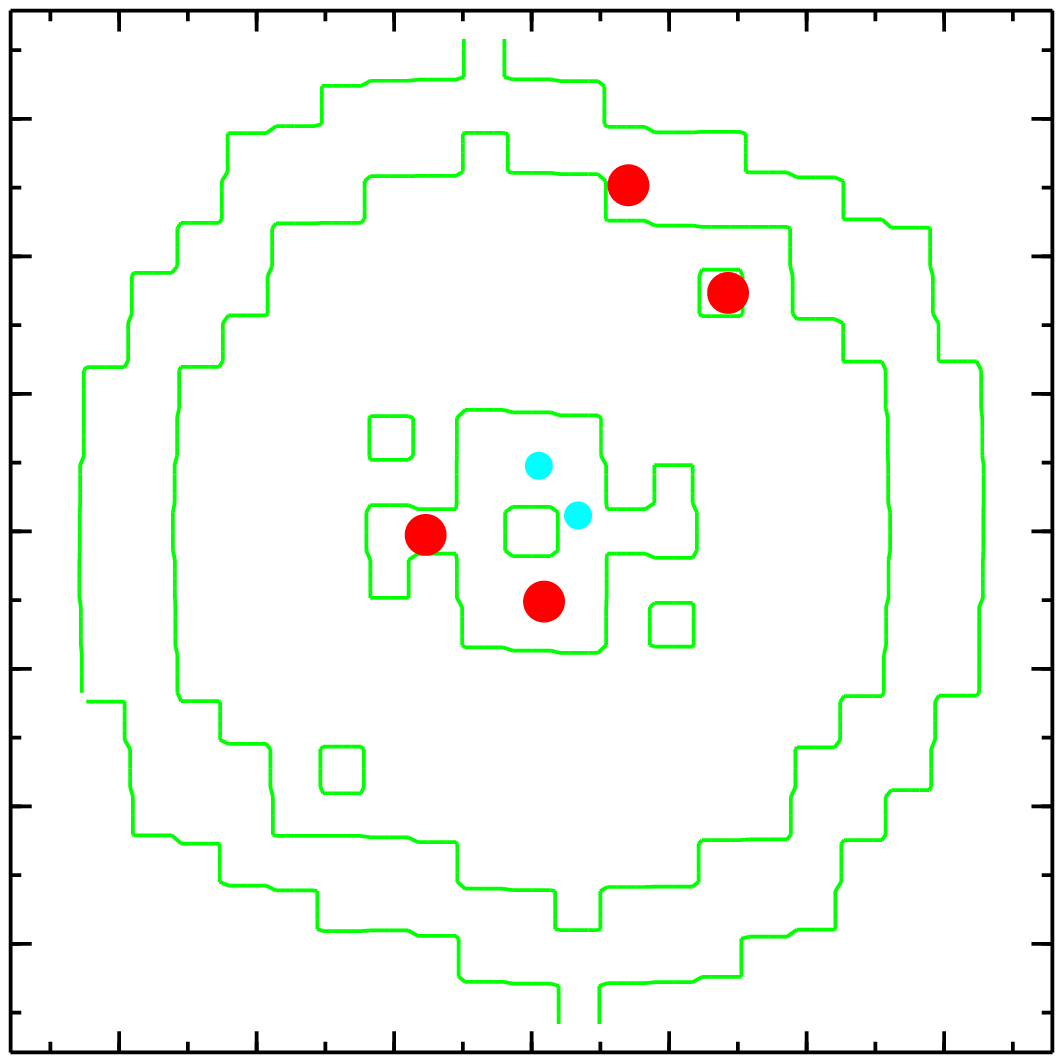}\\
\end{minipage}\begin{minipage}[l]{0.18\textwidth}
\small
\begin{tiny}
\begin{verbatim}
object PG1115+080
pixrad 10 
redshifts 0.31 1.72 
quad 
0.381 1.344
-0.947 -0.690  0 
-1.096 -0.232   0 
0.722 -0.617  0
g 13.7
\end{verbatim}
\end{tiny}\end{minipage}
\begin{minipage}[l]{0.15\textwidth}
\includegraphics[width=75pt]{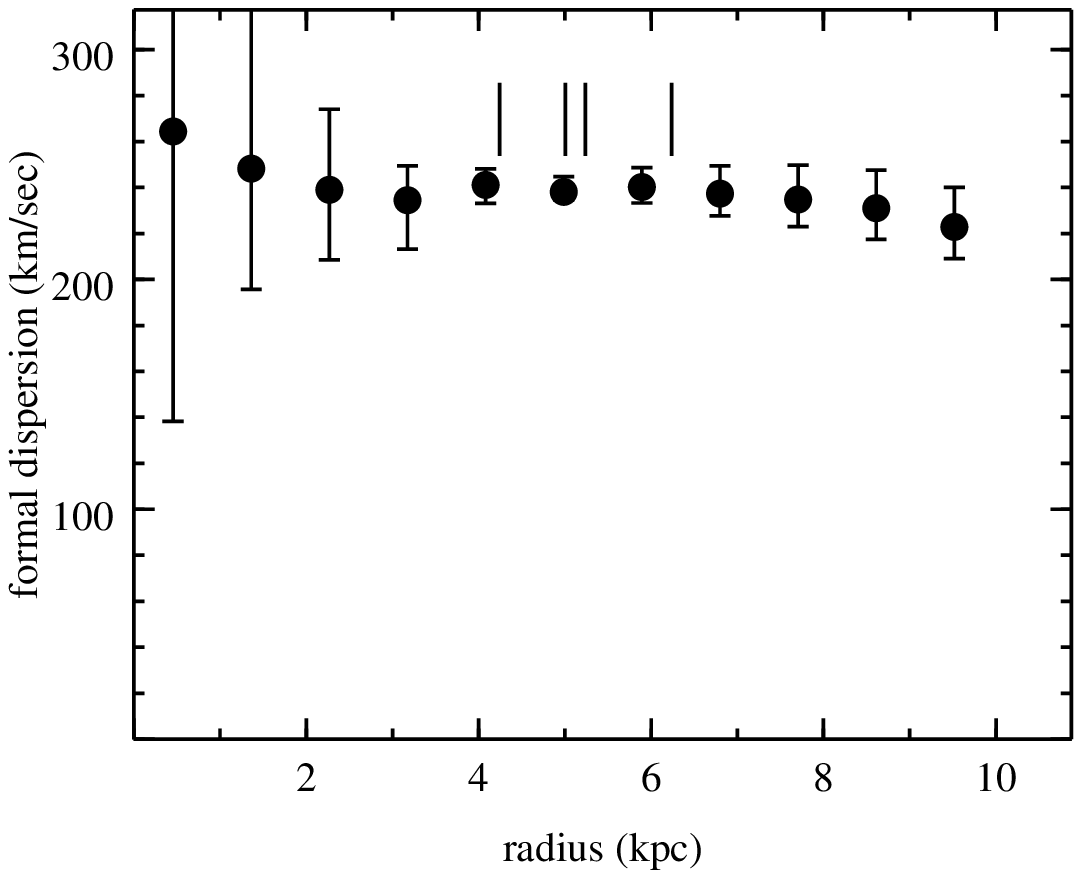}
\end{minipage}\begin{minipage}[r]{0.15\textwidth}
\includegraphics[width=52pt, bb = 0 15 320 320]{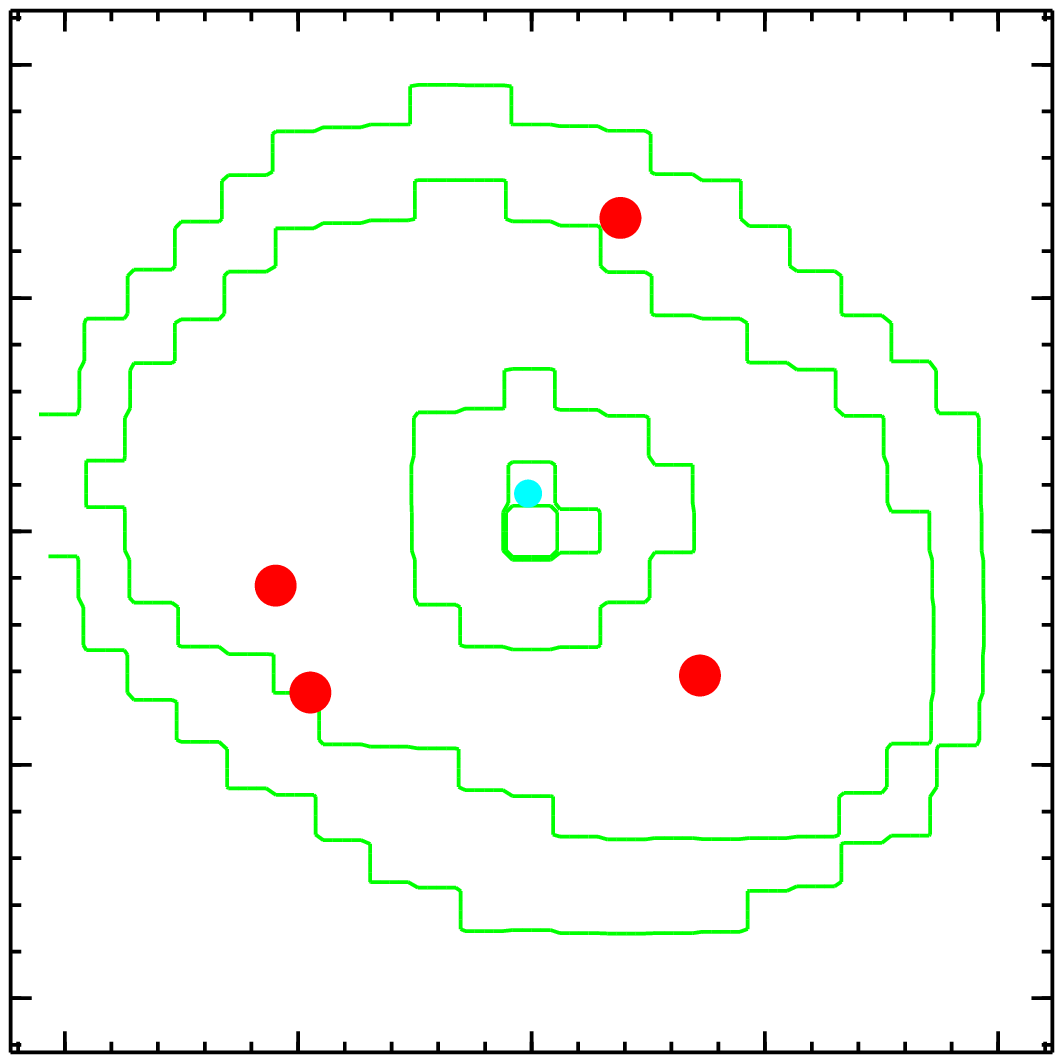}\\
\end{minipage}
\end{figure*}

\begin{figure*}
\begin{minipage}[l]{0.18\textwidth}
\small
\begin{tiny}
\begin{verbatim}
object HST14176+5226 
pixrad 10 
redshifts 0.81 3.4
quad 
 -1.288 -1.175 
-0.880 0.879 0 
0.792 1.332 0 
0.808 -0.794 0 
g 13.7
\end{verbatim}
\end{tiny}\end{minipage}
\begin{minipage}[l]{0.15\textwidth}
\includegraphics[width=75pt]{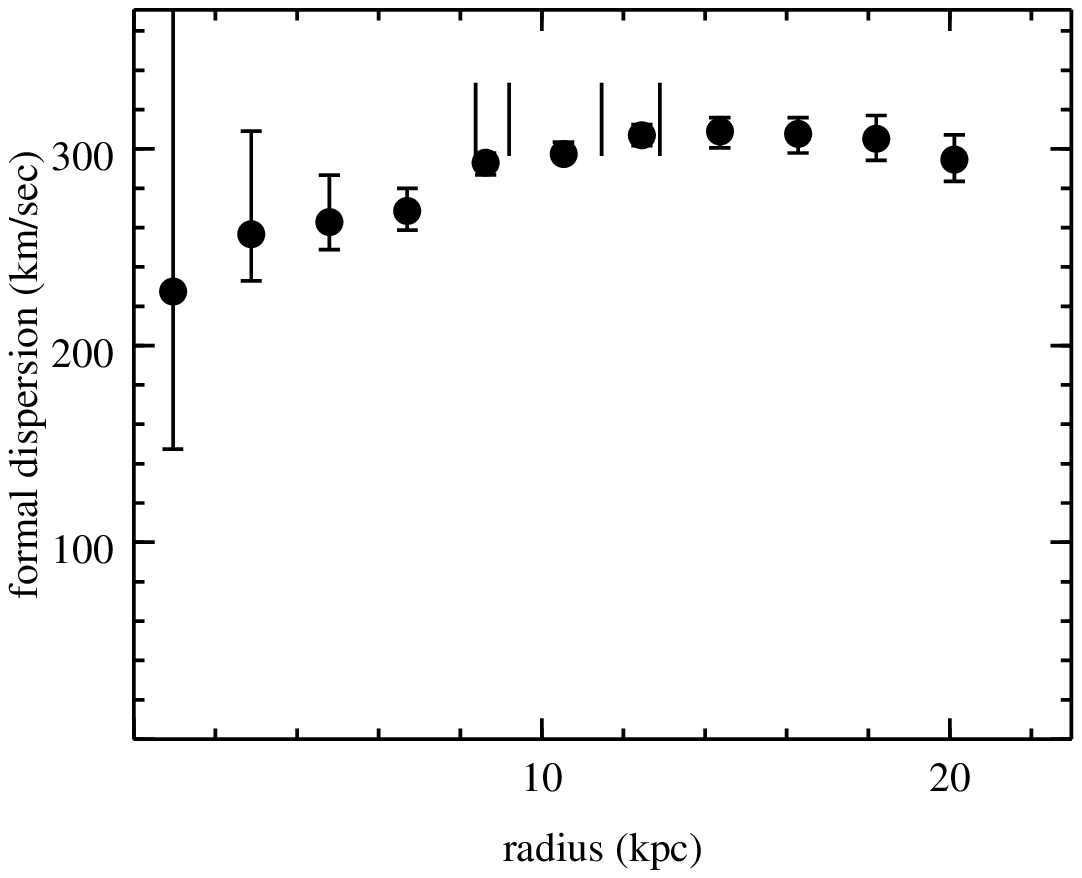}
\end{minipage}\begin{minipage}[r]{0.15\textwidth}
\includegraphics[width=52pt, bb = 0 15 320 320]{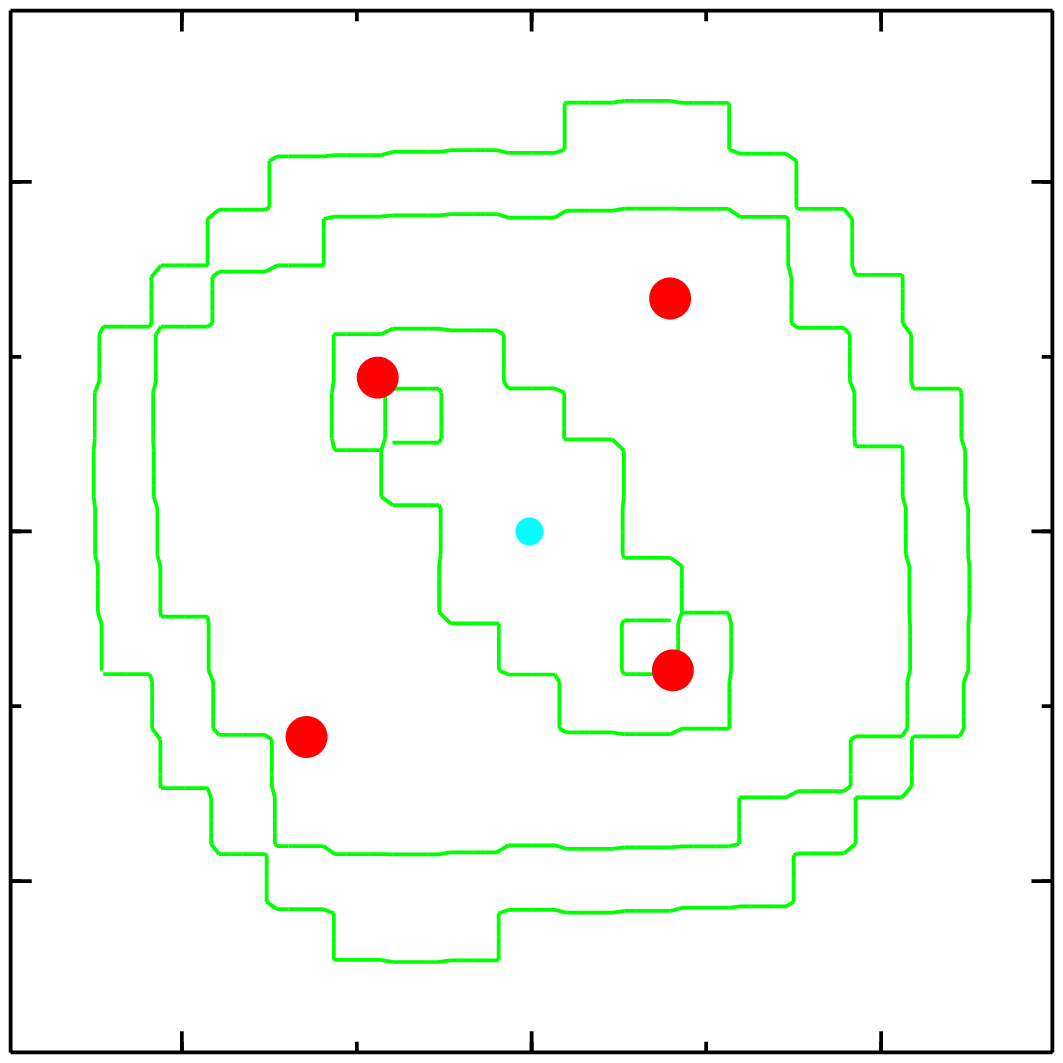}\\
\end{minipage}\begin{minipage}[l]{0.18\textwidth}
\small
\begin{tiny}
\begin{verbatim}
object HST15433+5352
pixrad 10 
redshifts 0.497 2.092 
quad 
0.590 0.200 
-0.110  -0.770  0 
0.360  -0.450   0 
-0.330 0.100  0
g 13.7 
\end{verbatim}
\end{tiny}\end{minipage}
\begin{minipage}[l]{0.15\textwidth}
\includegraphics[width=75pt]{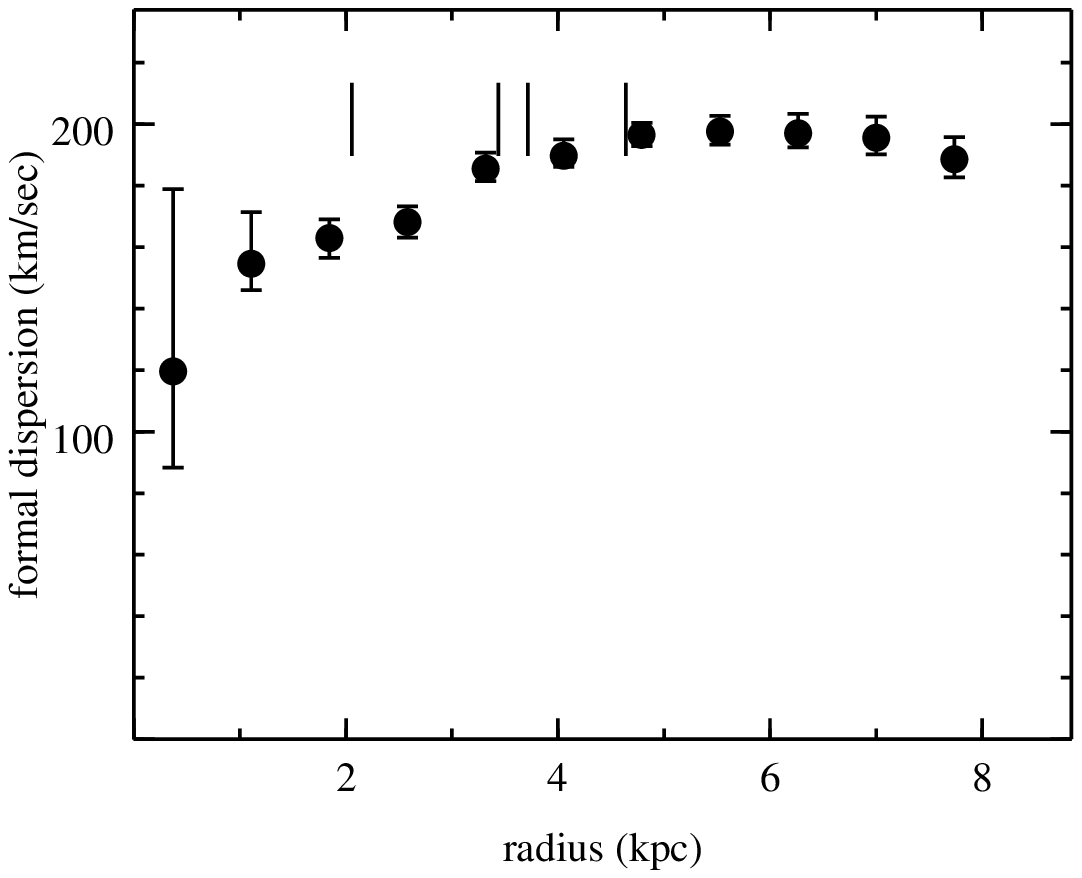}
\end{minipage}\begin{minipage}[r]{0.15\textwidth}
\includegraphics[width=52pt, bb = 0 15 320 320]{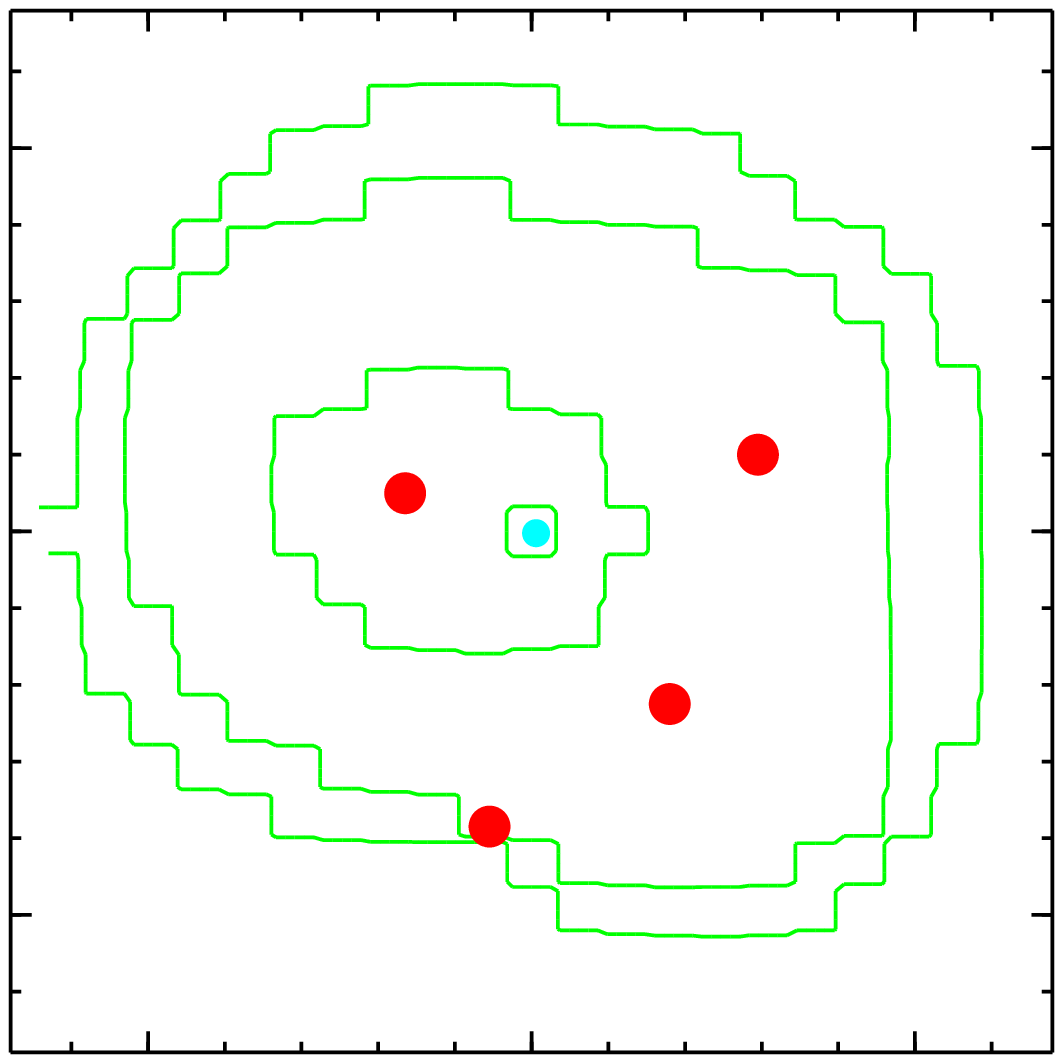}\\
\end{minipage}
\end{figure*}

\begin{figure*}
\begin{minipage}[l]{0.18\textwidth}
\small
\begin{tiny}
\begin{verbatim}
object B1608+656 
pixrad 10 
redshifts 0.630 1.394 
quad 
-1.300 -0.800 
-0.560  1.160  31 
-1.310  0.700   5 
 0.570 -0.080  40 
g 13.7
\end{verbatim}
\end{tiny}\end{minipage}
\begin{minipage}[l]{0.15\textwidth}
\includegraphics[width=75pt]{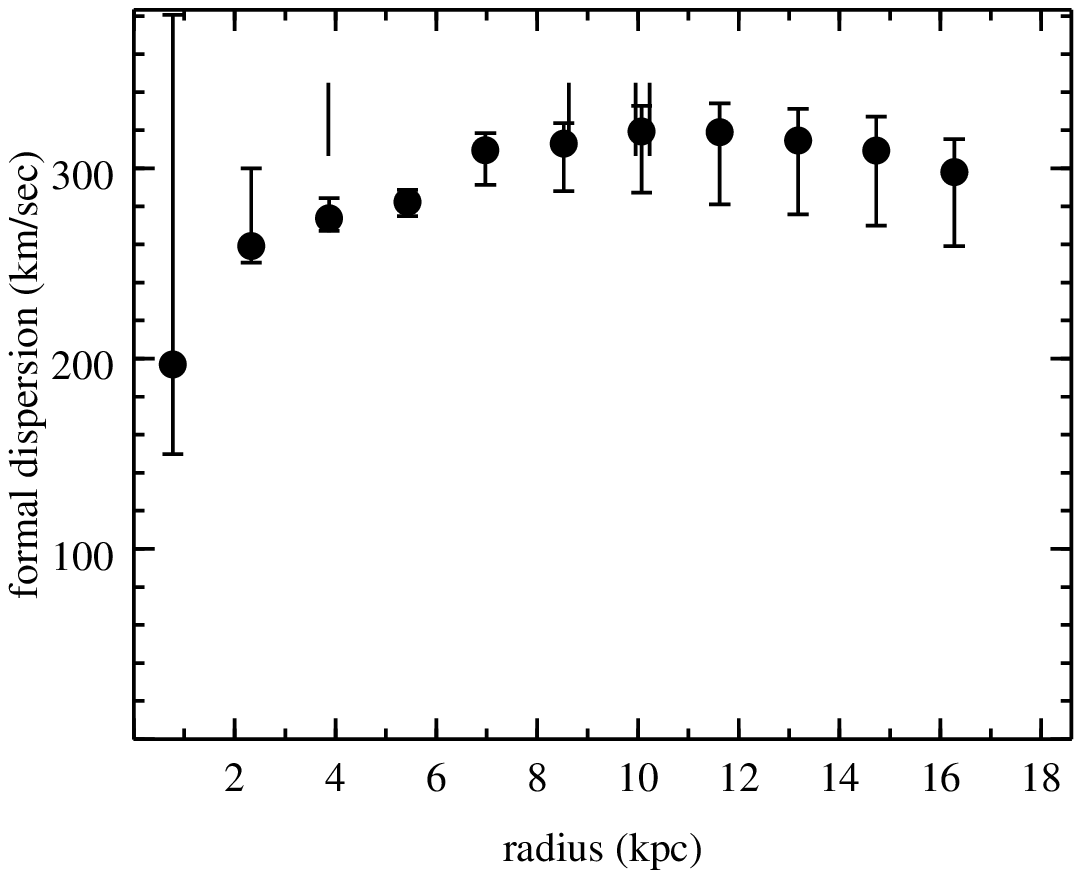}
\end{minipage}\begin{minipage}[r]{0.15\textwidth}
\includegraphics[width=52pt, bb = 0 15 320 320]{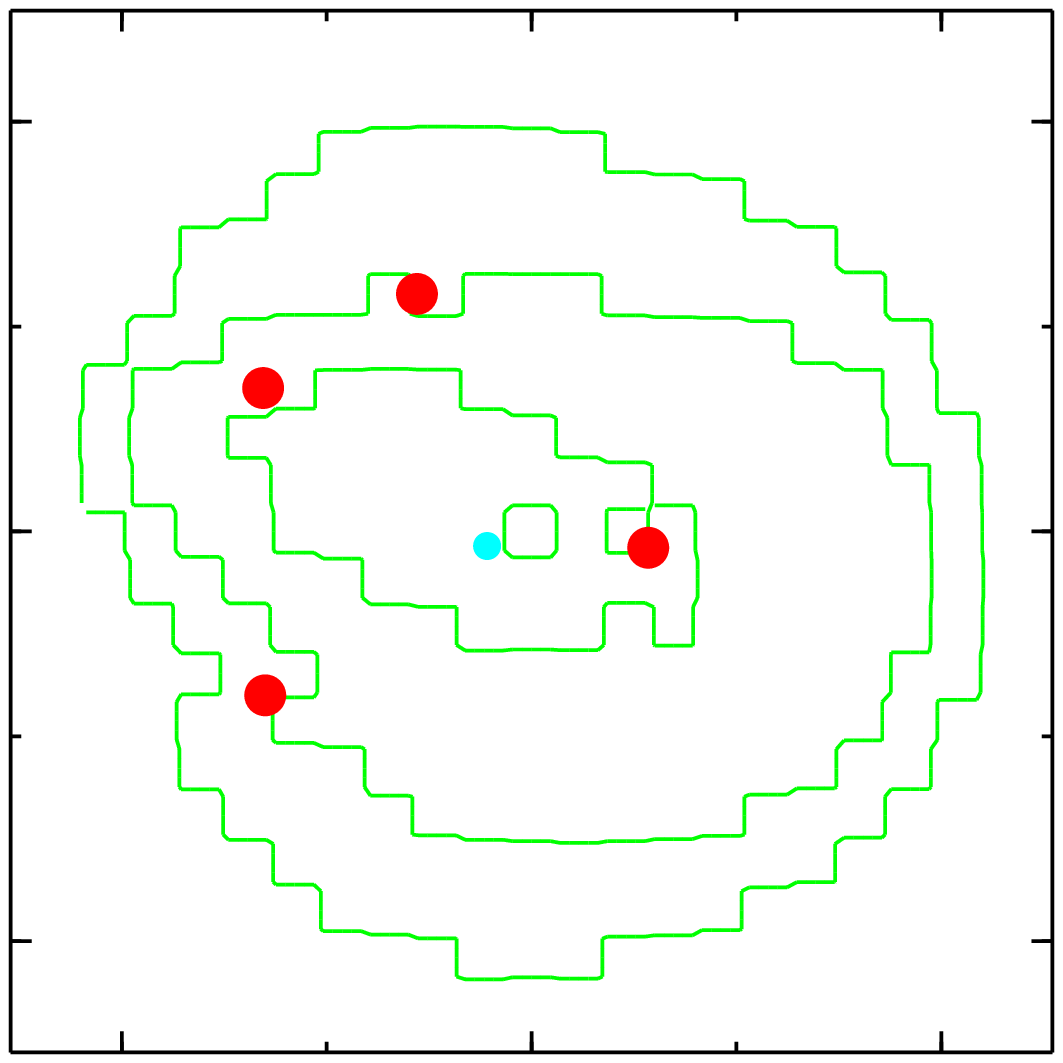}\\
\end{minipage}\begin{minipage}[l]{0.18\textwidth}
\small
\begin{tiny}
\begin{verbatim}
object MG2016+112
pixrad 10
redshifts 1.01 3.27
double
-1.735  1.778
1.269  0.274 0
g 13.7
\end{verbatim}
\end{tiny}\end{minipage}
\begin{minipage}[l]{0.15\textwidth}
\includegraphics[width=75pt]{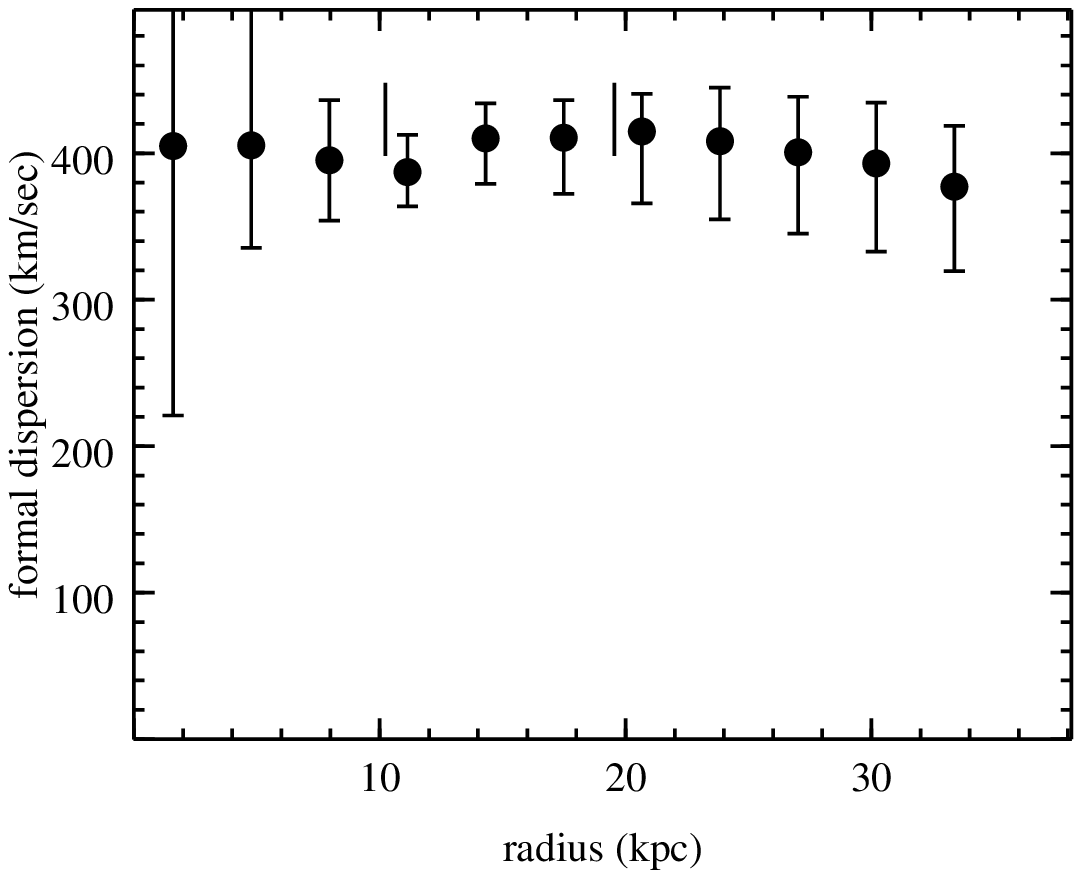}
\end{minipage}\begin{minipage}[r]{0.15\textwidth}
\includegraphics[width=52pt, bb = 0 15 320 320]{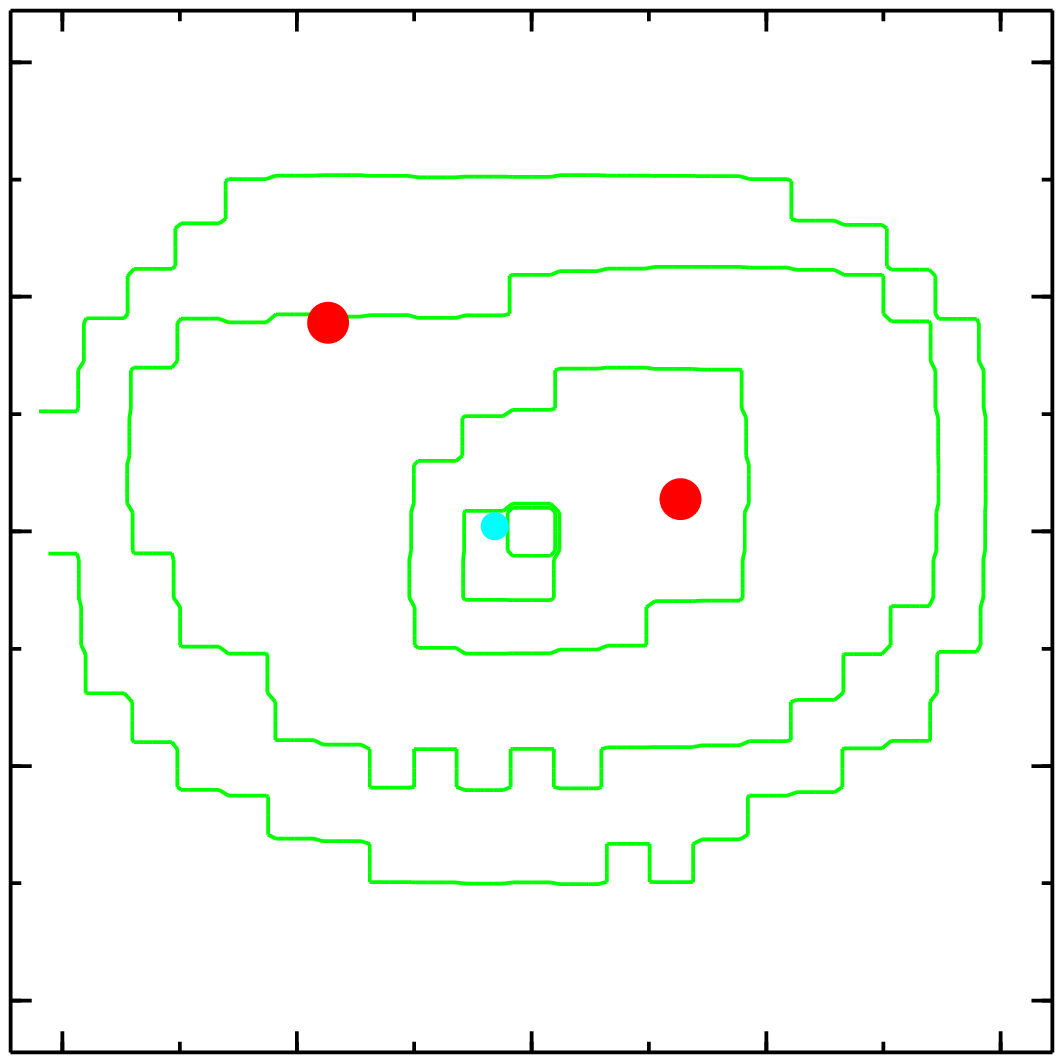}\\
\end{minipage}
\end{figure*}

\begin{figure*}
\begin{flushleft}
\hspace{0.15cm}
\begin{minipage}[l]{0.18\textwidth}
\small
\begin{tiny}
\begin{verbatim}
object Q2237+030
pixrad 10 
redshifts 0.04 1.69
quad 
0.598 0.758
-0.075 -0.939 0 
0.791 -0.411 0
-0.710 0.271 0
g 13.7
\end{verbatim}
\end{tiny}\end{minipage}
\begin{minipage}[l]{0.15\textwidth}
\includegraphics[width=75pt]{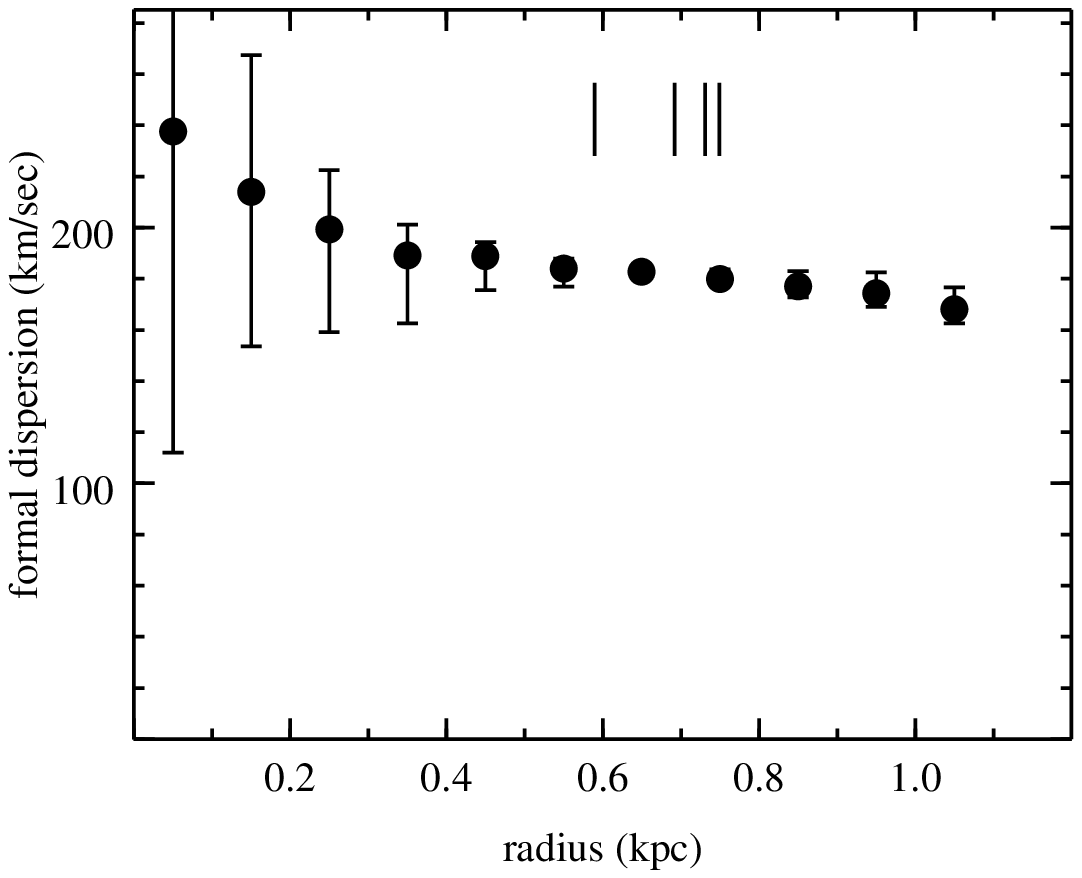}
\end{minipage}\begin{minipage}[r]{0.15\textwidth}
\includegraphics[width=52pt, bb = 0 15 320 320]{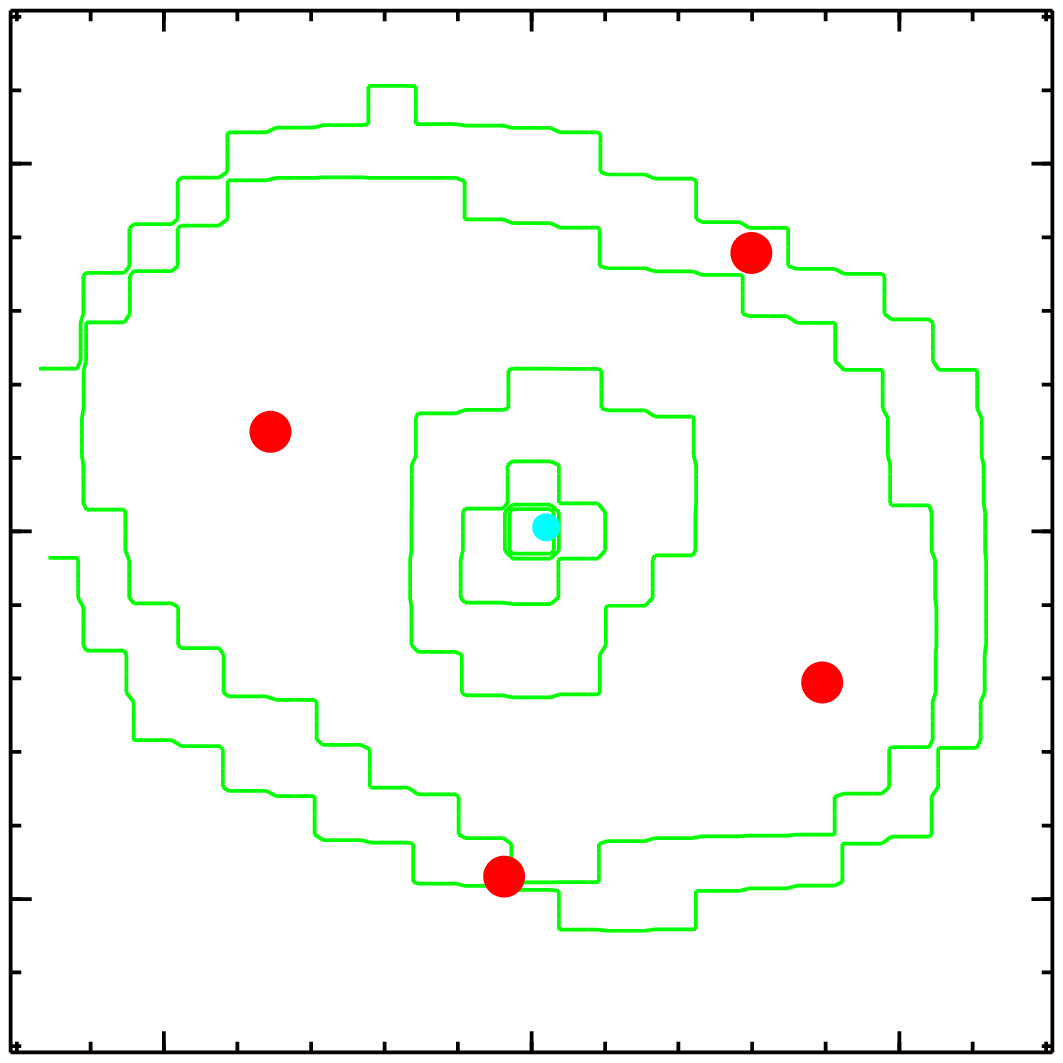}\\
\end{minipage}\begin{minipage}[l]{0.15\textwidth}
\small
\begin{tiny}
\end{tiny}\end{minipage}
\begin{minipage}[l]{0.15\textwidth}
\end{minipage}\begin{minipage}[r]{0.15\textwidth}
\end{minipage}
\end{flushleft}
\end{figure*}

\clearpage

\begin{figure*}
\caption{SLACS lenses used in this analysis. First column includes the Pixelens input files, the second column shows the formal velocity dispersion curves and in the third column the projected mass distributions (red dots mark the image positions, blue dots mark the source position) are presented. IMPORTANT NOTE: The y-axes of the velocity dispersion plots need to be multiplied by $\sqrt{2/\pi}\approx 0.8$ to yield the true $\sil$ values.}
\label{B3}
\vspace{0.5cm}
\begin{minipage}[l]{0.18\textwidth}
\small
\begin{tiny}
\begin{verbatim}
object J0037-094 
pixrad 10 
redshifts 0.1954 0.6322
double 
1.920 -0.730
-0.730 0.730 0
g 13.7
\end{verbatim}
\end{tiny}\end{minipage}
\begin{minipage}[l]{0.15\textwidth}
\includegraphics[width=75pt]{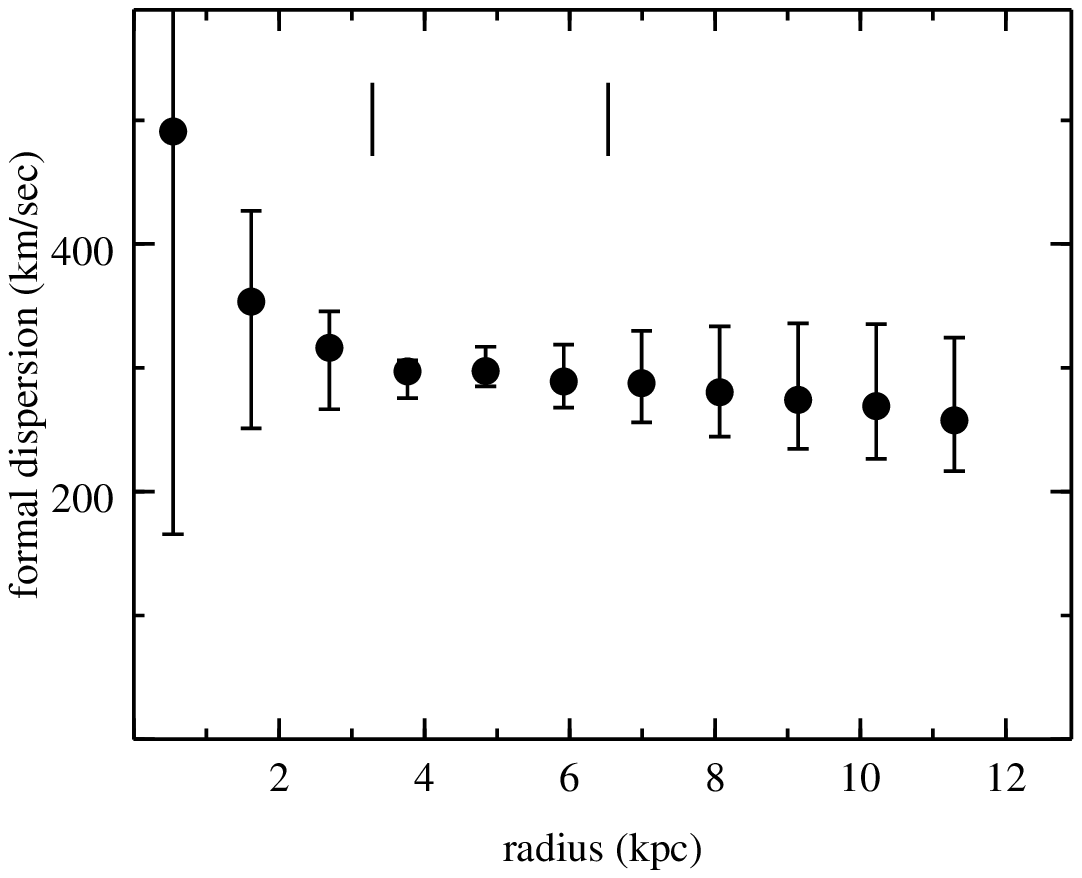}
\end{minipage}\begin{minipage}[r]{0.15\textwidth}
\includegraphics[width=52pt, bb = 0 15 320 320]{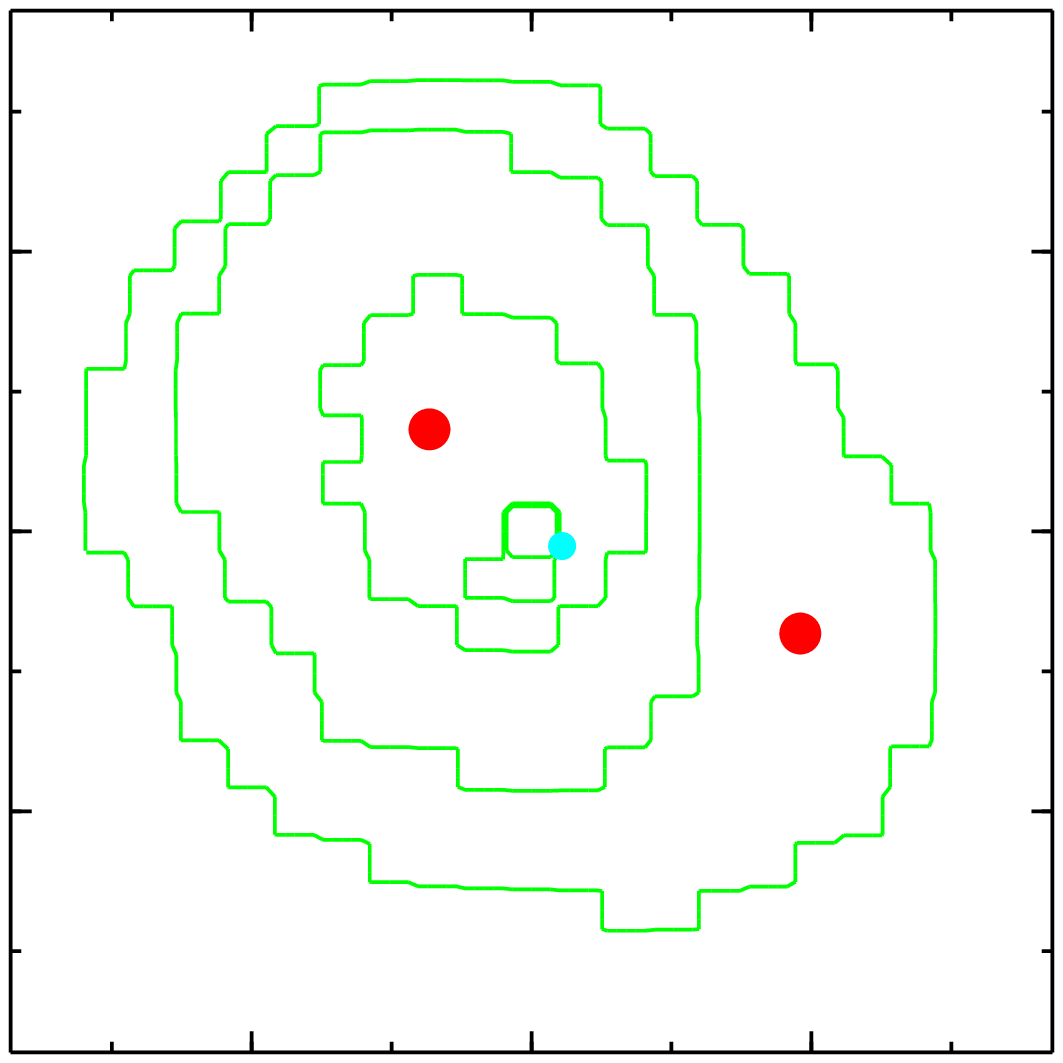}\\
\end{minipage}\begin{minipage}[l]{0.18\textwidth}
\small
\begin{tiny}
\begin{verbatim}
object J0737+321 
pixrad 10 
zlens 0.3223
multi 3 0.5812
-0.08 1.17   1 
0.52  -0.77  2 
-0.68 -0.57  2 
g 13.7
\end{verbatim}
\end{tiny}\end{minipage}
\begin{minipage}[l]{0.15\textwidth}
\includegraphics[width=75pt]{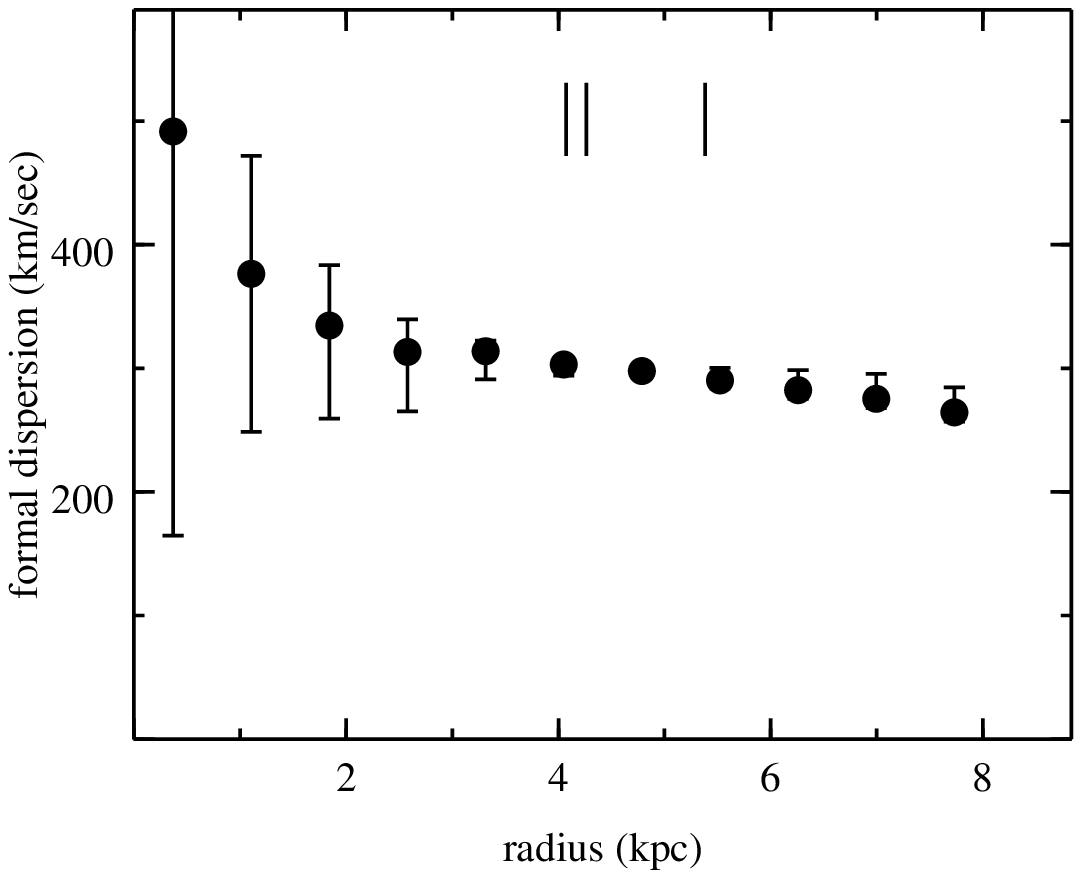}
\end{minipage}\begin{minipage}[r]{0.15\textwidth}
\includegraphics[width=52pt, bb = 0 15 320 320]{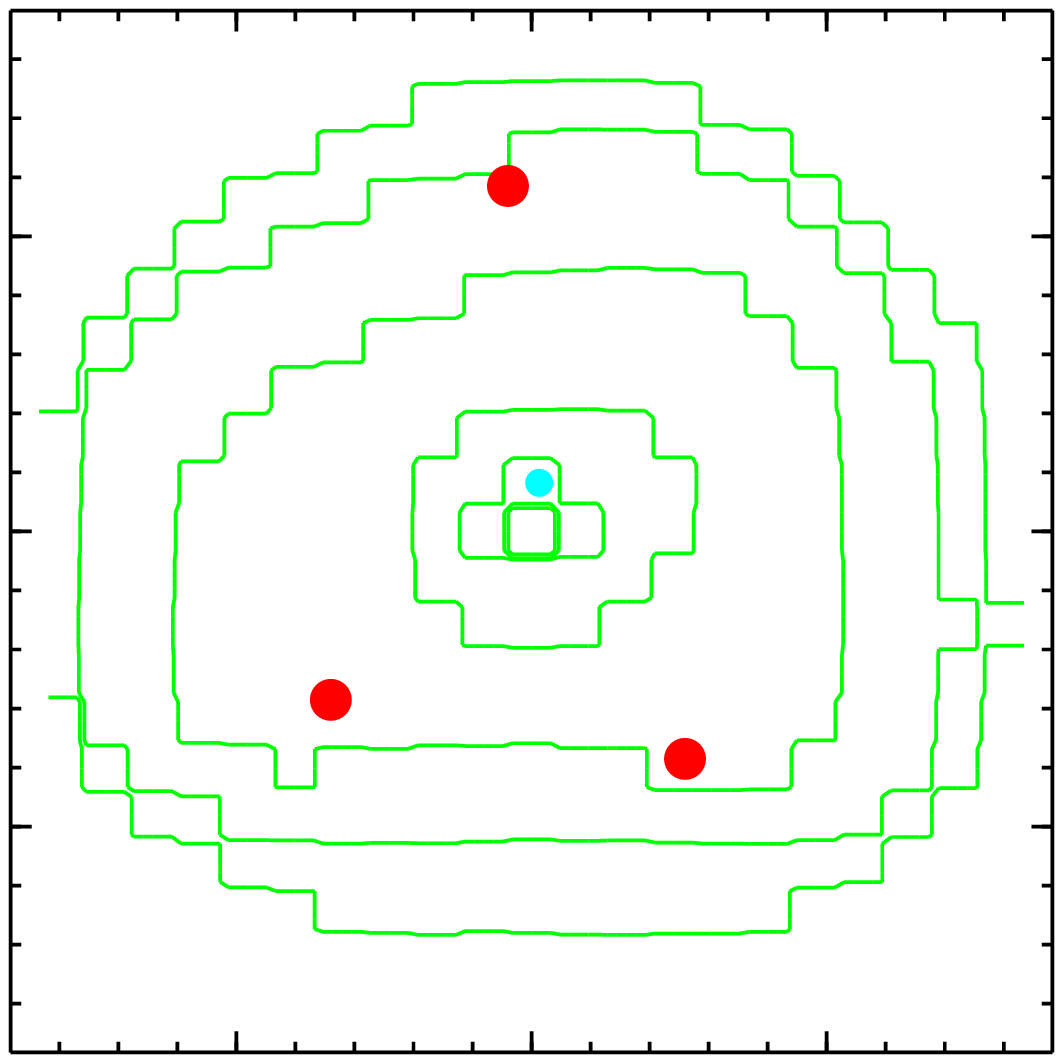}\\
\end{minipage}
\end{figure*}

\begin{figure*}
\begin{minipage}[l]{0.18\textwidth}
\small
\begin{tiny}
\begin{verbatim}
object J0912
pixrad 10
redshifts 0.1642 0.3239
double -1.275 1.025 0.375 -0.825 0
double -0.425 1.675 0.225 -0.975 0
double 0.975 1.575 -0.425 -1.075 0
double 1.475 1.225 -0.625 -0.575 0
g 13.7
\end{verbatim}
\end{tiny}\end{minipage}
\begin{minipage}[l]{0.15\textwidth}
\includegraphics[width=75pt]{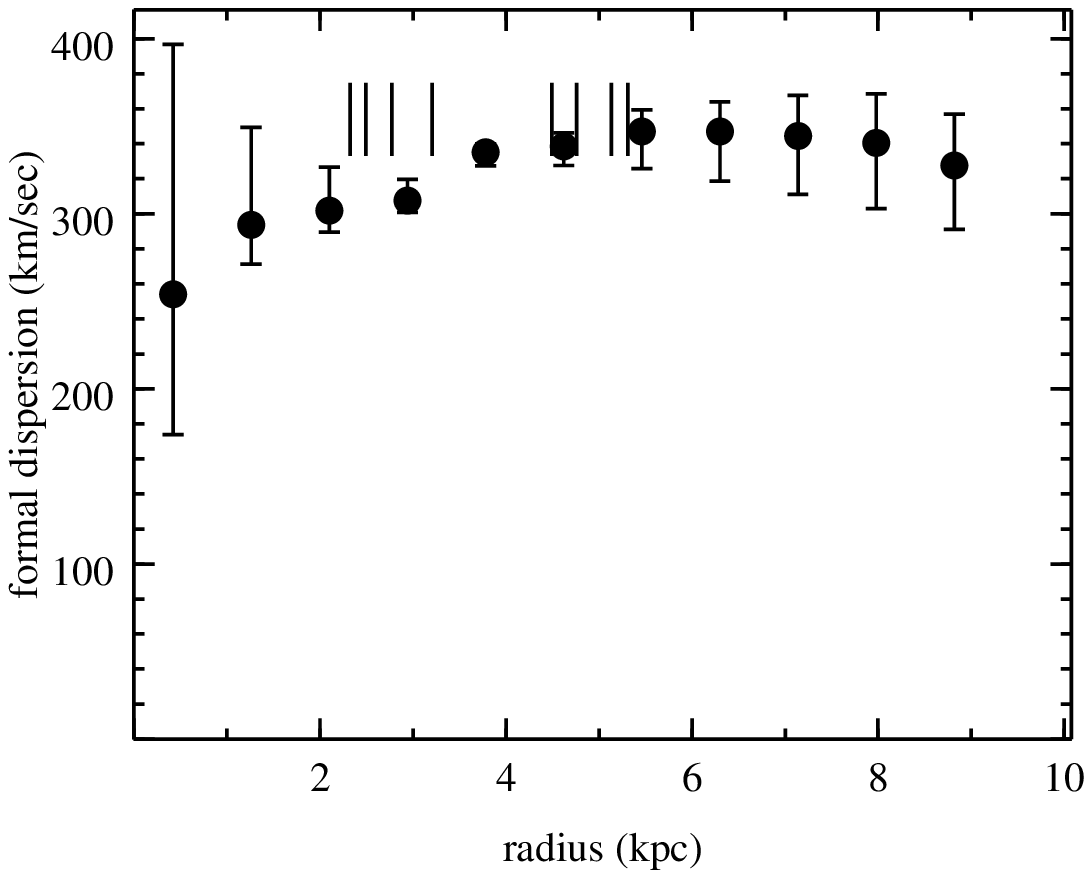}
\end{minipage}\begin{minipage}[r]{0.15\textwidth}
\includegraphics[width=52pt, bb = 0 15 320 320]{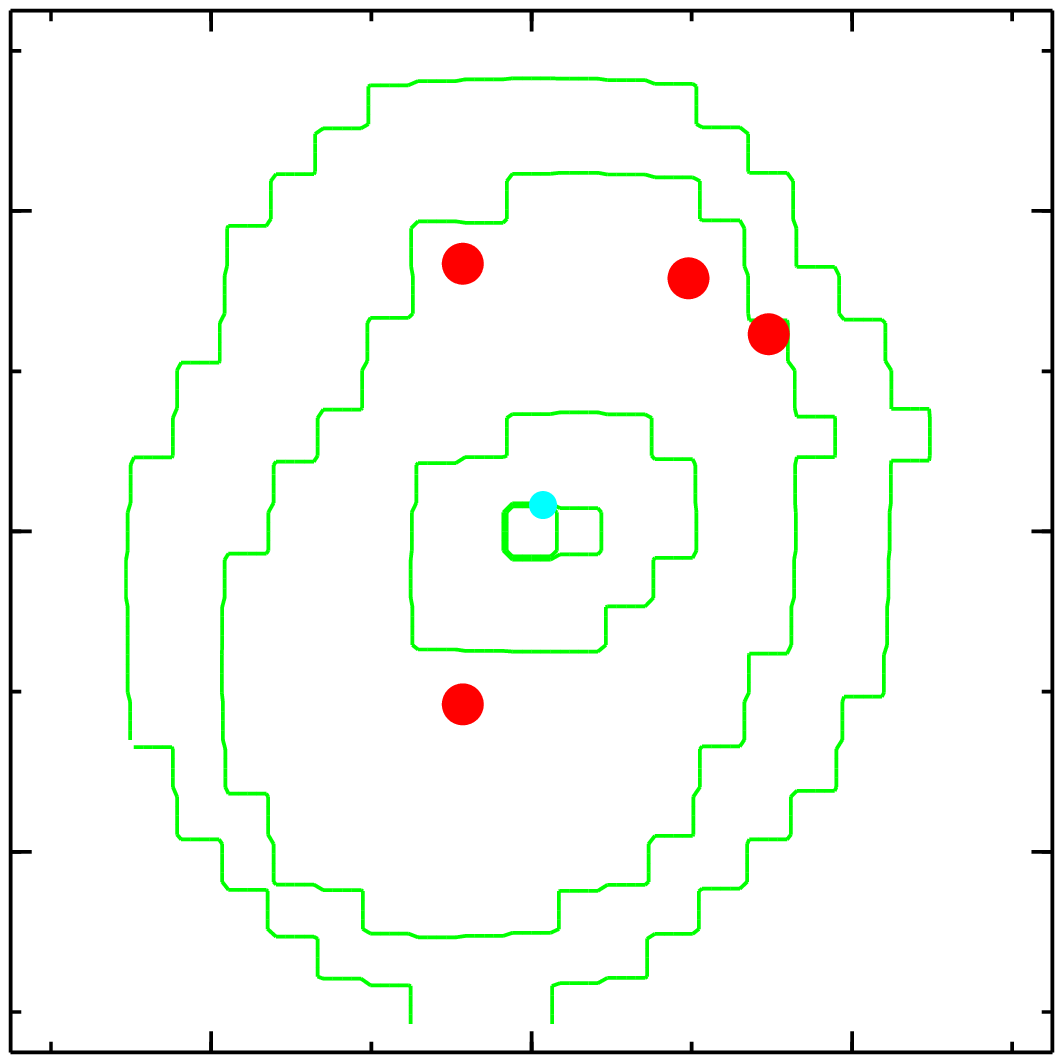}\\
\end{minipage}\begin{minipage}[l]{0.18\textwidth}
\small
\begin{tiny}
\begin{verbatim}
object J0956+510
pixrad 10 
zlens 0.2405
multi 3 0.47
-1.67	0.170 1	
-0.220	-1.270 2
0.730	0.620 2
g 13.7
\end{verbatim}
\end{tiny}\end{minipage}
\begin{minipage}[l]{0.15\textwidth}
\includegraphics[width=75pt]{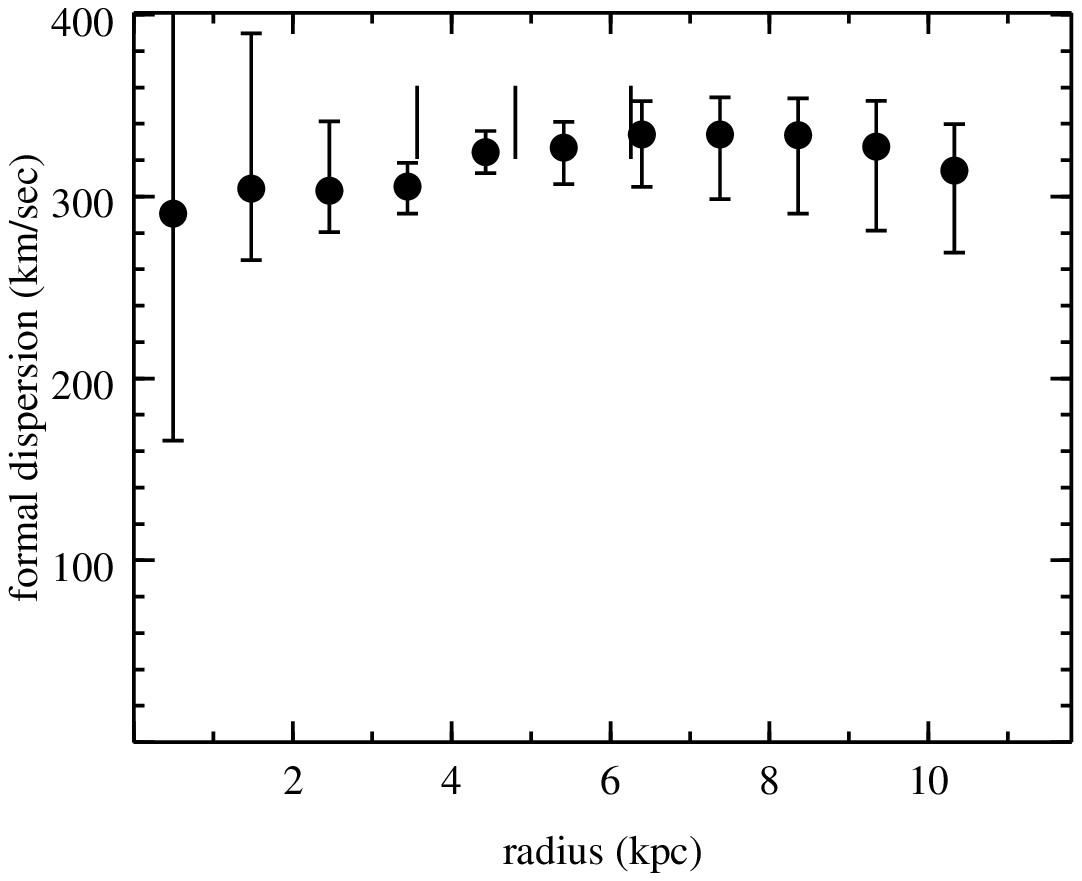}
\end{minipage}\begin{minipage}[r]{0.15\textwidth}
\includegraphics[width=52pt, bb = 0 15 320 320]{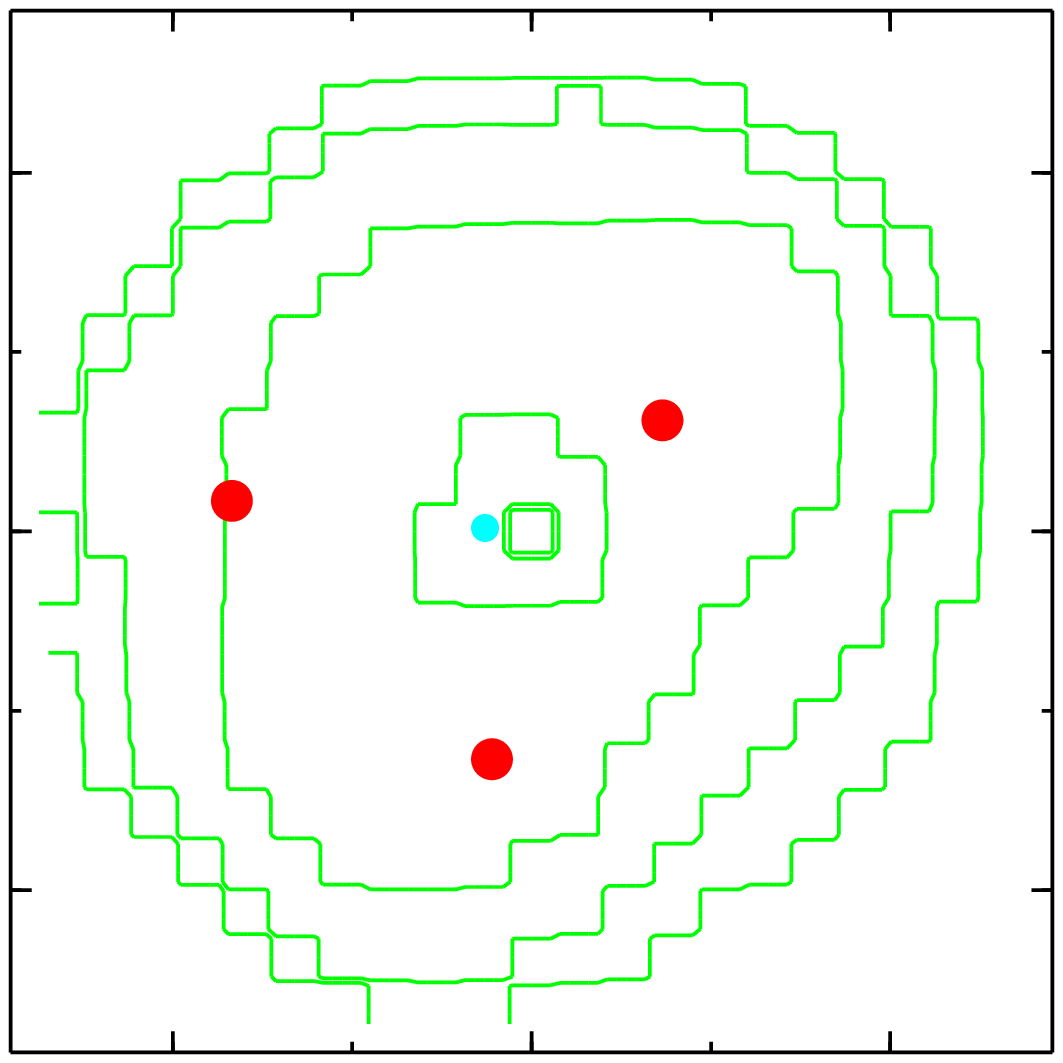}\\
\end{minipage}
\end{figure*}

\begin{figure*}
\begin{minipage}[l]{0.18\textwidth}
\small
\begin{tiny}
\begin{verbatim}
object J1205+491
pixrad 10 
redshifts 0.2150 0.4808
quad
1.480	-0.320 
-0.88	0.570 0
-0.520 0.930 0
-0.820 -0.470 0
g 13.7
\end{verbatim}
\end{tiny}\end{minipage}
\begin{minipage}[l]{0.15\textwidth}
\includegraphics[width=75pt]{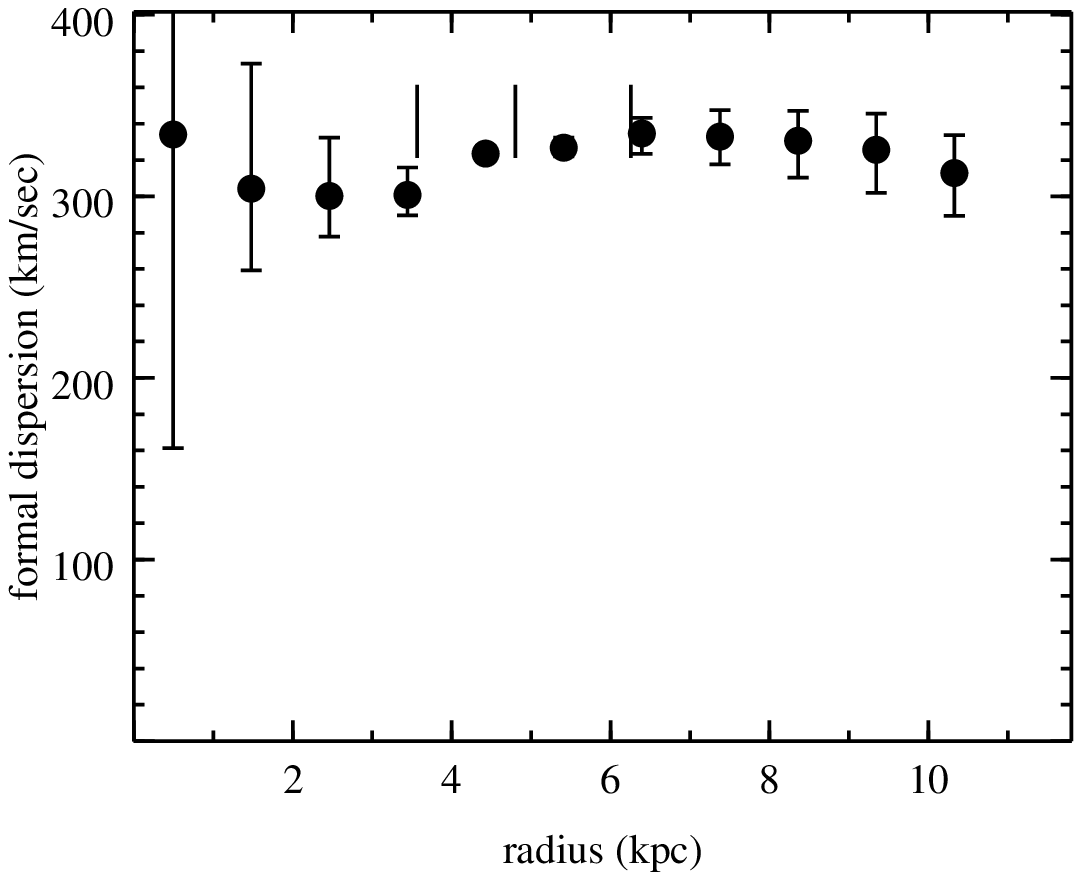}
\end{minipage}\begin{minipage}[r]{0.15\textwidth}
\includegraphics[width=52pt, bb = 0 15 320 320]{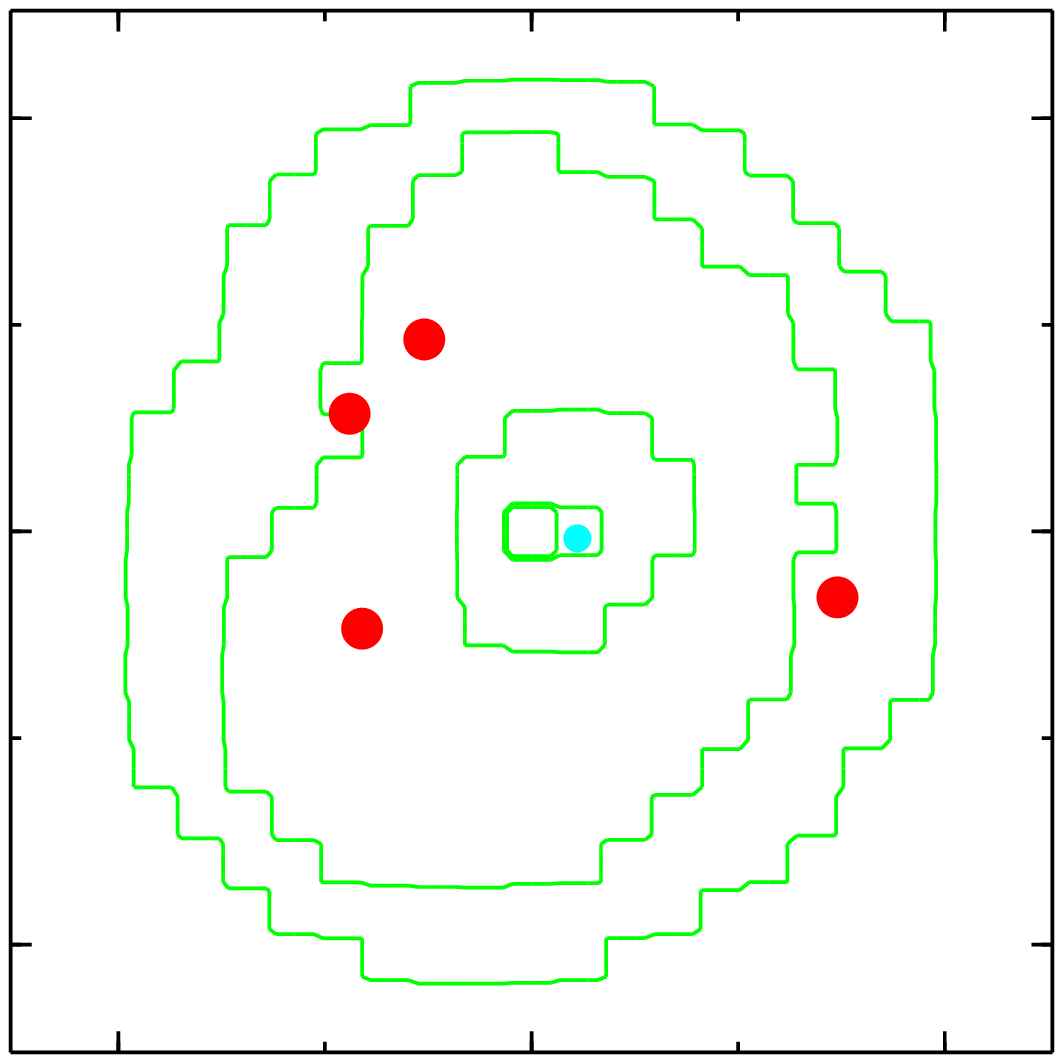}\\
\end{minipage}\begin{minipage}[l]{0.18\textwidth}
\small
\begin{tiny}
\begin{verbatim}
object J1330-014
symm pixrad 10 
redshifts 0.0808 0.7115
double
1.020	-0.520
-0.43	0.080 0
g 13.7
\end{verbatim}
\end{tiny}\end{minipage}
\begin{minipage}[l]{0.15\textwidth}
\includegraphics[width=75pt]{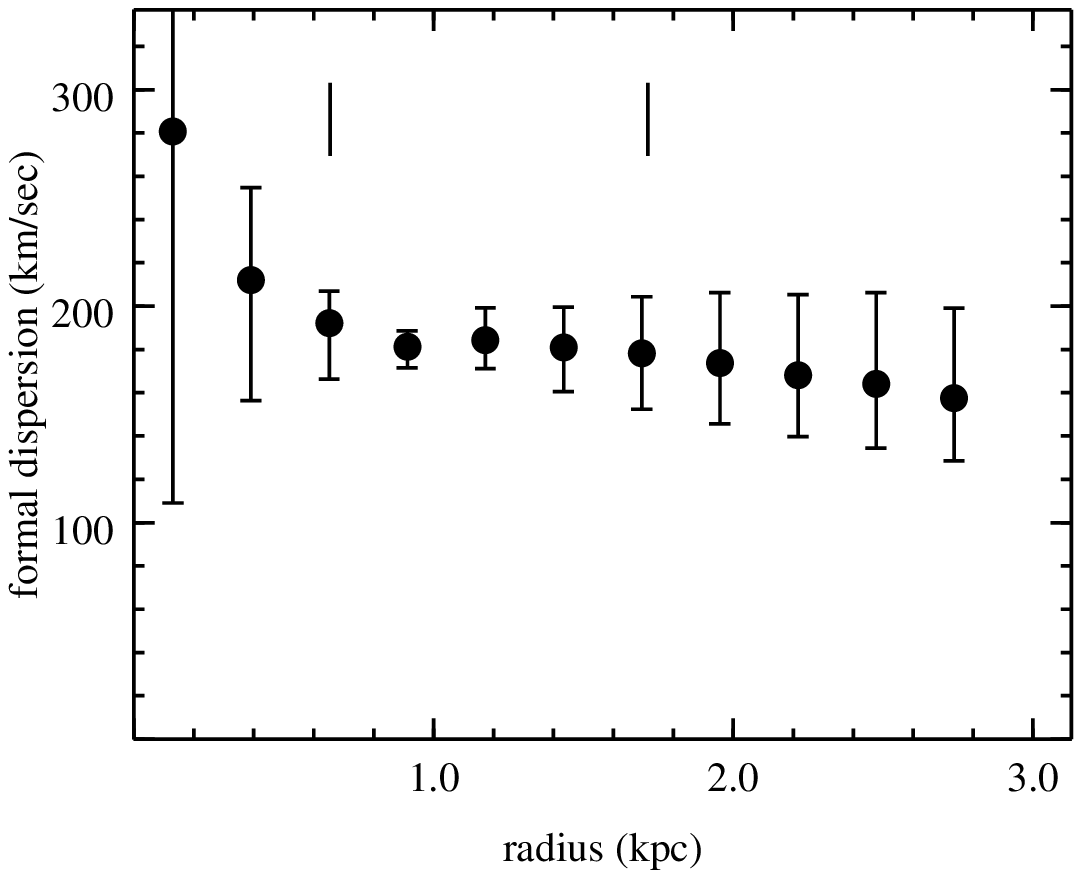}
\end{minipage}\begin{minipage}[r]{0.15\textwidth}
\includegraphics[width=52pt, bb = 0 15 320 320]{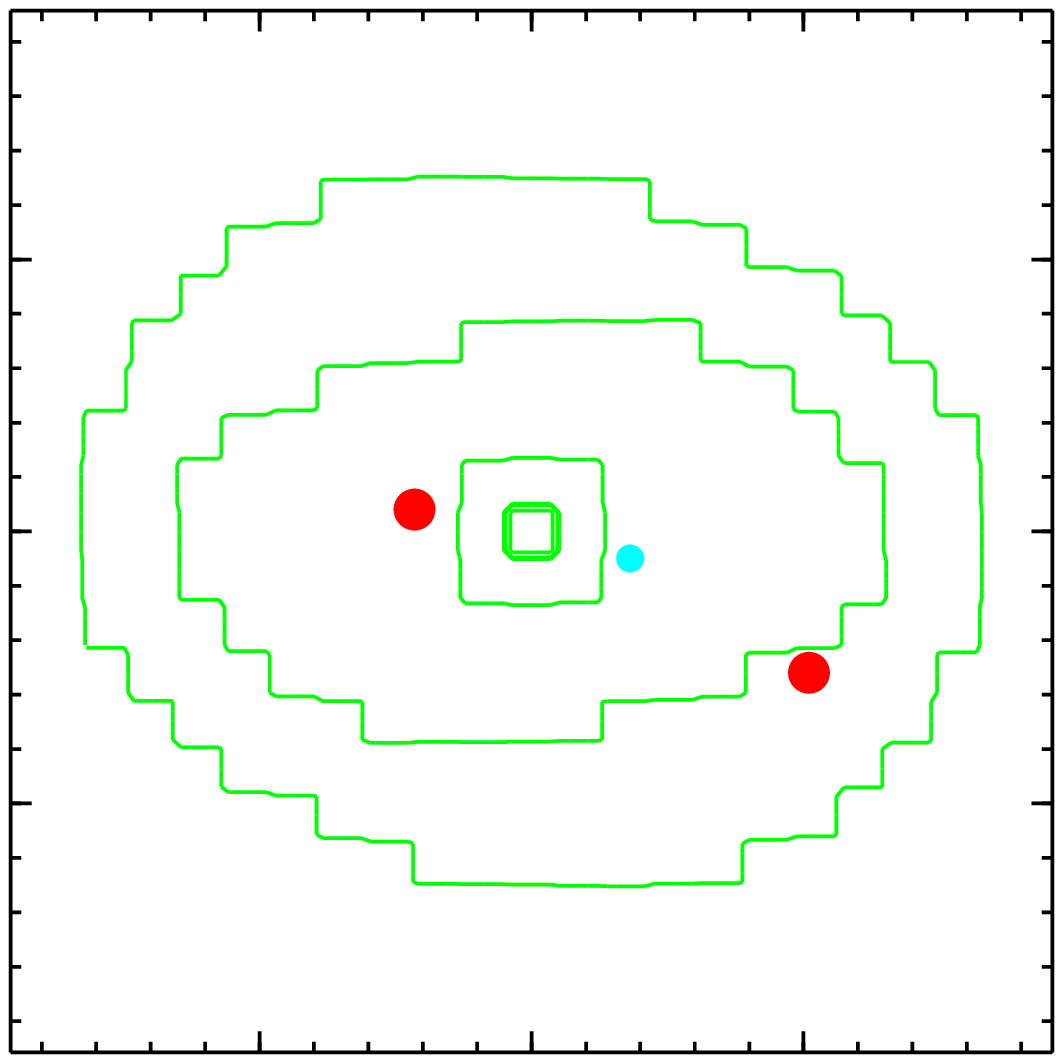}\\
\end{minipage}
\end{figure*}

\begin{figure*}
\begin{minipage}[l]{0.18\textwidth}
\small
\begin{tiny}
\begin{verbatim}
object J1636+470
pixrad 10 
redshifts 0.2282 0.6745	
quad
-0.080	1.730
0.88	-0.620 0
1.080	-0.280 0 
-0.770	-0.430 0 
g 13.7
\end{verbatim}
\end{tiny}\end{minipage}
\begin{minipage}[l]{0.15\textwidth}
\includegraphics[width=75pt]{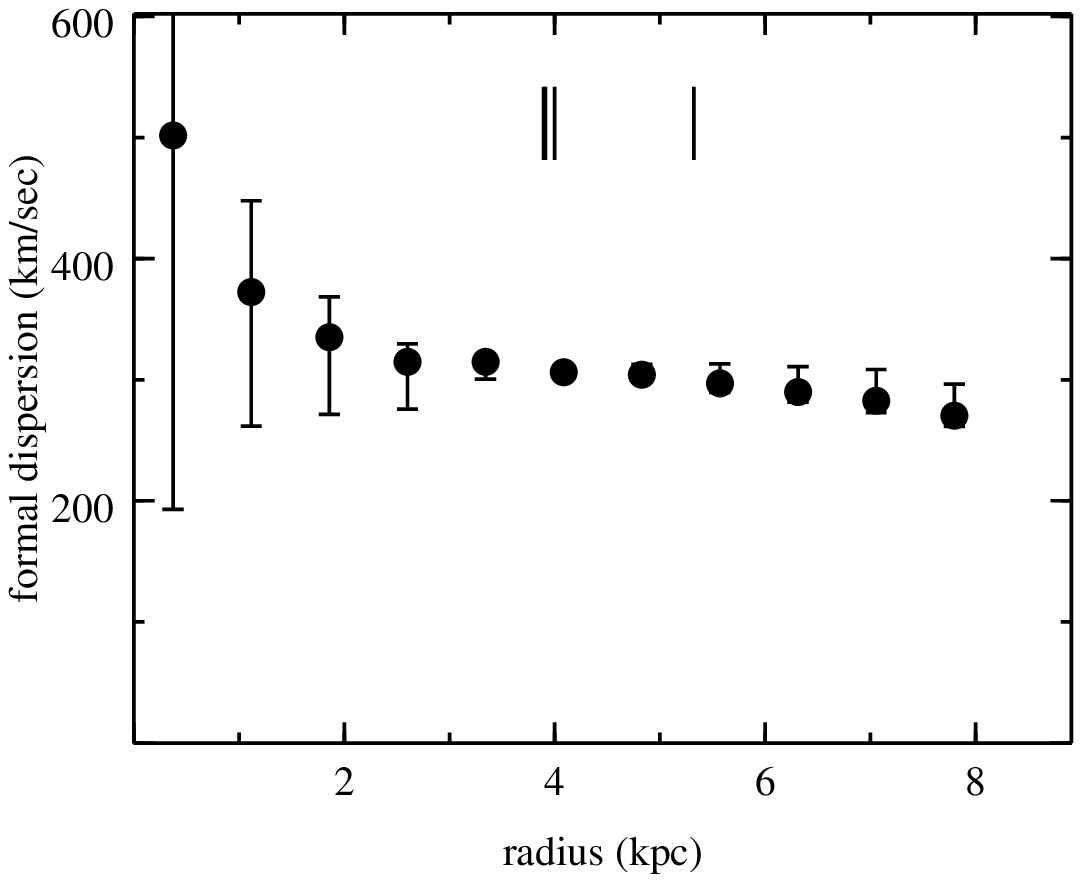}
\end{minipage}\begin{minipage}[r]{0.15\textwidth}
\includegraphics[width=52pt, bb = 0 15 320 320]{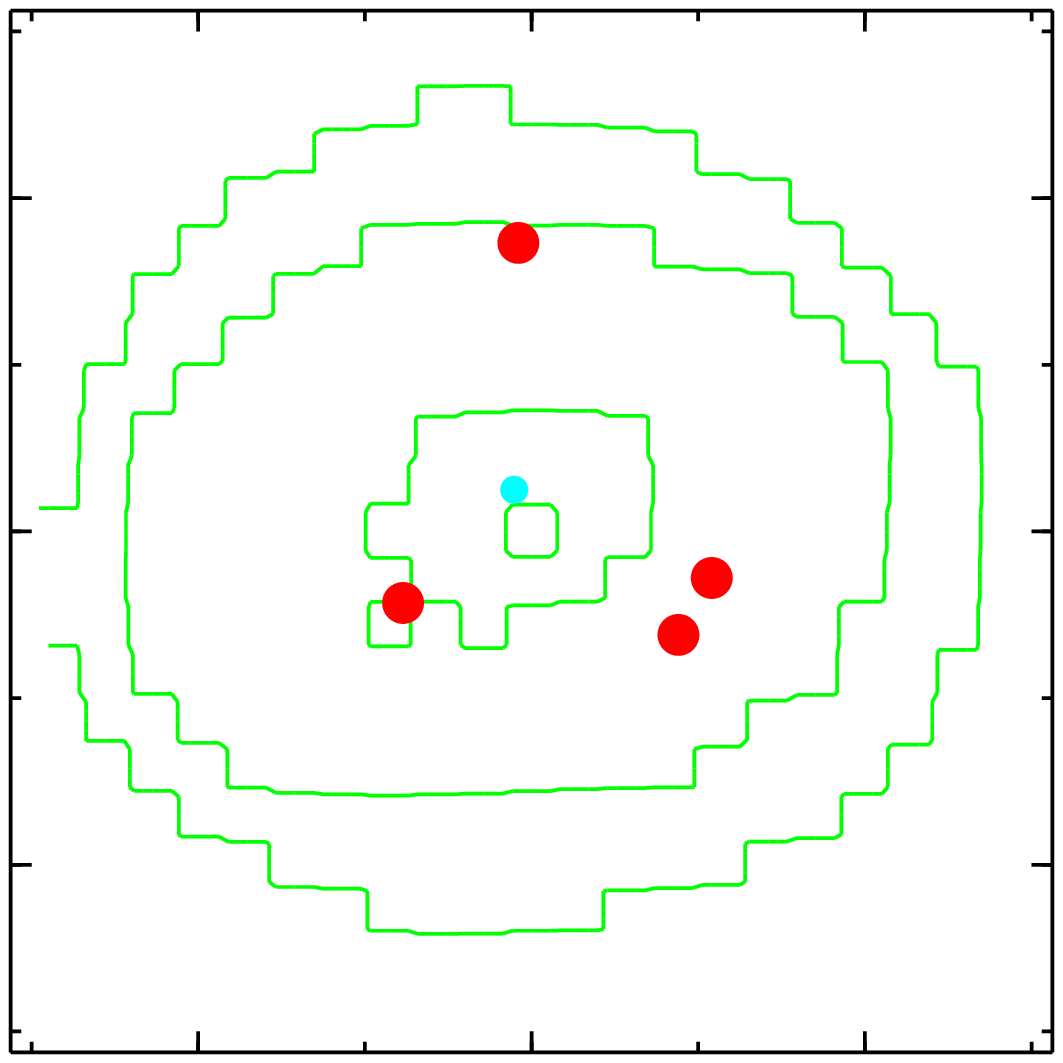}\\
\end{minipage}\begin{minipage}[l]{0.18\textwidth}
\small
\begin{tiny}
\begin{verbatim}
object J2300+002
symm pixrad 10 
redshifts 0.2285 0.4635	
quad
-0.120	1.480
-0.28	-1.080 0
-0.570	-0.930 0
0.120	-1.080 0
g 13.7
\end{verbatim}
\end{tiny}\end{minipage}
\begin{minipage}[l]{0.15\textwidth}
\includegraphics[width=75pt]{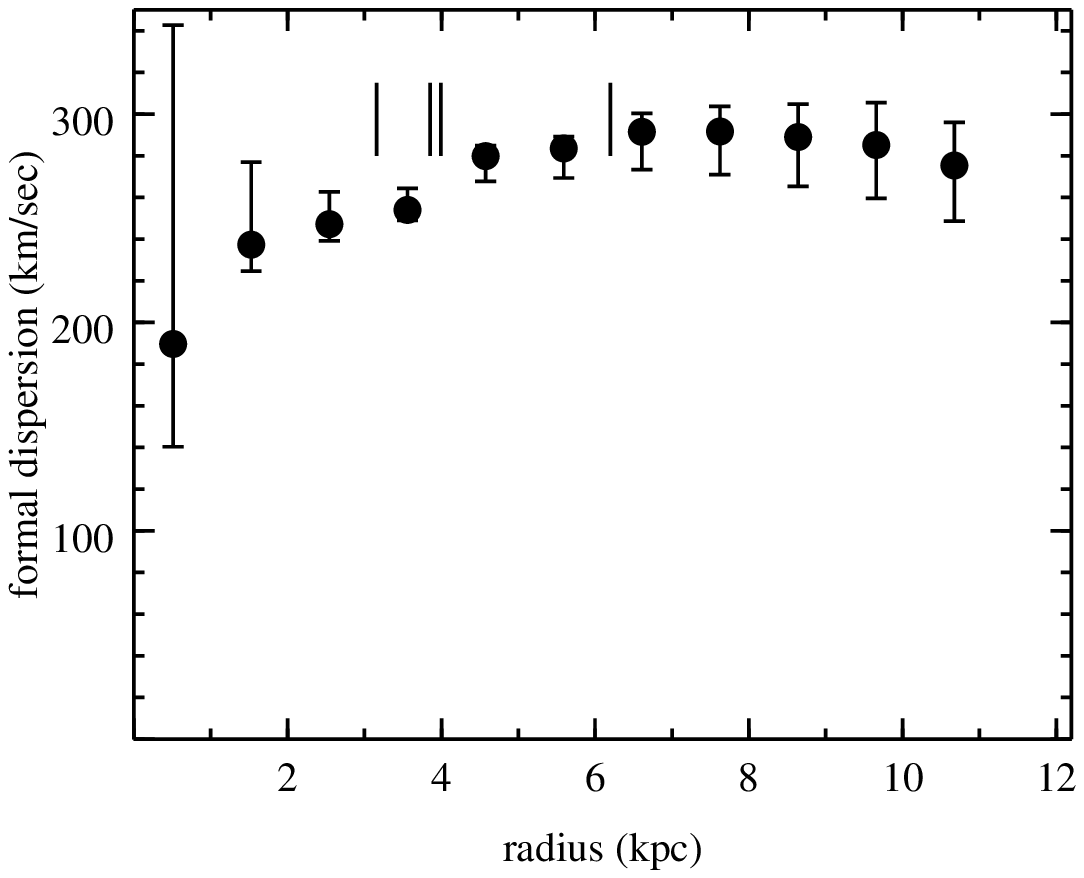}
\end{minipage}\begin{minipage}[r]{0.15\textwidth}
\includegraphics[width=52pt, bb = 0 15 320 320]{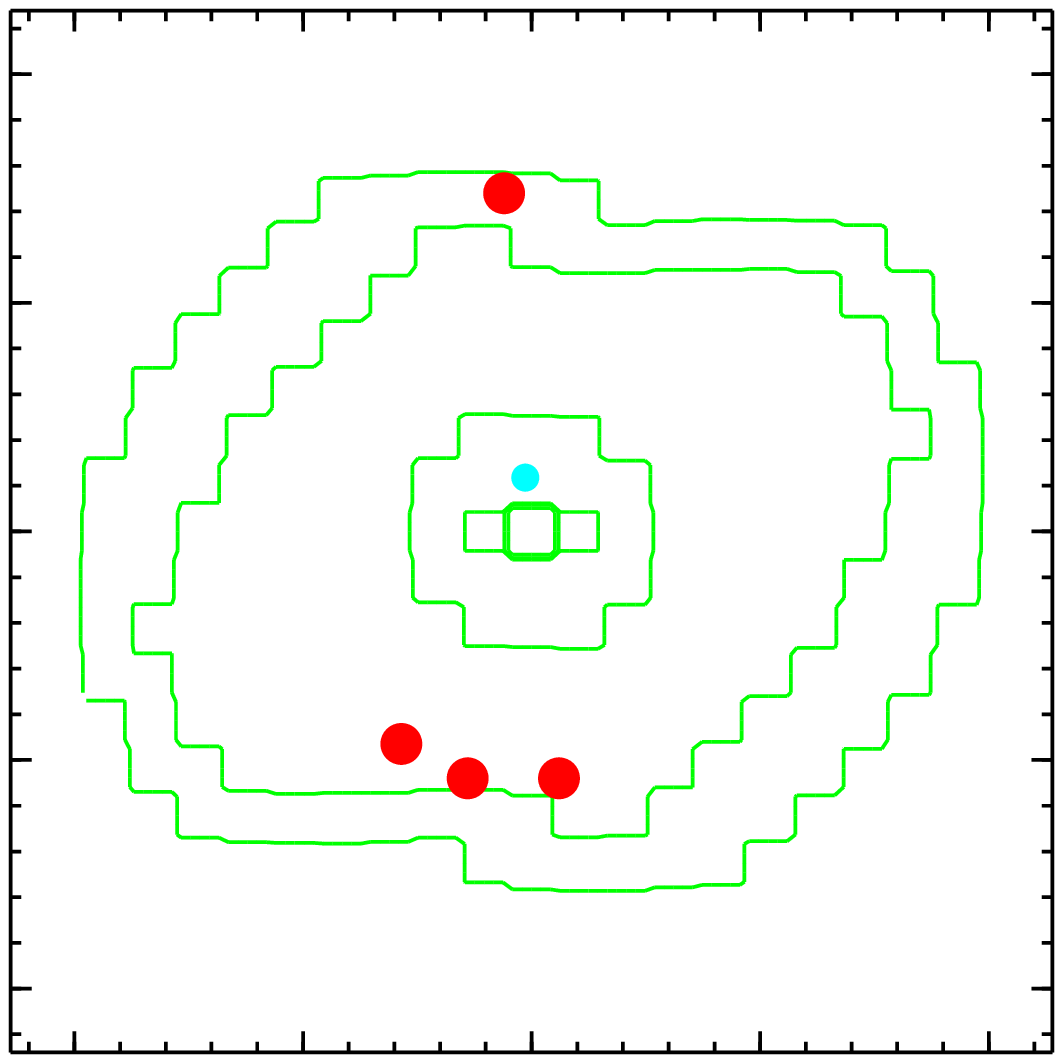}\\
\end{minipage}
\end{figure*}

\begin{figure*}
\begin{flushleft}
\hspace{0.15cm}
\begin{minipage}[l]{0.18\textwidth}
\small
\begin{tiny}
\begin{verbatim}
object J2303+142
pixrad 10 
redshifts 0.1553 0.517
double
-1.270	-1.580
-0.28	1.170 0
g 13.7
\end{verbatim}
\end{tiny}\end{minipage}
\begin{minipage}[l]{0.15\textwidth}
\includegraphics[width=75pt]{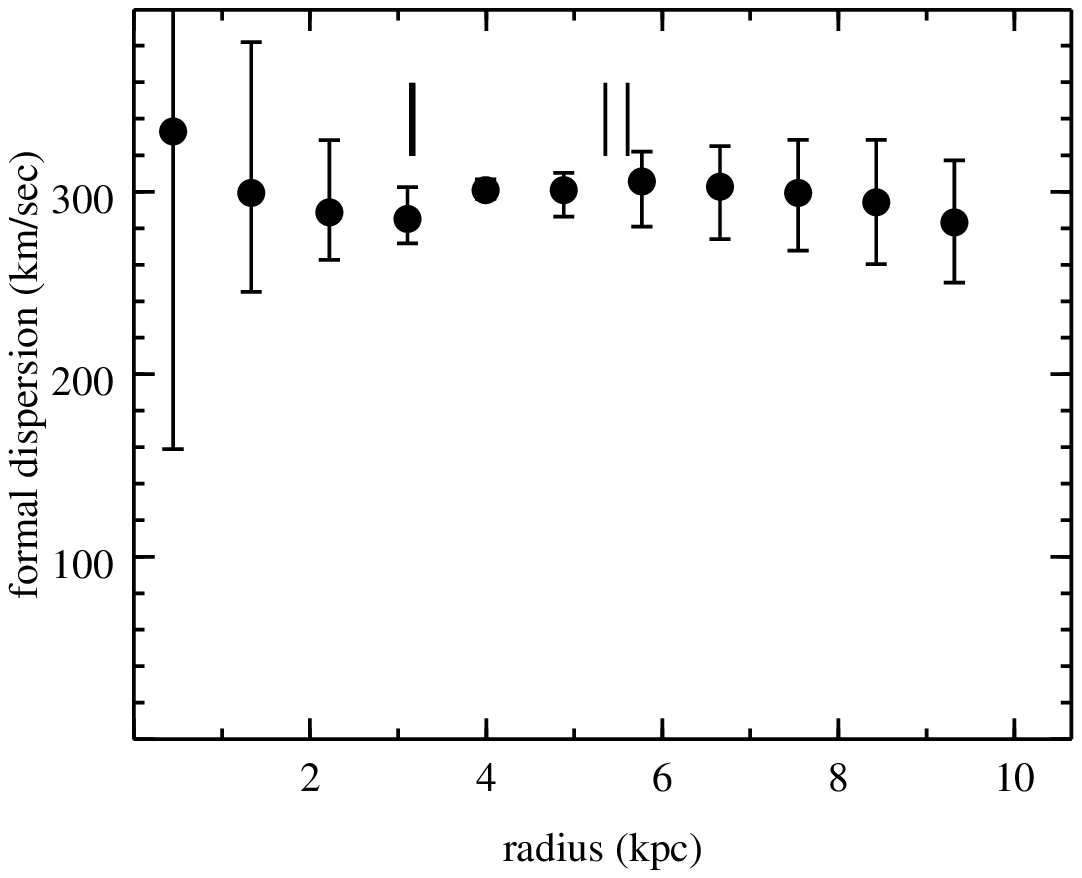}
\end{minipage}\begin{minipage}[r]{0.15\textwidth}
\includegraphics[width=52pt, bb = 0 15 320 320]{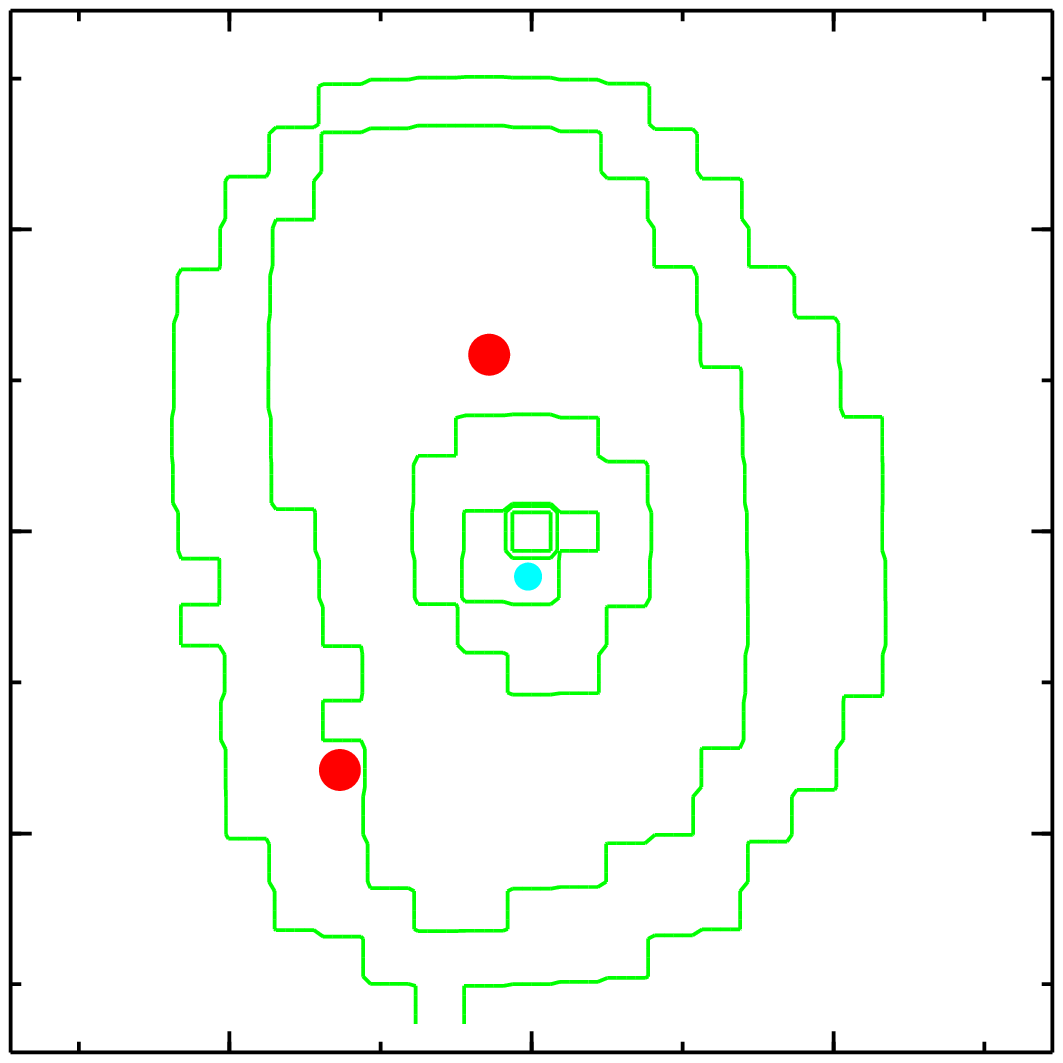}\\
\end{minipage}\begin{minipage}[l]{0.15\textwidth}
\small
\begin{tiny}
\end{tiny}\end{minipage}
\begin{minipage}[l]{0.15\textwidth}
\end{minipage}\begin{minipage}[r]{0.15\textwidth}
\end{minipage}
\end{flushleft}
\end{figure*}

\begin{figure*}
\caption{Lensing clusters used in this analysis. IMPORTANT NOTE: The y-axes of the velocity dispersion plots need to be multiplied by $\sqrt{2/\pi}\approx 0.8$ to yield the true $\sil$ values.}
\label{B4}
\vspace{0.5cm}
\begin{minipage}[l]{0.18\textwidth}
\small
\begin{tiny}
\begin{verbatim}
object ACO 1689
pixrad 18 
maprad 95 
minsteep 0 
shear 0 
g 13.7
models 100 
zlens 0.183 
\end{verbatim}
\end{tiny}\end{minipage}
\begin{minipage}[l]{0.15\textwidth}
\includegraphics[width=75pt]{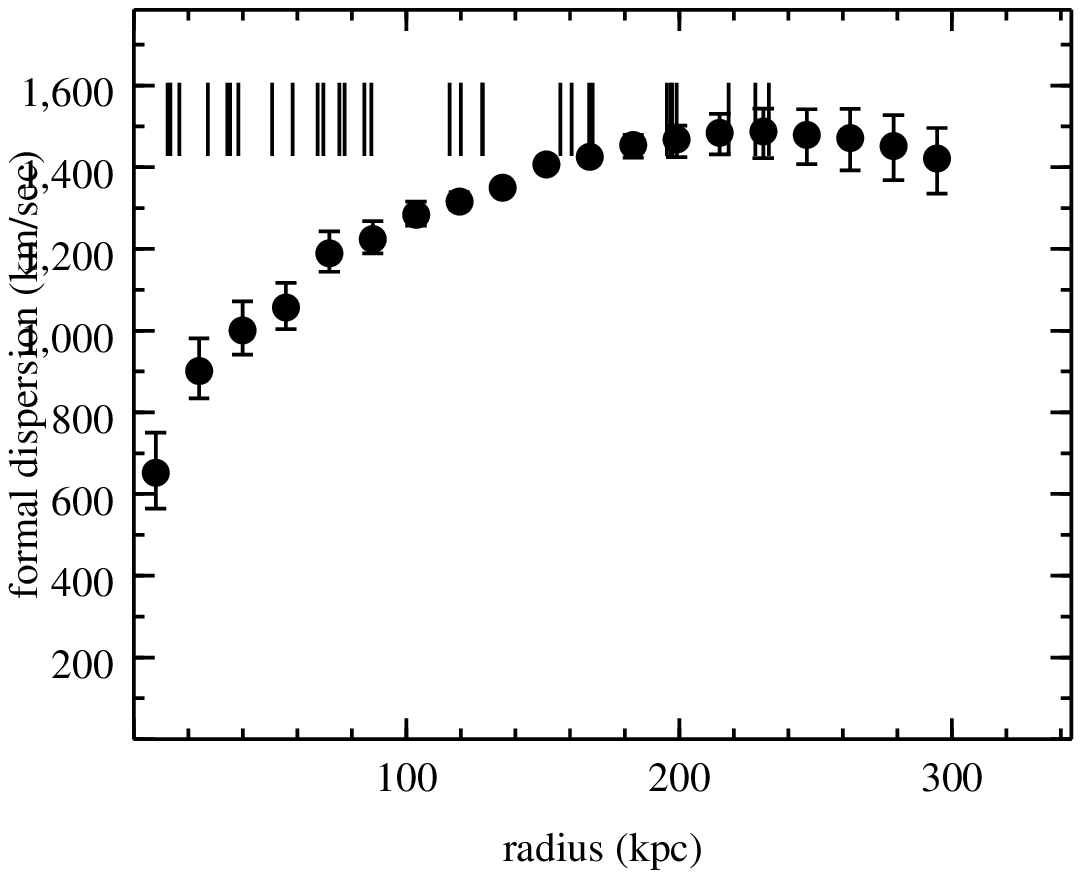}
\end{minipage}\begin{minipage}[r]{0.15\textwidth}
\includegraphics[width=52pt, bb = 0 15 320 320]{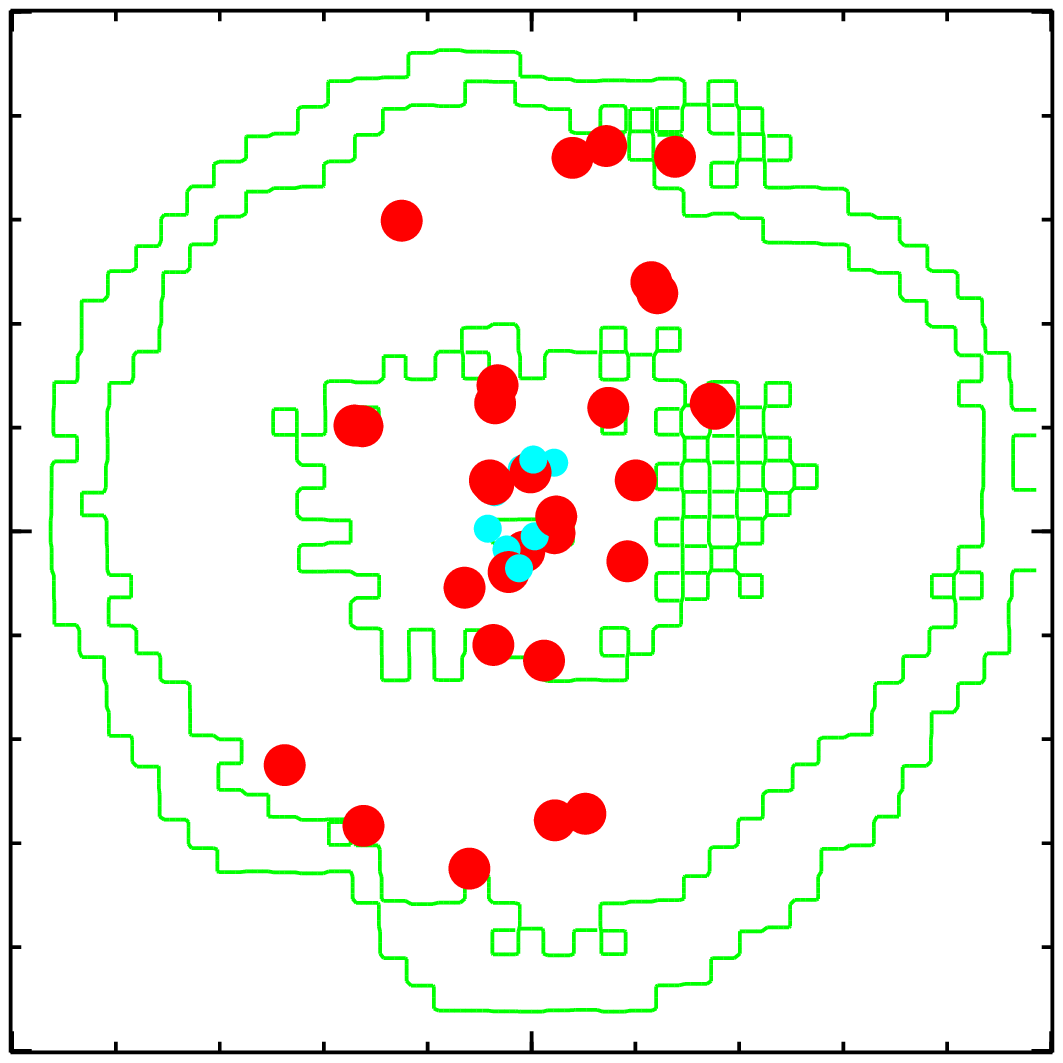}\\
\end{minipage}\begin{minipage}[l]{0.18\textwidth}
\small
\begin{tiny}
\begin{verbatim}
object ACO 2667
pixrad 10
minsteep 0
zlens 0.233
g 13.7
models 100
\end{verbatim}
\end{tiny}\end{minipage}
\begin{minipage}[l]{0.15\textwidth}
\includegraphics[width=75pt]{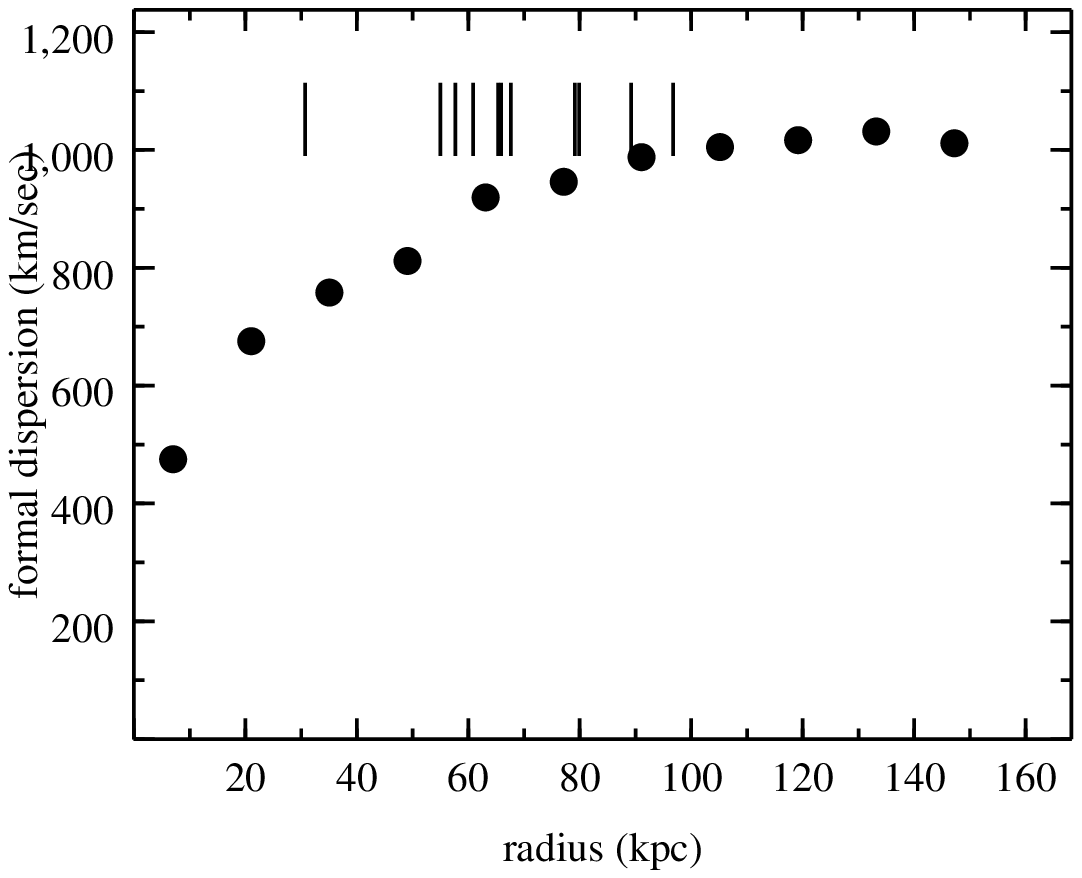}
\end{minipage}\begin{minipage}[r]{0.15\textwidth}
\includegraphics[width=52pt, bb = 0 15 320 320]{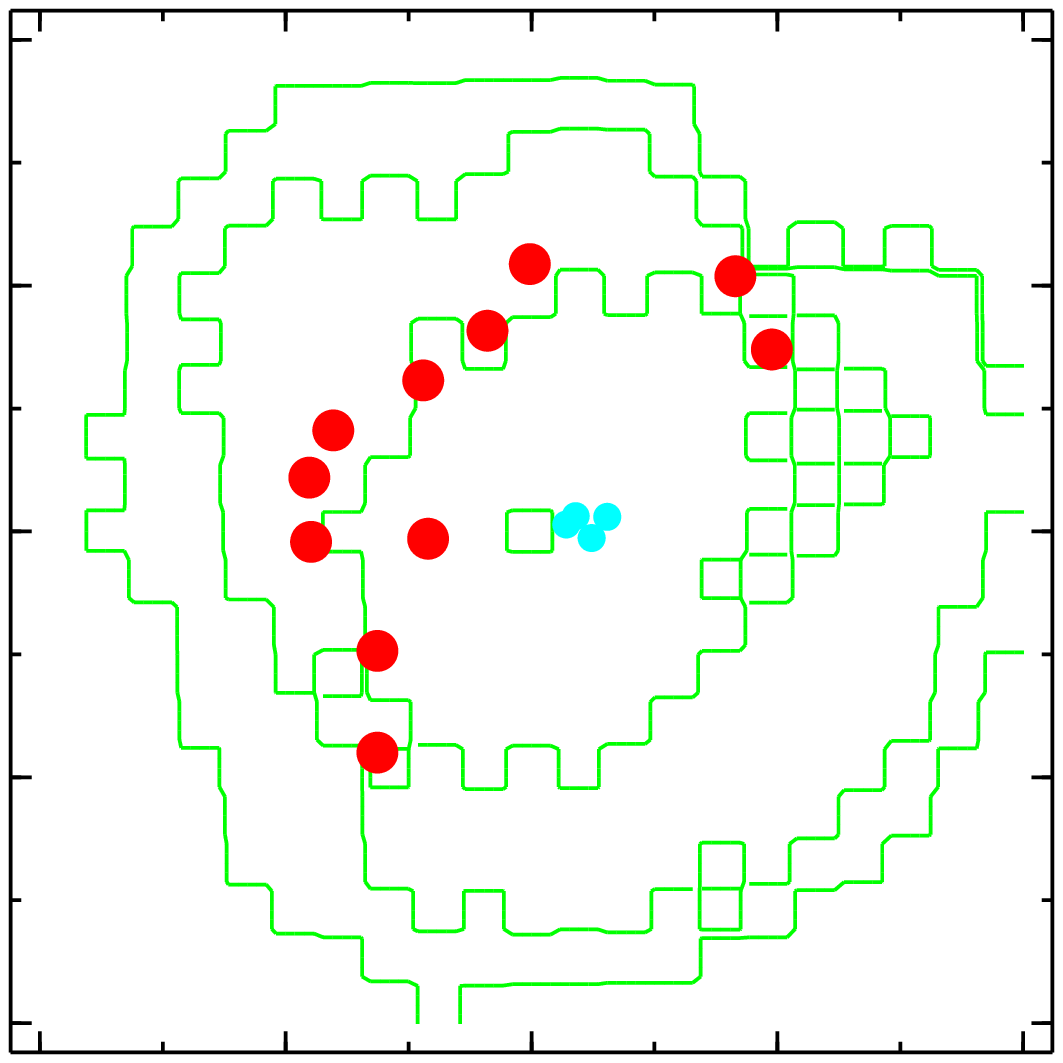}\\
\end{minipage}
\end{figure*}

\begin{figure*}
\begin{flushleft}
\hspace{0.15cm}
\begin{minipage}[l]{0.18\textwidth}
\small
\begin{tiny}
\begin{verbatim}
ACO 1689 - continued -
multi 5 2.54 
 10.38 -54.35  1 
 24.18  45.85  1 
 34.48  24.60  2 
-32.58  20.25  2 
 -8.03   9.85  3 
multi 3 1.99 
  7.88  71.90  1 
  2.38 -24.90  2 
 -1.43  -3.85  3 
multi 3 1.98 
-24.98  59.80  1 
 18.43  -5.75  2 
  4.42  -0.35  3 
\end{verbatim}
\end{tiny}\end{minipage}
\begin{minipage}[l]{0.15\textwidth}
\small
\begin{tiny}
\begin{verbatim}
multi 2 4.53 
 27.58  72.15  1 
-12.88 -10.80  2 
multi 3 1.74 
-32.33 -56.70  1 
 14.78  23.80  2 
 -0.18  11.35  3 
multi 5 2.99 
  4.48 -55.60  1 
 23.08  48.00  1 
 35.27  23.60  2 
-34.12  20.40  2 
 -7.33   9.10  3 
\end{verbatim}
\end{tiny}
\end{minipage}\begin{minipage}[l]{0.15\textwidth}
\small
\begin{tiny}
\begin{verbatim}
multi 3 4.92 
-47.53 -45.00  1 
 20.08   9.80  2 
  4.73   2.90  3 
multi 3 2.01 
 14.38  74.20  1 
 -7.38 -21.85  2 
 -4.38  -7.85  3 
multi 3 1.78 
-11.98 -64.90  1 
 -6.53  28.15  2 
 -7.03  24.65  3  
\end{verbatim}
\end{tiny}
\end{minipage}\begin{minipage}[l]{0.15\textwidth}
\small
\begin{tiny}
\begin{verbatim}
ACO 2667 - continued - 
multi 3 1.033
-17.94  -0.84 1
-3.60 16.33 1
-8.81 12.28  2
multi 3 1.20
16.59 20.77 1
-12.54 -9.71 2
-8.42 -0.58 3
multi 3 1.578
-0.13 21.74 1
-18.07 4.37 1
-16.14 8.23 2
multi 2 3.0
19.55 14.79 1
-12.54 -18.01 2
\end{verbatim}
\end{tiny}\end{minipage}
\begin{minipage}[l]{0.15\textwidth}
\end{minipage}\begin{minipage}[l]{0.15\textwidth}
\end{minipage}
\end{flushleft}
\end{figure*}

\end{document}